\documentclass[12pt,openany,oneside]{book}
\usepackage{amsmath,amsthm,amssymb} % Useful AMS stuff
\usepackage{graphicx}
\usepackage{ragged2e}
\usepackage{setspace}
\usepackage[top=1in, bottom=1.25in, left=1.5in, right=1in]{geometry}             
\usepackage[parfill]{parskip}   
\usepackage{fancyhdr}
\fancypagestyle{plain}{\pagestyle{fancy}}

\usepackage{hyperref}
\hypersetup{
	colorlinks=true,
	linkcolor=blue,
	filecolor=magenta,      
	urlcolor=cyan,
}

\usepackage{tabularx,ragged2e,booktabs,caption}
\newtheorem{theorem}{Theorem}[chapter]
\newtheorem{proposition}[theorem]{Proposition}

\newcommand{\cref}[1]{Corollary~\textup{\ref{#1}}}

\usepackage{float}

\makeatletter
\renewcommand\l@part{\@dottedtocline{-1}{0em}{2.3em}}
\newcommand{\chapfnt}{\fontsize{20}{20}}
\newcommand{\secfnt}{\fontsize{14}{17}}
\newcommand{\ssecfnt}{\fontsize{12}{14}}
\renewcommand\numberline[1]{\hb@xt@\@tempdima{#1.\hfil}}

\renewcommand\addvspace[1]{}

\makeatother
\makeatletter
\def\@makechapterhead#1{%
	\vspace*{15\p@}%
	{\parindent \z@ \raggedright \normalfont
		\ifnum \c@secnumdepth >\m@ne
		\chapfnt\bfseries \@chapapp\space \thechapter
		\par\nobreak
		\vskip 15\p@
		\fi
		\interlinepenalty\@M
		\chapfnt \bfseries #1\par\nobreak
		\vskip 20\p@
}}
\renewcommand\section{\@startsection {section}{1}{\z@}%
	{-3.5ex \@plus -1ex \@minus -.2ex}%
	{2.3ex \@plus.2ex}%
	{\normalfont\secfnt\bfseries}}
\renewcommand\subsection{\@startsection{subsection}{2}{\z@}%
	{-3.25ex\@plus -1ex \@minus -.2ex}%
	{1.5ex \@plus .2ex}%
	{\normalfont\ssecfnt\bfseries}}
\makeatother

\makeatletter
\newcommand{\thesisdate}[3]{\def\@year{#1}\def\@month{#2}}
\newcommand{\chair}[1]{\def\@chair{#1}}
\newcounter{members}% Counter for the number of committee members
\newcommand\member[1]{
	\ifcase\value{members}
	\def\@memberA{#1} \or \def\@memberB{#1} \or \def\@memberC{#1}  \or \def\@memberD{#1} 
	\fi
	\addtocounter{members}{1}
}

\newcommand{\maketitlepage}{
	\thispagestyle{empty}
	
	\vspace*{0pt}%Weird. This moves the thesis title 30pt down. Wonder why.
	\begin {center}
	\large
	\textbf{University of Nevada, Reno}
	\vskip 48pt plus0pt minus18pt
	
	\textbf{Generalized Univariate Distributions and A New Asymmetric Laplace Model}
	
	\vskip 48pt plus0pt minus18pt

	A thesis submitted in partial fulfillment of the \\
	requirements for the degree of Master of Science in \\
	Mathematics (Concentration in Statistics)   
	\vfill
	By\\    
	\@author \\
	Dr.~Tomasz Kozubowski/Thesis Advisor\\
	\@month\ \@year 
	\end {center}
}

\newcommand\makecopyright{
	\thispagestyle{empty}
	\newpage
	\thispagestyle{empty}
	\ \vfill
	\begin{center}
		\copyright~\@year \\
		\@author\\
		ALL RIGHTS RESERVED
\end{center}}

\newcommand{\makeacknowledgments}{
	\newpage
	\addcontentsline{toc}{part}{Acknowledgments}
	\begin{center}
		ACKNOWLEDGMENTS\\ 
	\end{center}\par}

\newcommand{\makeabstract}{
	\newpage
	\addcontentsline{toc}{part}{Abstract}
	\begin{center}
		ABSTRACT\\
		\@title\\
		By\\
		\@author
	\end{center}
	\par}

\makeatother

\author{Palash Sharma}% 
\title{\textbf{Generalized Univariate Distributions and A New Asymmetric Laplace Model}}% Mixed upper and lower case here
\thesisdate{2017}{June}{1}% {Year}{Month}{Date}
\member{~Dr.~Tomasz J. Kozubowski}
\member{~Dr.~Minggen Lu}
\member{~Dr.~Ilya  Zaliapin}
\member{~Dr.~Anna K. Panorska}

\begin{document}
	
	%%%%%%%%%%%%%%%%%%%%%%%%%%%%%%%%%%%%%%%%%%%
	% Approval Page GS-13
	%%%%%%%%%%%%%%%%%%%%%%%%%%%%%%%%%%%%%%%%%%%
	\maketitlepage
	\makecopyright
	\begin{titlepage}
		\begin{singlespace}
			
			\newpage
			\thispagestyle{empty}
			
			\includegraphics[scale=.1]{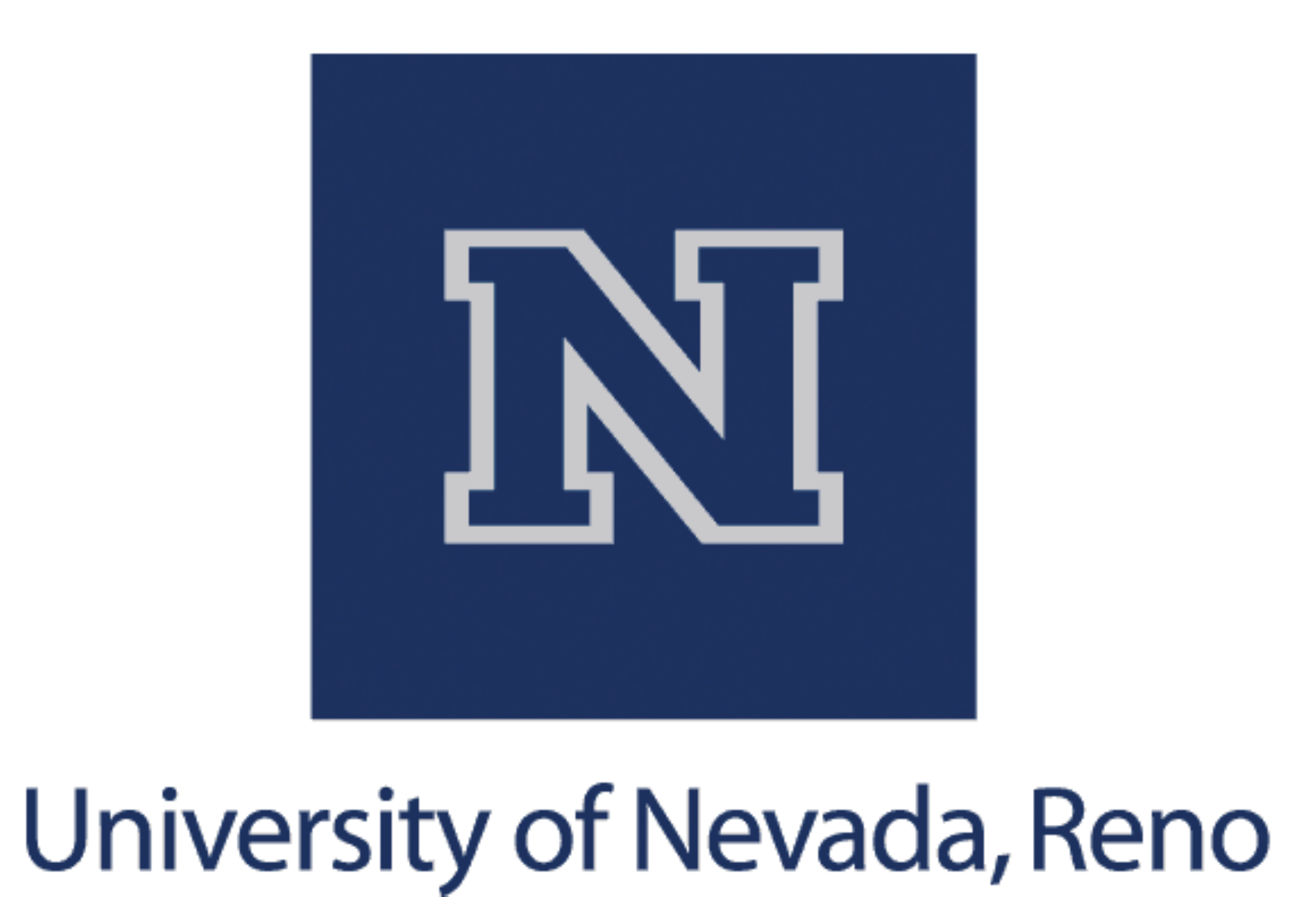}
			\hspace{1.cm} \large   \textbf{ \quad THE GRADUATE SCHOOL}
			
			\begin{center}
				\vspace{.5cm}
				We recommend that the thesis \\
				prepared under our supervision by
				
				\vspace{.5cm}
				\textbf{Palash Sharma}
				
				\vspace{.5cm}
				entitled
				
				\vspace{.5cm}
				\textbf{Generalized Univariate Distributions and A New Asymmetric Laplace Model}
				
				\vspace{.5cm}
				be accepted in partial fulfillment of the \\
				requirements for the degree of
				
				\vspace{.45cm}
				MASTER OF SCIENCE
				
				\vspace{.45cm}
				Tomasz J. Kozubowski, Ph.D., Advisor
				
				\vspace{.45cm}
				Anna Panorska, Ph.D., Committee Member
				
				\vspace{.45cm}
				Minggen Lu, Ph.D., Graduate School Representative
				
				\vspace{.45cm}
				David Zeh, Ph.D., Dean, Graduate School
				
				\vspace{.1cm}
				June, $1^{st}$ 2017
				
			\end{center}
		\end{singlespace}
	\end{titlepage}

	%%%%%%%%%%%%%%%%%%%%%%%%%%%%%%%%%%%%%%%%%%%
	% Abstract
	%%%%%%%%%%%%%%%%%%%%%%%%%%%%%%%%%%%%%%%%%%%
	\frontmatter% starts lowercase roman page numbering
	\makeabstract
	This work provides a survey of general class of distributions generated
	from the mixture of the beta random variables. We provide an extensive review of the literature, concerning generating new distributions via the inverse CDF transformation. In particular, we accounted for beta generated and Kumaraswamy generated families of distributions. We provide a brief summary of each of their families of distributions. We also propose a new asymmetric mixture distribution, which is an alternative to beta generated distributions. We provide basic properties of this new class of distributions generated from the Laplace model. We also address the issue of parameter estimation of this new skew generalized Laplace model.\\
	
	
	%\frontmatter% starts lowercase roman page numbering

	%%%%%%%%%%%%%%%%%%%%%%%%%%%%%%%%%%%%%%%%%%%
	% Copyright Page
	%%%%%%%%%%%%%%%%%%%%%%%%%%%%%%%%%%%%%%%%%%%

	%%%%%%%%%%%%%%%%%%%%%%%%%%%%%%%%%%%%%%%%%%%
	% Acknowledgments
	%%%%%%%%%%%%%%%%%%%%%%%%%%%%%%%%%%%%%%%%%%%
	
	\makeacknowledgments
	
	At first, I would like to thank my honorable thesis advisor, Professor Tomasz J. Kozubowski, who showed me a great interest in the field of statistics and probability theory. Professor Kozubowski not only helped me to complete the thesis but also encouraged me, supported me, guided me with great patience, which accelerated me to complete my graduate study in the department of Mathematics and Statistics at University of Nevada, Reno (UNR). I am really grateful to him. I am also very grateful to the graduate school representative, Dr. Minggen Lu, for his great support. A special thanks also goes to Professor Anna K. Panorska, who not only gave me the best suggestion but also carefully helped me academically to complete my graduate study at University of Nevada,Reno. I also would like to acknowledge my parent's support and encouragement. Moreover, I am very much thankful to all of the faculty members, graduate students, and staffs of the department of Mathematics and Statistics at University of Nevada, Reno.

	%%%%%%%%%%%%%%%%%%%%%%%%%%%%%%%%%%%%%%%%%%%
	% Table of Contents
	%%%%%%%%%%%%%%%%%%%%%%%%%%%%%%%%%%%%%%%%%%%
	
	\tableofcontents
	
	%%%%%%%%%%%%%%%%%%%%%%%%%%%%%%%%%%%%%%%%%%%
	% List of Tables
	% 
	% Delete the following two lines if your thesis has no tables.
	%%%%%%%%%%%%%%%%%%%%%%%%%%%%%%%%%%%%%%%%%%%
	\newpage\addcontentsline{toc}{part}{List of Tables}%
	\listoftables
	\addtocontents{lot}{\noindent Table\par}% Puts the word  "Table" in List of Tables
	
	%%%%%%%%%%%%%%%%%%%%%%%%%%%%%%%%%%%%%%%%%%%
	% List of Figures
	% 
	% Delete the following three lines if your thesis has no figures.
	%%%%%%%%%%%%%%%%%%%%%%%%%%%%%%%%%%%%%%%%%%%
	\newpage\addcontentsline{toc}{part}{List of Figures}%
	\listoffigures
	\addtocontents{lof}{\noindent Figure\par}% Puts the word  "Figure" in List of Figures

	%				\newpage\addcontentsline{toc}{part}{APPENDICES}
	%				\appendix 
	%				\addtocontents{lof}{\noindent appendix\par}
	%%%%%%%%%%%%%%%%%%%%%%%%%%%%%%%%%%%%%%%%%%%
	% Chapter 1
	%%%%%%%%%%%%%%%%%%%%%%%%%%%%%%%%%%%%%%%%%%%
	
	\mainmatter% starts arabic  page numbering 
	
	\addtocontents{toc}{\noindent Chapter\par}% Puts the word  "Chapter" in Table of Contents
	
	\chapter {\textbf{Introduction}}
	
	This thesis work offers a survey of recently developed families of probability distributions, which provide more flexibility in modeling data. The generalized distributions we shall review are obtained from standard probability distributions and a generating mechanism connected with a probability distribution on a unit interval, as described below.
	
	Let X be a continuous random variable with the cumulative distribution function (CDF) $F(x)$ and the probability density function (PDF) $f(x)$. Then, $X$ admits the well known representation,
	\begin{equation}
	\label{eq1.1}
	X \stackrel{d}{=} F^{-1} (U) ,
	\end{equation}
	known as the \textit{Probability Integral Transform Theorem}. The quantity $U$ in (\ref{eq1.1}) has a standard uniform distribution. Following this idea, one can generalize the distribution of $ X $ by defined a new random variable via
	\begin{equation}
	\label{eq1.2}
	Y=F^{-1} (T) ,
	\end{equation}
	where $T$ has a distribution on $(0,1)$ which is not necessarily a uniform distribution. The generalized PDF and CDF of Y will be denoted as $g(y)$ and $G(Y)$ respectively. Thus, we obtain a generalization of $ X $ via a generating mechanism connected with the random variable $T$ on $(0,1)$. If $T$ is a uniform random variable, then the generalization coincides with $X$. Such generalizations appeared in the literature in recent years, particularly with $T$  having beta, Kumaraswamy or truncated exponential distributions. Indeed, numerous new classes of distributions were obtained via (\ref{eq1.2}), which generalize many standard distributions, such as normal, exponential, Weibull, Pareto, among many others. The particular two schemes used in this connection are based upon two specific probability distributions of $T$ : the beta distribution and the Kumaraswamy distribution. One of the aims of this thesis is to review both of these schemes and gather information regarding specific distributions that are scattered in the literature. 
	
	Beta distribution is perhaps the most well-known and widely used continuous probability distribution on finite domain. It is most conveniently defined through its PDF, which is of the form
	\begin{equation}
	\label{eq1.3}
	h(x)=\frac{1}{B(\alpha,\beta)} x^{\alpha-1}(1-x)^{\beta-1} ,\; x \in (0,1),
	\end{equation}
	where $\alpha>0$ and $\beta>0$ are two shape parameters, while
	\begin{equation}
	\label{eq1.4}
	B(\alpha,\beta)=\frac{\Gamma(\alpha) \Gamma(\beta)}{\Gamma(\alpha+\beta)} , \;\alpha, \; \beta>0,
	\end{equation}
	is the beta function. The corresponding cumulative distribution function (CDF) generally does not admit a closed form, and is expressed through the \textit{incomplete beta function ratio},
	\begin{equation}
	\label{eq1.5}
	H(x)=I_x(\alpha,\beta)=\frac{1}{B(\alpha,\beta)} \int_{0}^{x} t^{\alpha-1}(1-t)^{\beta-1} dt, \; x \in[0,1],
	\end{equation}
	see Chapter 25 of Johnson et al. (1995). Since, as we should see in Chapter 2, the CDF of the generalized random variable $Y$ in (\ref{eq1.2}) is given by
	\begin{equation}
	\label{eq1.6}
	G(x)=H(F(x)) ,
	\end{equation} 
	the beta generated family of distributions, popularized by Eugene et al. (2002), will be of the form
	\begin{equation}
	\label{eq1.7}
	G(x)=\int_{0}^{F(x)} h(t)dt,
	\end{equation}
	where $h(t)$ is the PDF of the beta random variable (\ref{eq1.3}) and $F(x)$ is the base CDF. The PDF corresponding to (\ref{eq1.7}) is given by
	\begin{equation}
	\label{eq1.8}
	g(x)=\frac{1}{B(\alpha,\beta)} f(x) [F(x)]^{\alpha-1}[1-F(x)]^{\beta-1},
	\end{equation} 
	where $f(\cdot)$ is the PDF of the based distribution $F(\cdot)$. The beta generated method provides a convenient way to generate new distributions, and many distributions were obtained using this scheme, as we shall see in Chapter 3. These beta generating distributions shall be termed as $BG$ distributions.				
	
	Another popular method of generating new classes of distributions, which we shall review in Chapter 4, is connected with $T$ in (\ref{eq1.2}) having the so called Kumaraswamy distribution, given by the CDF
	\begin{equation}
	\label{eq1.9}
	H(x)=1-\left(1-x^a\right)^b , \; x\in [0,1],
	\end{equation} 
	where $a,b>0$ are two shape parameters (see, Kumaraswamy, 1980). The corresponding PDF is of the form
	\begin{equation}
	\label{eq1.10}
	h(x)=ab\; x^{\alpha-1} \;(1-x)^{\beta-1}, \; x\in (0,1).
	\end{equation} 
	Among several advantages of this model over the beta distribution, discussed in Jones (2009), is the fact that no special functions are required to describe its CDF. Consequently, the Kumaraswamy generalized distributions obtained via (\ref{eq1.2}) and popularized by Cordeiro and de Castro (2011), generally do not require special functions to describe their CDFs, given by 
	\begin{equation}
	\label{eq1.11}
	G(x)=1-[1-F^a(x)]^b,
	\end{equation}
	where  $a>0$ and $b>0$. The corresponding PDF will then be of the form
	\begin{equation}
	\label{eq1.12}
	g(x)=ab \;f(x)F^{a-1}(x)\;[1-F^a(x)]^{b-1},
	\end{equation}
	where $f(\cdot)$ is the PDF of the based distribution $F(\cdot)$. The Kumaraswamy generated method provides a convenient way to generate new family of distributions as we shall see in Chapter 4. These Kumaraswamy generated distributions shall be termed as $Kum-G$ distributions.	
	
	Our thesis is organized as follows. In Chapter 2, we review the general method discussed above for generating new classes of distributions via the inverse CDF transformation. In Chapter 3, we present numerous beta generated family of distributions. Then, in Chapter 4, we present Kumaraswamy generated family of distributions. Finally, in Chapter 5, we present a new scheme, where the generator T is a mixture of two special beta distributions. We discuss general properties of the obtained families, and focus on one particular case involving a new generalized asymmetric Laplace distribution.

	\chapter{\textbf{Generalized Distributions Via Inverse CDF Transformation}}
	Let $X$ be a continuous random variable with CDF $F(X)$ and PDF $f(x)$. Then, a new distribution can be defined as 
	\begin{equation}
	\label{2.1}
	Y=F^{-1}(T),
	\end{equation}
	where $T$ is a random variable on $(0,1)$ with the CDF $H(\cdot)$. This general scheme was discussed in Ferreira and Steel (2006), with the view towards obtaining skew generalizations of symmetric $F(\cdot)$.
	
	We present several properties of this construction, assuming that the base CDF $F(\cdot)$ is continuous and increasing on the support of the distribution and the distribution of the generating $T$ is absolutely continuous with PDF $h(\cdot)$.
	\section{Basic Properties of Generalized Distributions }
	We start with the following, fundamental results.
	\begin{proposition}
		\label{prop2.1}
		In the above notation, the CDF of $Y$ in (\ref{2.1}) is given by
		\begin{equation}
		\label{2.2}
		G(y)=H(F(y)),
		\end{equation}
		while the PDF of $Y$ in (\ref{2.1}) is 
		\begin{equation}
		\label{2.3}
		g(y)=h(F(y))f(y) ,
		\end{equation}
		where $f(y)$ is the PDF corresponding to $F$. 
	\end{proposition}
	The above result easily follows when we note that
	$G(y)  = Pr(Y \le y) =Pr(F^{-1}(T) \le y)=Pr(T \le F(y)) =H(F(y)).$
	The PDF of $Y,$ is obtained by straightforward differentiation,
	$g(y) = h(F(y)) f(y) .$
	The following result, discussed in Ferreira and Steel (2006), is straightforward to establish as well.
	\begin{proposition}
		\label{prop2.2}
		If F is  a symmetric distribution about zero and H is a symmetric distribution on (0,1) about 1/2, then the generalized distribution G is also symmetric about 0.
	\end{proposition}
	Indeed, assume that $X$ is a continuous random variable with PDF and CDF $f(x)$ and $F(x)$, respectively. Since $F$ is symmetric about zero, we have $f(x)=f(-x)$. Also, we know that $H$ is symmetric on $(0,1)$, so that $h(t)=h(1-t)$ for $t \in (0,1)$.
	
	We need to show that generalized distribution is also symmetric, i.e. $g(x)=g(-x)$.
	
	Since $ g(x)=h(F(x))f(x)$, we have $g(-x)=h(F(-x))f(-x)$. But  
	$X$ is a symmetric random variable, so that  $Pr(X \le -x)=1-Pr(X \le x)$, which means that $F(-x)=1-F(x)$.
	
	Thus, $g(-x)=h(1-F(x))f(x)=h(F(x))f(x)=g(x)$. It follows that the generalized distribution $G$ is also a symmetric distribution about $0$, which concludes the argument.
	
	\textbf{Remark:}  If we want to generalize a symmetric distribution into a skew one via (\ref{eq1.2}), we need to take an asymmetric generator $T$.
	
	The third result taken from Ferreira and Steel (2006) as well, concerns modality of the generalized distribution.
	\begin{proposition}
		\label{2.3}
		If F has a symmetric and unimodal distribution with the mode $y_{0}$, and if H is unimodal distribution with the mode at 1/2, then generalized distribution G is also unimodal and its mode is at $y_{0}$.
	\end{proposition}
	As discussed in the above paper, this property is useful is case of mode preserving skewing mechanisms. The unique mode of the generalized distribution $G$ equals to that of the distribution of $F$. This mode is unaffected by its degree of skewness and helps modeling because the location parameter or regression function can be interpreted as the mode. A choice of $H$ with a unique mode at $1/2$ will ensure that.	
	
	The next result, taken from Ferreira and Steel (2006) as well, discuss the existence of moments.
	\begin{proposition}
		\label{prop2.4}
		If limits of h(x) when x tends to 0 and 1 are finite and nonzero, then the moment existence of G is equal to that of F.
	\end{proposition}
	While all of the above properties were discussed in Ferreira and Steel (2006), the following facts appear to be new and play an important role in the theory of generalized distributions defined by (\ref{eq1.2}).
	
	%	\begin{proposition}
	%		\label{prop2.5}
	%		Generalized distribution holds commute property.
	%	\end{proposition}
	%
	%	Proof: Let X has a distribution with PDF f(x) and CDF F(x). Also assume that $T_1,T_2,.......T_n $ be n random variables on the support (0,1). Then the generalized form of $T_1,T_2,.......T_n $ using $F(.)$ can be written as $G_1,G_2,.......G_n $ where each
	%	$$ G_{i}(y)=H_i(F(y))$$ Then the mixture of this generalized distribution can be written as 
	%	\begin{equation}
	%	\label{2.4}
	%	\begin{split}
	%	G(y) & =\sum_{i=1}^{n}p_{i} G_{i}(y) \\
	%	& = \sum_{i=1}^{n}p_{i} H_{i}(F(y)) 
	%	\end{split}
	%	\end{equation} 
	%
	%	Where $\sum_{i=1}^{n}p_{i}=1 $ and $G_{i}$ is the generalized form of $T_{i}.$ 
	%	
	%	Similarly, Let $H_1,H_2,.......H_n $ be the CDF of $T_1,T_2,.......T_n $. Then the mixture of $T_1,T_2,.......T_n $ can be written as in the following form:
	%	
	%	\begin{equation}
	%	\label{2.5}
	%	H(F(y))=\sum_{i=1}^{n}p_{i} H_{i}(F(y)) .
	%	\end{equation}
	%
	%	Then according to the scheme, The generalization of T using $F(.)$ is
	%	
	%	\begin{equation}
	%	\label{2.6} 
	%	\begin{split}
	%	G(y) & =H(F(y)) \\
	%	& = \sum_{i=1}^{n}p_{i} H_{i}(F(y))  
	%	\end{split}
	%	\end{equation} 
	%
	%	So commute property holds for generalization of the distributions.
	%	% \; and \quad is used for spacing
	%	
	Namely, we shall consider the effect of mixing on the generalized distribution obtained by (\ref{eq1.2}), or equivalently through (\ref{eq1.6}).
	
	Let $T_1,T_2,\cdots, T_n$ be $n$ random variables on $(0,1)$, with respect to CDFs $H_1,H_2,\cdots, H_n$ and PDFs $h_1,h_2,\cdots, h_n$. Next, define a new CDF via
	\begin{equation}
	\label{a1}
	H(x)=\sum_{i=1}^{n} p_i \; H_i(x) \; ,
	\end{equation}
	where $p_i>0$ and  
	\begin{equation}
	\label{a2}
	\sum_{i=1}^{n} p_i=1 \; ,
	\end{equation}
	called a mixture of the $\{H_i\}$. With this set-up, we now consider generalized distributions based on a given CDF $F$ obtained via (\ref{eq1.6}) through each of the $\{H_i\}$, that is we obtain $n$ new generalizations of $F$ as follows :
	\begin{equation}
	\label{a3}
	G_i(y)=H_i(F(y)),\; i=1, \cdots,n.
	\end{equation}
	On the other hand, we can also generalize $F$ via (\ref{eq1.6}) using a random variable $T$ with the CDF $H$ given by (\ref{a1}). In this case, the generalized CDF will be of the form
	\begin{equation}
	\label{a4}
	G(y)= H(F(y))\; .
	\end{equation}
	However, when we take into account the structure of $H$ provide by (\ref{a1}), we can write the CDF $G$ is (\ref{a4}) as follows :
	\begin{equation}
	\label{a5}
	G(y)=\sum_{i=1}^{n} p_i \; H_i(F(y))\;.
	\end{equation}
	When we now recall the relation (\ref{a3}), we consider that
	\begin{equation}
	\label{a6}
	G(y)=\sum_{i=1}^{n} p_i \; G_i(y)\;.
	\end{equation}
	In other words, the operations of obtaining a generalized version of $F$ via (\ref{eq1.6}) form a given generator $T \sim H$ and the operation of mixing, commute. That is a generalized distributions obtained from $T$ that is a mixture, is also a mixture of generalized distributions, with the same weights. We formulate this results below.  
	\begin{proposition}
		\label{2.5}
		Let $T_1,T_2,\cdots, T_n$ be $n$ random variables on $(0,1)$, with CDFs $H_1,H_2,\cdots, H_n,$ and define a new random variable $T$ to be a mixture of the $\{T_i\}$, where the CDF of $T$ is given by $(\ref{a1})-(\ref{a6})$. Then, a generalization of $F$ obtained via $(\ref{eq1.6})$ through $T \sim H$ is a mixture of $Y_i \sim G_i$, where each $Y_i$ is a generalized version of $X\sim
		F$ obtained via (\ref{eq1.6}) through $T_i \sim H_i$.
	\end{proposition}
	This result implies that one can generate new classes of generalized distributions via mixing two or more given families of generalized distributions. We shall follow this approach in Chapter 5, where we develop a new skew Laplace model by this method.

	\chapter{\textbf{Generalized Beta Family of Distributions}}
	As discussed in the introduction, beta distribution is a continuous probability distribution with two positive shape parameters, $\alpha$ and $\beta$. It is the conjugate prior of the binomial distribution. It is also a natural extension of the uniform distribution. One of the attractive features of beta distribution is that one is able to rescale and shift the beta distribution to create a new distributions with a wide range of shapes and, as a result, it has been used for a variety of applications. The PDF and the CDF of beta distribution are
	\begin{equation}
	\label{3.1}
	h(x)=\frac{1}{B(\alpha,\beta)}x^{\alpha-1}(1-x)^{\beta-1}, \;  0<x<1,\;  \alpha >0 , \; \beta>0 \; ,
	\end{equation}
	and
	\begin{equation}
	\label{3.2}
	H(x)=\frac{1}{B(\alpha,\beta)}\int_{0}^{x}t^{\alpha-1}(1-t)^{\beta-1}dt,  \;0<t<1 ,\; \alpha>0, \; \beta>0 \; ,
	\end{equation}
	respectively, where $B(\alpha,\beta)$ is  defined by (\ref{eq1.4}). The CDF for beta generalized distributions, as discussed in Eugene et al. (2002), is of the form
	\begin{equation}
	\label{3.3}
	G(x)=\frac{1}{B(\alpha,\beta)}\int_{0}^{F(x)}t^{\alpha-1}(1-t)^{\beta-1} dt \; .
	\end{equation}
	The corresponding  PDF, obtained by taking the derivative in (\ref{3.3}), can be expressed as follows :
	\begin{equation}
	\label{3.4}
	g(x)  = \frac{f(x)}{B(\alpha,\beta)}F(x)^{\alpha-1}[1-F(x)]^{\beta-1}\; ,
	\end{equation}
	where $f(x)$ is the density function of the parent distribution $F(\cdot)$. The above work by Eugene et al. (2002) introduced this general family that triggered other authors to consider generalized distributions with some statistical applications. We will now review major contributions to this area and present relevant results from the literature.
	\begin{table}[ht]
		\caption{Summary of the literatures related to beta generated family of distributions}
		% title of Table
		\centering 
		% used for centering table
		\begin{tabular}{|p{1.5cm}|p{6cm}| p{8cm}|}
			% centered columns (4 columns)
			\hline
			\hline                        %inserts double horizontal lines
			Number & Name of distribution & Author(s) name \\[0.5ex]
			% inserts table 
			%heading
			\hline
			% inserts single horizontal line
			1 & Beta normal  & Eugene et al. (2002)\\ 
			&  &  Famoye et al. (2004)  \\
			&  &  Gupta and Nadarajah (2004) \\
			&  & Jones (2004) \\
			&  &  R\'ego et al. (2012) \\
			2 & Beta exponential  &  Nadarajah and Kotz (2006) \\ 
			3 & Beta gamma  &  Kong et al. (2007) \\
			
			4 & Beta Gumbel  &  Nadarajah and Kotz (2004) \\ 
			
			5 & Beta Fr\'echet  &  Nadarajah and Gupta (2004) \\ 
			&  &  Barreto-Souza et al. (2011) \\
			6 & Beta Weibull  &  Famoye et al. (2005) \\
			&  & Lee et al. (2007) \\
			&  &  Mahmoud and Mandouh (2012) \\
			7 & Beta Bessel  & Gupta and Nadarajah (2006) \\
			8 & Beta Pareto &  Akinsete et al. (2008) \\
			9 & Beta Rayleigh  &  Akinsete and Lowe (2009) \\
			10 & Beta Laplace  &  Kozubowski and Nadarajah (2008) \\
			&  & Cordeiro and Lemonte (2011) \\
			11 & Beta generalized logistic-IV  &  Morais et al. (2013) \\
			12 & Beta modifid Weibull  &  Silva et al. (2010) \\
			&  &  Nadarajah et al. (2012) \\
			13 & Beta generalized half-normal  &  Pescim et al. (2010) \\
			14 & Beta generalized exponential  &  Barreto-Souza et al. (2010) \\
			15 & Beta Maxwell  &  Amusan (2010) \\
			16 & Beta hyperbolic secant  &  Fischer and Vaughan (2010) \\
			17 & Beta inverse Weibull  &  Hanook et al. (2013) \\
			18 & Beta Cauchy  &  Alshawarbeh et al. (2012) \\
			19 & Beta half-Cauchy  &  Cordeiro and Lemonte (2011) \\
			20 & Beta Burr XII  &  Parana  et al. (2011) \\
			21 & Beta generalized Pareto  &  Mahmoudi (2011) \\
			&  &  Nassar and Nada (2011) \\
			22 & Beta Birnbaum-Sanders  &  Cordeiro and Lemonte (2011) \\
			23 & Beta skew-normal  &  Mameli (2012) \\
			24 & Beta exponential-geometric  &  Bidram (2012) \\
			25 & Beta Moyal & Cordeiro et al. (2012) \\
			26 & Beta generalized Weibull  &  Singla et al. (2012) \\
			27 & Beta exponentiated Pareto  &  Zea et al. (2012) \\
			28 & Beta power distribution &  Cordeiro and Brito (2012) \\
			29 & Beta linear failure rate  &  Jafari and Mahmoudi (2012) \\
			30 & Beta extended Weibull  &  Cordeiro et al. (2012) \\
			31 & Beta truncated Pareto  & Lourenzutti et al. (2012) \\
			32 & Beta Weibull-geometric  &  Cordeiro et al. (2013) \\
			&  &  Bidram et al. (2013) 
			\\[1ex]
			% [1ex] adds vertical space
			\hline
			%inserts single line
		\end{tabular}
		\label{table2}
		% is usedto refer this table in the text
	\end{table}

	\begin{table}[ht]
		\caption{Table (\ref{table2}) (continue)}
		% title of Table
		\centering 
		% used for centering table
		\begin{tabular}{|p{1.5cm}|p{6cm}| p{8cm}|}
			% centered columns (4 columns)
			\hline                       %inserts double horizontal lines
			Number & Name of distribution & Author(s) name \\[0.5ex]
			% inserts table 
			%heading
			\hline
			33 & Beta generalized gamma  &  Cordeiro et al. (2013) \\
			34 & Beta log-normal &  Castellars et al. (2013) \\
			35 & Beta generalized Rayleigh  &  Cordeiro et al. (2013) \\
			36 & Beta generalized logistic  &  Morais et al. (2013) \\
			37 & Beta exponentiated Weibull  & Cordeiro et al. (2013) \\
			38 & Beta Nakagami  &  Shittu and Adepoju (2013) \\
			39 & Beta Burr III  &  Gomes et al. (2013) \\
			40 & Beta Dagum  &  Domma and Condino (2013) \\
			41 & Beta Stoppa  &  Mansoor (2013) \\
			42 & Beta inverse Rayleigh  & Le\~ao et al. (2013) \\ 
			43 & Beta generalized inverse Weibull  & Baharith et al. (2014) \\
			44 & Beta extended half-normal  & Cordeiro et al. (2014) \\
			45 & Beta log-logistic  &  Lemonte (2014) \\
			46 & Beta Lindley  & Merovci and Sharma (2014) \\
			47 &  Beta-Fisher Snedecor  & Adepoju et al. (2015)\\
			48 & Beta Nadarajah-Haghighi  & Dias et al. (2016) \\
			49 & Beta-Gompertz  & Jafari et al. (2014) \\
			50 & Beta-weighted Weibull  &  Idowu et al. (2013) \\
			51 & Beta Gompertz Mekaham  & Chukwu and Ogunde (2015) \\
			52 & beta-geometric  & Weinberg and Gladen (1986) \\
			&  &  Oluyede et al. (2016) \\
			53 & Beta Lomax   & Rajab et al. (2013) \\
			& & Javanshiri and Maadooliat (2014) \\
			54 & beta Burr type X  & Merovci et al. (2016) 
			\\[1ex]
			% [1ex] adds vertical space
			\hline
			%inserts single line
		\end{tabular}
		\label{table3}
		% is usedto refer this table in the text
	\end{table}

	\section{Beta-Normal Distribution}
	The normal distribution is perhaps the most widely used continuous probability distribution in statistics, with numerous applications across many areas. In turn, the beta-normal distribution provides more flexibility in modeling symmetric, heavy-tailed distributions, in addition to skewed and bimodal distributions.
	
	Beta-normal distribution was studied by Eugene et al. (2002) [see also Famoye et al., 2004; Gupta and Nadarajah 2004; Jones 2004; R\'ego et al., 2012]. The PDF of beta-normal distribution can be expressed as 
	\begin{equation}
	\label{3.5}
	g(x)=\frac{\Gamma(\alpha+\beta)}{\Gamma(\alpha) \; \Gamma(\beta)}\left[\Phi \left(\frac{x-\mu}{\sigma}\right)\right]^{\alpha-1} \left[1-\Phi \left(\frac{x-\mu}{\sigma}\right)\right]^{\beta-1} \sigma^{-1}\Phi\left(\frac{x-\mu}{\sigma}\right), \; x \in R \;,
	\end{equation}
	where 
	\begin{equation}
	\label{3.6}
	\Phi(x)=\frac{1}{\sqrt{2 \pi}} e^{-\frac{x^2}{2}}, \quad x\in R \; ,
	\end{equation}
	is the standard normal PDF while
	\begin{equation}
	\label{3.7}
	\Phi(x)=\frac{1}{\sqrt{2 \pi}} \int_{- \infty}^{x} \; e^{-\frac{t^2}{2}} dt, \quad x\in R \;,
	\end{equation}
	is the standard normal CDF. For more details on this distribution please see the above references. 
	%	
	%	
	%	\begin{lemma}
	%		A mode of the beta normal $(\alpha,\beta,\mu,\sigma)$ is any point $x_0=x_0(\alpha,\beta)$ that satisfies 
	%		
	%		\begin{equation*}
	%		\label{lemma1}
	%		x_0=\frac{\sigma\; \left(\frac{x_0-\mu}{\sigma}\right)}{1-\left(\frac{x_0-\mu}{\sigma}\right)}[2-\alpha-\beta]+\frac{(\alpha-1)\; \sigma \; \left(\frac{x_0-\mu}{\sigma}\right)}{\left(\frac{x_0-\mu}{\sigma}\right)\; \left[1-\left(\frac{x_0-\mu}{\sigma}\right)\right]}+\mu
	%		\end{equation*}
	%	\end{lemma}
	%	
	%	\begin{lemma}
	%		\label{lemma2}
	%		If $\alpha=\beta$ and one mode of beta normal $(\alpha,\beta,\mu,\sigma)$ is at $x_0,$ then the other mode is at the point $2\mu-x_0.$
	%	\end{lemma}
	%	
	%	\begin{lemma}
	%		\label{lemma3}
	%		If beta normal $(\alpha,\beta,\mu,\sigma)$ has a mode at $x_0,$ then beta normal $(\alpha,\beta,\mu,\sigma)$ has a mode is at the point $2\mu-x_0.$
	%	\end{lemma}
	%	
	%	\begin{lemma}
	%		\label{lemma4}
	%		The modal point $x_0)\alpha,\beta$ is an increasing function of $\alpha$ and a decreasing function of $\beta$.
	%	\end{lemma}

	\section{Beta-Gumbel Distribution}
	Gumbel distribution is widely used in hydrological engineering design, where it has been used in modeling flood events. The beta Gumbel distribution provides more general and flexible framework for statistical analysis.
	
	The CDF of Gumbel distribution, also known as the extreme value distribution of the Type I, can be expressed as 
	\begin{equation}
	\label{3.8}
	F(x)=e^{-e^{(-\frac{x-\mu}{\sigma})}}, \;-\infty<x<\infty \;,
	\end{equation}
	where $\mu \in R $ and $\sigma >0$. The PDF of Gumbel distribution is of the form 
	\begin{equation}
	\label{3.9}
	f(x)=\frac{1}{\sigma}e^{-u-e^{-u}}, \; x\in R,
	\end{equation}
	where $u=\frac{x-\mu}{\sigma}$. According to Nadarajah and Kotz (2004), the PDF of beta Gumbel distribution is given by
	\begin{equation}
	\label{3.10}
	g(x)=\frac{1}{\sigma B(\alpha,\beta)}u e^{-\alpha u}[1-e^{-u}]^{\beta-1}, \;   -\infty < x < \infty\; . 
	\end{equation}
	The CDF of beta-Gumbel distribution is given by
	\begin{equation}
	\label{3.11}
	\begin{split}
	G(x) & =I_{exp(-\mu)}(\alpha,\beta)\\
	& =\sum_{i=\alpha}^{n}\binom{n}{i}\{1-e^{-\mu}\}^{n-i} e^{-i\mu}\; .
	\end{split}
	\end{equation}
	The $n^{th}$ moment of $X$ can be expressed as
	\begin{equation}
	\label{3.12}
	E(X^n)=\frac{\Gamma(\alpha+\beta)\Gamma(n+1)\mu^n}{\Gamma(\alpha)}\sum_{k=0}^{n}\sum_{l=0}^{\infty}\frac{(-1)^{k+1}\left(\frac{\sigma}{\mu}\right)^k}{k!l!\Gamma(n-k+1)\Gamma(\beta-1)}\; a_{(k,l)}\; ,
	\end{equation}
	where $$a_{(k,l)}=\left(\frac{\partial}{\partial\delta}\right)^k[(\alpha+l)^{\delta}\Gamma(\delta)]|_{\delta=1}\; . $$
	In Particular, the mean can be written as
	\begin{equation}
	\label{3.13}
	E(X)=\frac{\Gamma(\alpha+\beta)}{\Gamma(\alpha)}\sum_{0}^{\infty}\frac{(-1)^l\{\mu+C\sigma+\sigma(l+\alpha)\}}{l!(l+a)\Gamma(\beta-1)}\; .
	\end{equation}
	More information on this model can be found in Nadarajah and Kotz (2004).
	\section{Beta-Weibull Distribution}
	While Weibull distribution has been widely used for modeling data in reliability and the sciences, the beta Weibull generalized class of distributions provides more general and flexible framework for statistical analysis.\\
	The PDF of Weibull distribution is given by
	\begin{equation}
	\label{3.14}
	f(x)=c \lambda^c x^{c-1}e^{-(\lambda x)^c}, \;x>0\; .
	\end{equation}
	The CDF of a Weibull distribution with parameters $c$ and $\lambda$, is given by
	\begin{equation}
	\label{3.15}
	F(x)=1-e^{-(\lambda x)^c}, \;x>0 \; .
	\end{equation}
	According to Famoya et al. (2005) [ see also Lee et al., 2007 ; Cordeiro et al., 2011], the PDF of beta Weibull distribution is given by
	\begin{equation}
	\label{3.16}
	g(x)=\frac{c \lambda^c}{B(\alpha,\beta)}x^{c-1}e^{-\beta (\lambda x)^c}[1-e^{-(\lambda x)^c}]^{\alpha-1}, \;x>0 \; . 
	\end{equation}
	In turn, the CDF of beta Weibull distribution can be expressed as
	\begin{equation*}
	G(x) =I_{1-e^{-(\lambda x)^c}} (\alpha,\beta) , \; x>0 \; ,
	\end{equation*}
	which for non-integer values of $\alpha$ can be written as
	\begin{equation}
	\label{3.17}
	\begin{split}
	G(x) &=\frac{1}{B(\alpha,\beta)}\sum_{i=0}^{\infty}\frac{(-1)^i \Gamma(\alpha)}{\Gamma(\alpha-i) i! (\beta+i)}\{1-e^{-(\beta+i) (\lambda x)^c}\}\\
	&=\frac{\Gamma(\alpha+\beta)}{\Gamma(\beta)}\sum_{i=0}^{\infty}\frac{(-1)^i}{\Gamma(\alpha-i) i! (\beta+i)}\{1-e^{-(\beta+i) (\lambda x)^c}\} ,  \; x>0 \;.
	\end{split} 
	\end{equation}
	The $r^{th}$  moment of $X$, for positive real non-integer $ \alpha $, can be written as
	\begin{equation}
	\label{3.18}
	E(X^r)=\frac{\Gamma(\alpha)\Gamma(\frac{r}{c}+1)}{\lambda^r B(\alpha,\beta)}\sum_{i=0}^{\infty}\frac{(-1)^i}{\Gamma(\alpha-i)i!(\beta+i)^{\frac{r}{c}+1}} \;. 
	\end{equation}
	If $\alpha>0$ is an integer, then we have
	\begin{equation}
	\label{3.19}
	E(X^r)=\frac{\Gamma(\frac{r}{c}+1)}{\lambda^r B(\alpha,\beta)}\sum_{i=0}^{\alpha-1}\binom{\alpha-1}{i}\frac{(-1)^i}{(\beta+i)^{\frac{r}{c}+1}} \; .
	\end{equation}
	More information regarding this distribution can be found in the above references.

	\section{Beta-Exponential Distribution}
	Beta-exponential distribution is a generalization of the exponential distribution, which is one of the most widely used continuous distributions on the positive half line. For example, in queuing theory, the service times of agents are often modeled by the exponential distribution. In hydrology, beta-exponential distribution is used to analyze extreme values such as daily rainfall and river discharge volumes.\\
	According to Nadarajah and Kotz (2006), the PDF of beta-exponential distribution is given by
	\begin{equation}
	\label{3.20}
	g(x)=\frac{\lambda}{B(\alpha,\beta)}e^{-\beta \lambda x}\left(1-e^{-\lambda x}\right)^{\alpha-1} ,\; x>0 \;.
	\end{equation}
	The CDF of beta-exponential distribution is given by
	\begin{equation}
	\label{3.21}
	%	G(x)=\sum_{i=a}^{n}\binom{n}{i}e^{-(n-i) \lambda x}\{1-e^{-\lambda x}\}^i.
	G(x)=I_{1-e^{-\lambda x}}(\alpha,\beta).
	\end{equation}
	Its moment generating function, defined by $M(t)=E(e^{tX}),$ is given by 
	\begin{equation}
	\label{3.22}
	M(t)=\frac{\lambda}{B(\alpha,\beta)}\int_{0}^{\infty}e^{(t-\beta \lambda) x}\{1-e^{-\lambda x}\}^{\alpha-1}dx,
	\end{equation}
	by substituting $y=e^{-\lambda x}$, the integral on the right can be reduced to
	$$\frac{1}{\lambda}\int_{0}^{1}y^{\beta-\frac{t}{\lambda-1}}(1-y)^{\alpha-1}dy=\frac{1}{\lambda}B(\beta-\frac{t}{\lambda},\alpha)\; ,$$ 
	so that
	\begin{equation}
	\label{3.23}
	M(t)=\frac{B(\beta-\frac{t}{\lambda},\alpha)}{B(\alpha,\beta)} \;.
	\end{equation}
	The $n^{th} $ moment of $X$ can be written as
	\begin{equation}
	\label{3.24}
	E(X^n)=\frac{(-1)^n}{\lambda^n B(\alpha, \beta)}\frac{\partial^n}{\partial p^n}B(\alpha,1+p-\alpha)|_{p=\alpha+\beta-1}\; .
	\end{equation}
	For more information on this model, see Nadarajah and Kotz, (2006).

	\section{Beta-Laplace Distribution}
	The Laplace distribution has numerous applications across many fields of science and engineering (see, e.g., Kotz et al., 2001). It is also used in Bayesian regression analysis as a Laplacian prior. On the other hand, beta Laplace distribution offers more flexibility compared with standard Laplace distribution.  \\
	A random variable $X$ has Laplace distribution with location parameter $\mu $ and scale parameter $\sigma >0 $, if its PDF is given by
	\begin{equation}
	\label{3.25}
	f(x)=\frac{1}{2\sigma}e^{|\frac{x-\mu}{\sigma}|}, \quad -\infty<x<\infty\, .
	\end{equation}
	In the standard case $\mu=0$ and $\sigma=1$, the CDF can be expressed as
	\begin{equation}
	\label{3.26}
	F(x)=\begin{cases}
	\frac{1}{2}e^x, & x<0\\
	1-\frac{1}{2}e^{-x}, & x\ge 0.\\
	\end{cases} 
	\end{equation}
	The standard beta-Laplace distribution, first described in Kozubowski and Nadarajah (2008) and subsequently studied by Cordeiro and Lemonte (2011), can be described by its CDF as follows :
	\begin{equation}
	\label{3.27}
	G(x)=\begin{cases}
	I_{\frac{1}{2}e^x}(\alpha,\beta), & x<0  \\
	I_{1-\frac{1}{2}e^{-x}} (\alpha,\beta), & x\ge 0 . \\
	\end{cases} 
	\end{equation}
	The corresponding PDF  is given by
	\begin{equation}
	\label{3.28}
	g(x)=\begin{cases}
	\frac{1}{2^\alpha B(\alpha,\beta)}e^{-|x|}e^{-|x|(\alpha-1)}(1-e^{-\frac{|x|}{2}})^{\beta-1}\; , & x<0  \\
	\frac{1}{2^\beta B(\alpha,\beta)}e^{-|x|}e^{-|x|(\beta-1)}(1-e^{-\frac{|x|}{2}})^{\alpha-1}\; , & x<0  .\\
	\end{cases}
	\end{equation}
	More information regarding this distribution can be found in the above references.		
	
	\section{Beta-Rayleigh Distribution}
	Rayleigh distribution is vastly used in modeling of lifetime data as well as in reliability analysis. The PDF of Rayleigh distribution is given by
	\begin{equation}
	\label{3.29}
	f(x)=\frac{x}{\sigma^2}e^{-\frac{1}{2}(\frac{x}{\sigma})^2},\; x\ge 0,
	\end{equation}
	where $ \sigma>0 $ is a scale parameter. In turn, the CDF of Rayleigh distribution is given by
	\begin{equation}
	\label{3.30}
	F(x)=1-e^{-\frac{1}{2}(\frac{x}{\sigma})^2} , \; x\ge 0. 
	\end{equation}
	The beta-Rayleigh distribution was studied in Cordeiro et al. (2013). The PDF of beta-Rayleigh distribution can be expressed as 
	\begin{equation}
	\label{3.31}
	g(x)=\frac{x}{\sigma^2 B(\alpha,\beta)}e^{-\frac{1}{2}(\frac{x}{\sigma})^2\beta}(1-e^{-\frac{1}{2}(\frac{x}{\sigma})^2})^{\alpha-1} , \; x\ge 0 .
	\end{equation}
	The CDF of beta-Rayleigh distribution is given by
	\begin{equation}
	\label{3.32}
	G(x)=1-\frac{e^{-\frac{\alpha}{2}(\frac{x}{\sigma})^2}}{\alpha B(\alpha,\beta)}
	{_2F_1(\alpha,1-\beta;1+\alpha; e^{-\frac{\alpha}{2}(\frac{x}{\sigma})^2}}) , \; x\ge 0,
	\end{equation}
	where ${_2F_1(\alpha,1-\beta;1+\alpha; e^{-\frac{\alpha}{2}(\frac{x}{\sigma})^2}})$ is the Gauss Hypergeometric special function. \\
	The $n^{th}$ moment of a random variable $X$ with beta-Rayleigh distribution is given by
	\begin{equation}
	\label{3.33}
	E[X^n]=\frac{(\sigma \sqrt{2})^n (\frac{n}{2})!}{B(\alpha,\beta)}\sum_{k=0}^{\infty}(-1)^k\binom{\alpha-1}{k}\frac{1}{(\beta+k)^{\frac{n}{2}+1}} \;.
	\end{equation}
	More information on this distribution can be found in the above literature.

	\section{Beta-Maxwell Distribution}
	The Maxwell distribution, also known as the Maxwell-Boltzmann
	distribution, is a continuous probability distribution with applications
	in physics and chemistry. This distribution is commonly used in statistical
	mechanics to determine the speeds of molecules. On the other hand, beta Maxwell distribution has similar type of applications and offers more flexibility.\\
	The PDF of Maxwell distribution can be described as
	\begin{equation}
	\label{3.34}
	f(x)=\sqrt{\frac{2}{\pi}}\frac{x^2e^{-\frac{x^2}{2a^2}}}{a^3}, \quad 0\le x < \infty,
	\end{equation}
	where $a>0$ is a scale parameter. The CDF of Maxwell distribution is
	\begin{equation}
	\label{3.35}
	F(x)=\frac{2\gamma(\frac{3}{2},\frac{x^2}{2a^2})}{\sqrt{\pi}}, \quad 0 \le x < \infty ,
	\end{equation}
	where  $\gamma(a,b)$ is the lower incomplete gamma function, defined as
	$$\gamma(a,b)=\int_{0}^{b}t^{a-1}e^{-t}dt. $$
	In applications, where the random variable $X$ denotes the speed of a molecule while  $a=\sqrt{\frac{KT}{m}}$ where $K$ is the Boltzmann constant, $T$ is the temperature, and $m$ is the mass of a molecule. \\
	According to Amusan (2010), the PDF of beta-Maxwell distribution can be written as
	\begin{equation}
	\label{3.36}
	g(x)=\frac{1}{B(\alpha,\beta)}\left[\frac{2}{\sqrt{\pi}}\gamma (\frac{3}{2},\frac{x^2}{2a^2})\right]^{a-1} \left[1-\frac{2}{\sqrt{\pi}}\gamma (\frac{3}{2},\frac{x^2}{2a^2})\right]^{\beta-1} \sqrt{\frac{2}{\pi}}\frac{x^2e^{-\frac{x^2}{2a^2}}}{a^3}.
	\end{equation}
	Given that $X \sim BM(\alpha,\beta,a), $ its CDF can be expressed as
	\begin{equation}
	\label{3.37}
	G(x)=\int_{0}^{x}\frac{1}{B(\alpha,\beta)}\left[\frac{2}{\sqrt{\pi}}\gamma (\frac{3}{2},\frac{x^2}{2a^2})\right]^{a-1} \left[1-\frac{2}{\sqrt{\pi}}\gamma (\frac{3}{2},\frac{x^2}{2a^2})\right]^{\beta-1} \sqrt{\frac{2}{\pi}}\frac{x^2e^{-\frac{x^2}{2a^2}}}{a^3} ,
	\end{equation}
	which, with a substitution of 
	$$z=\frac{2}{\sqrt{\pi}}\gamma (\frac{3}{2},\frac{t^2}{2a^2}),$$ 
	and
	$$\frac{dz}{dt}=\sqrt{\frac{2}{\pi}}\frac{x^2e^{-\frac{x^2}{2a^2}}}{a^3},$$ 
	can simplified as follows :
	\begin{equation}
	\label{3.38}
	G(x)=\int_{0}^{A}\frac{1}{B(\alpha,\beta)z^{\alpha-1(1-z)^{\beta-1}}}dz \;= \;
	\frac{B(A;\alpha,\beta)}{B(\alpha,\beta)} \; ,
	\end{equation}
	where $B(A;\alpha,\beta)$ is an incomplete beta function with $$A=\frac{2}{\sqrt{\pi}}\gamma (\frac{3}{2},\frac{t^2}{2a^2}).$$
	The well known relationship between the incomplete beta function and series enables us to express the CDF alternatively as
	\begin{equation}
	\label{3.39}
	G(x)=\frac{A^{\alpha}}{B(\alpha,\beta)}\left\{\frac{1}{\alpha}+\frac{1-\beta}{\alpha +1}A+ \cdots +\frac{(1-\beta)(2-\beta)\cdots(n-\beta)A^n}{n!(\alpha+n)}\right\}.
	\end{equation}
	More information on this distribution can be found in the above literature.

	\section{Beta-Pareto Distribution}
	
	Pareto distribution is widely applied across numerous areas of applications to account for heavy tailed data with outliers.\\
	One version of Pareto CDF can be expressed as follows :
	\begin{equation}
	\label{3.40}
	F(x)=1-\left(\frac{x}{\theta}\right)^{-k}, \quad x\ge \theta>0,
	\end{equation}
	where $k>0$ is a tail parameter. According to Akinsete et al. (2008), the PDF of beta-Pareto distribution is given by
	\begin{equation}
	\label{3.41}
	g(x)=\frac{k}{\theta B(\alpha,\beta)}\left\{1-\left(\frac{x}{\theta}\right)^{-k}\right\}^{\alpha-1} \left\{\frac{t}{\theta}\right\}^{-k \beta-1}, \quad x\ge \theta>0,
	\end{equation}
	where $ \alpha,\beta,\theta,k>0 $. The CDF of beta-Pareto random variable is given by 
	\begin{equation}
	\label{3.42}
	G(x)=1-\frac{z^{\beta}}{B(\alpha,\beta)}\left\{\frac{1}{\beta}+\frac{1-\alpha}{\beta +1}z+\cdots +\frac{(1-\alpha)(2-\alpha)\cdots(n-\alpha)}{n!(\beta+n)}z^n+\cdots\right\}.
	\end{equation}
	General moments for the beta-Pareto distribution are given by
	\begin{equation}
	\label{3.43}
	E\left(\frac{X}{\theta}\right)^r=\frac{k}{\theta B(\alpha,\beta)} \int_{\theta}^{\infty}\left\{1-\left(\frac{x}{\theta}\right)^{-k}\right\}^{\alpha-1} \left\{\frac{t}{\theta}\right\}^{-k \beta+r-1}. 
	\end{equation}
	In particular, the mean of beta-Pareto distribution may be expressed as
	\begin{equation}
	\label{3.44}
	\begin{split}
	E(X) & =\theta\left\{\frac{B(\alpha,\beta-\frac{1}{k})}{B(\alpha,\beta)}\right\} \\
	& = \frac{\theta \Gamma(\alpha+\beta)}{\Gamma(\beta)}\frac{\Gamma(\beta-\frac{1}{k})}{\Gamma(\alpha+\beta-\frac{1}{k})}.
	\end{split} 
	\end{equation}
	See Akinsete et al. (2008) for more information regarding this distribution.

	\section{Beta-Nakagami Distribution}
	Nakagami distribution is a popular model in engineering. A random variable $X$ has the Nakagami distribution if its PDF  is as follows :
	\begin{equation}
	\label{3.45}
	f(x)=\frac{2\mu^{\mu}x^{2\mu-1}e^{-\frac{\mu x^2}{\omega}}}{\Gamma(\mu) \omega^{\mu}}, \; x>0,
	\end{equation}
	where $\mu \; (\mu>0)$ is a shape parameter and $\omega \; (\omega>0)$ is a scale parameter.
	The CDF of Nakagami distribution is
	\begin{equation}
	\label{3.46}
	F(x)=\frac{2\mu^{\mu}}{\Gamma(\mu)\omega^{\mu}}\int_{0}^{x} \left[\frac{\omega y}{\mu}\right]^{\frac{2 \mu-1}{2}} \frac{e^{-y}\omega dy}{2\left(\frac{\omega y}{\mu}\right)^{\frac{1}{2}} \mu} , \quad x>0,
	\end{equation}
	where $\mu \ge 0.5,\; \omega > 0, \; x>0. $ The beta Nakagami distribution was first studied by Shittu and Adepoju (2013). Its PDF is of the form
	\begin{equation}
	\label{3.47}
	g(x)=\frac{\Gamma(\alpha +\beta)}{\Gamma(\alpha) \Gamma(\beta)} F(x)^{\alpha-1}(1-F(x))^{\beta-1} f(x), \; x>0 ,
	\end{equation}
	with $F(x)$ as in (\ref{3.46}). As shown in Shittu and Adepoju (2013), we have
	$$F(x)=\frac{\gamma(\mu,\frac{\mu x^2}{\omega})}{\Gamma(\mu)},$$
	where
	$$\gamma(\mu,y)=\int_{0}^{y} t^{\mu-1} e^{-t} dt ,$$
	is the incomplete gamma function. In turn, the CDF becomes
	\begin{equation}
	\label{3.48}
	%\begin{split}
	G(x) = \int_{0}^{x}\frac{\Gamma(\alpha+\beta)}{\Gamma(\alpha)\Gamma(\beta)} \frac{\gamma(\mu,\frac{\mu x^2}{\omega})}{\Gamma(\mu)}^{\alpha-1}\left[1-\frac{\gamma(\mu,\frac{\mu x^2}{\omega})}{\Gamma(\mu)}\right]^{\beta-1}\frac{2\mu^{\mu}x^{2\mu-2}e^{-\frac{\mu x^2}{\omega}}}{\Gamma(\mu) \omega^{\mu}} dx\; . 
	%\end{split}
	\end{equation}
	According to Jones (2004), the above expression can be written as
	\begin{equation}
	\label{3.49}
	G(x)=\frac{F(x)^{\alpha}}{B(\alpha,\beta)}\left\{\frac{1}{\alpha}+\frac{1-\beta}{\alpha +1}F(x)+\cdots+\frac{(1-\beta)(2-\beta)\cdots(n-\beta)F(x)^{n}}{n!(\alpha+n)}+\cdots\right\}. 
	\end{equation}
	More information on this model can be found in the above literature.
	
	\section{Beta-Fr\'echet Distribution}
	The Fr\'echet distribution has been widely used for modeling wind speeds and track race records, earthquakes, floods, rainfall, sea currents analysis. The beta Fr\'echet distribution provides more general and flexible framework for statistical analysis of positive data. \\
	A random variable $X$ has Fr\'echet  distribution, if it has the PDF is given by
	\begin{equation}
	\label{3.50}
	f(x)=\lambda \sigma^{\lambda}x^{-(\lambda+1)} e^{-(\frac{\sigma}{x})^{\lambda}}, \quad x>0,
	\end{equation}
	where $\sigma > 0 $ is a scale parameter and $\lambda > 0$ is a shape parameter. The corresponding CDF of Fr\'echet distribution can be written as 
	\begin{equation}
	\label{3.51}
	F(x)=e^{-(\frac{\sigma}{x})^{\lambda}}, \quad x>0. 
	\end{equation}
	Barreto Souza et al. (2008) [see also Nadarajah and Gupta, 2004], studied beta- Fr\'echet distribution with parameters $\alpha >0 ,\beta >0, \sigma >0 $ and $\lambda >0 $, which is given by the CDF 
	\begin{equation}
	\label{3.52}
	G(x)= I_{e^{-(\frac{\sigma}{x})^{\lambda}}} (\alpha,\beta)=\frac{1}{B(\alpha,\beta)} \int_{0}^{e^{-(\frac{\sigma}{x})^{\lambda}}} x^{\alpha-1}(1-x)^{\beta-1}, \quad x>0. 
	\end{equation}
	The corresponding PDF of this distribution is 
	\begin{equation}
	\label{3.53}
	g(x)=\frac{\lambda (\sigma)^{\lambda}}{B(\alpha,\beta)}
	x^{-(\lambda+1)} e^{-\alpha (\frac{\sigma}{x})^{\lambda}}(1-e^{-(\frac{\sigma}{x})^{\lambda}})^{\beta-1} , \quad x>0 .
	\end{equation}
	The hazard function for this distribution can be written as
	\begin{equation}
	\label{3.54}
	h(x)=\frac{\lambda \sigma^{\lambda}x^{-(\lambda+1)} e^{-\alpha (\frac{\sigma}{x})^{\lambda}}(1-e^{-(\frac{\sigma}{x})^{\lambda}})^{\beta-1}}{B(\alpha,\beta)\{1-I_{e^{-(\frac{\sigma}{x})^{\lambda}}} (\alpha,\beta)\}} , \quad x>0.
	\end{equation}
	More detailed information on this distribution can be found in the above literatures.
	
	\section{Beta-Gompertz Distribution}
	Gompertz distribution is an extreme value distribution  that can be skewed to the right or to the left. This distribution is often used in lifetime data analysis as well as in actuarial science. In turn, beta Gompertz distribution offers lot more flexibility.\\
	A random variable $X$ has Gompertz distribution if its PDF is of the form
	\begin{equation}
	\label{3.55}
	f(x)=\theta e^{\gamma x} e^{\frac{-\theta}{\gamma} (e^{\gamma x}-1)}, \quad x>0, 
	\end{equation}
	where $\theta>0$ and $\gamma>0 $. The corresponding CDF becomes
	\begin{equation}
	\label{3.56}
	F(x)=1-e^{\frac{-\theta}{\gamma} (e^{\gamma x}-1)} , \quad x>0.
	\end{equation}
	A four parameter beta-Gompertz distribution based on this model was studied by Jafari et al. (2014). The CDF of this distribution can be written in the following form 
	\begin{equation}
	\label{3.57}
	G(x)=I_{F(x)}(\alpha,\beta) , \quad x>0,
	\end{equation}
	with the corresponding PDF 
	\begin{equation}
	\label{3.58}
	g(x)=\frac{\theta e^{\gamma x} e^{\frac{-\beta \theta}{\gamma} (e^{\gamma x}-1)}}{B(\alpha,\beta)}[1-e^{\frac{-\theta}{\gamma} (e^{\gamma x}-1)}]^{\alpha-1} , \quad x>0.
	\end{equation}
	We refer to the above literature for more detailed information.

	\section{Beta-Lomax or Beta-Burr Type XII Distribution}
	Lomax (1954) proposed Pareto Type – II distribution, also known as Lomax
	Distribution, and used it for  modeling business failure life time data analysis. The
	beta Lomax distribution is also widely applicable in reliability and life testing problems.\\
	The PDF of Lomax distribution with parameters $ \alpha, \lambda,\mu $ can be written as
	\begin{equation}
	\label{3.59}
	f(x)=\frac{\gamma}{\lambda} \left[1+\frac{x-\mu}{\lambda}\right]^{-(\gamma+1)} , \quad x\ge \mu,
	\end{equation}
	where $\gamma \;(\gamma>0) $ is a shape parameter and $\lambda\; (\lambda>0)$ is a scale parameter.
	The CDF associated with this distribution is of the form
	\begin{equation}
	\label{3.60}
	F(x)=1-\left\{1+\frac{x-\mu}{\lambda}\right\}^{-\gamma} ,\quad x\ge \mu. 
	\end{equation}
	A five-parameter beta Lomax distribution was described by Javanshiri and Maadooliat (2014) [see also Rajab et al., 2013]. According to (\ref{3.3}) - (\ref{3.4}), the PDF of this distribution becomes
	\begin{equation}
	\label{3.61}
	g(x)=\frac{\gamma}{\lambda B(\alpha,\beta)}\left[1-\left\{1+\frac{x-\mu}{\lambda}\right\}^{-\gamma} \right]^{\alpha-1}\left[1+\frac{x-\mu}{\lambda}\right]^{-(\gamma \beta+1)}, \quad x\ge \mu.
	\end{equation}
	In case of integer-valued parametrs $\alpha$, the CDF corresponding to the proposed five-parameters beta Lomax distribution can be expressed as
	\begin{equation}
	\label{3.62}
	G(x)=\frac{\gamma}{B(\alpha,\beta)} \sum_{i=0}^{\alpha-1}(-1)^i \binom{\gamma-1}{i} \frac{1}{\gamma \beta+i\gamma} \left[1-\left\{1+\frac{x-\mu}{\lambda}\right\}^{-(\gamma \beta+i\gamma)}\right] , \quad x \ge \mu.
	\end{equation}
	The hazard rate function of this distribution is of the form
	\begin{equation}
	\label{3.63}
	\gamma(x)=\frac{\frac{\gamma}{\lambda B(\alpha,\beta)}[1-\{1+(\frac{x-\mu}{\lambda})\}^{-\gamma} ]^{\alpha-1}[1+(\frac{x-\mu}{\lambda})]^{-(\gamma \beta+1)} }{1-\frac{\gamma}{B(\alpha,\beta)} \sum_{i=0}^{\alpha-1}(-1)^i \binom{\gamma-1}{i} \frac{1}{\gamma \beta+i\gamma} [1-\{1+\frac{x-\mu}{\lambda}\}^{-(\gamma \beta+i\gamma)}] } .
	\end{equation}
	See the above reference for more detailed information related to this distribution.

	\section{Beta-Lognormal Distribution}
	The lognormal distribution is quite flexible for analyzing positive data, and is particularly useful for modeling asymmetric data. The PDF of lognormal distribution is defined as follows :
	\begin{equation}
	\label{3.64}
	f(x)=\frac{1}{\sigma x\sqrt{2 \pi }} e^{-\frac{(\ln x-\mu)^2}{2\sigma^2}}, \quad x>0.
	\end{equation}
	The CDF of lognormal distribution becomes
	\begin{equation}
	\label{3.65}
	F(x)=\Phi \left[\frac{\ln(x)-\mu}{\sigma}\right]   ,\quad  x>0, 
	\end{equation}
	where $$\Phi(x)=\frac{1}{\sqrt{2 \pi}} \; \int_{- \infty}^{x} e^{- t^2/2} \; dt,$$
	is the CDF of the standard normal distribution. The corresponding beta-lognormal distribution was studied by Castellares et al. (2011). Its PDF is given by,
	\begin{equation}
	\label{3.66}
	g(x)=\frac{1}{B(\alpha,\beta)\sigma x\sqrt{2 \pi }} e^{-\frac{(\ln x-\mu)^2}{2\sigma^2}}\Phi \left(\frac{\ln x-\mu}{\sigma}\right)^{\alpha-1} \left[1-\Phi \left(\frac{\ln x-\mu}{\sigma}\right)\right]^{\beta-1}, \quad x>0,
	\end{equation}
	with $\Phi$ as above. The CDF can be expressed as
	\begin{equation}
	\label{3.67}
	G(x)=I_{\Phi(\frac{lnx-\mu}{\sigma})}(\alpha,\beta) ,
	\end{equation}
	and the hazard function of the four-parameter beta-lognormal distribution is
	\begin{equation}
	\label{3.68}
	h(x)=\frac{e^{-\frac{(lnx-\mu)^2}{2\sigma^2}}\Phi(\frac{lnx-\mu}{\sigma})^{\alpha-1} [1-\Phi (\frac{lnx-\mu}{\sigma})]^{\beta-1}}{B(\alpha,\beta)\sigma x\sqrt{2 \pi } (1-I_{\Phi(\frac{lnx-\mu}{\sigma})}(\alpha,\beta))}.
	\end{equation}
	More information regarding this distribution can be found in Castellares et al. (2011).

	\section{Beta-Burr Type X Distribution}
	The Burr type X distribution has been used in reliability analysis as well as modeling for life time of random phenomena, health, agriculture, and biology. The PDF of beta-Burr Type $X$ distribution can be written as
	\begin{equation}
	\label{3.69}
	f(x)=2 \theta \lambda^2 x e^{-(\lambda x)^2}[1-e^{-(\lambda x)}]^{\theta-1}, \quad x>0 .
	\end{equation}
	The CDF of Burr-Type $X$ distribution is
	\begin{equation}
	\label{3.70}
	F(x)=[1-e^{-(\lambda x)^2}]^{\theta} , \quad x>0 .
	\end{equation}
	The beta-Burr Type $X$ distribution was first studied by Merovci et all. (2016). The PDF of beta-Burr Type $X$ is of the form 
	\begin{equation}
	\label{3.71}
	g(x)=\frac{2 \theta \lambda^2 x}{B(\alpha,\beta)}\left\{[1-e^{-(\lambda x)^2}]\right\}^{\theta \alpha-1} \left\{1-[1-e^{-(\lambda x)^2}]^{\theta}\right\}^{\beta-1} e^{-(\lambda x)^2} , \quad x>0.
	\end{equation}
	where $\alpha >0,\; \beta >0,\; \lambda,\theta>0$. The CDF of beta-Burr Type $ X$ distribution becomes
	\begin{equation}
	\label{3.72}
	G(x)=I_{[1-e^{-(\lambda x)^2}]^{\theta}}(\alpha,\beta) =\frac{1}{B(\alpha,\beta)}\int_{0}^{[1-e^{-(\lambda x)^2}]^{\theta}} x^{\alpha-1} (1-x)^{\beta-1} dx, \quad x>0.
	\end{equation}
	The hazard rate function, defined as the ratio of the density to its survival function, is given by
	\begin{equation}
	\label{3.73}
	h(x)=\frac{2 \theta \lambda^2 x  e^{-(\lambda x)^2} }{B(\alpha,\beta) \left[1-I_{[1-e^{-(\lambda x)^2}]^{\theta}} (\alpha,\beta)\right]}\left\{1-e^{-(\lambda x)^2}\right\}^{\theta \alpha-1} \left\{1-[e^{-(\lambda x)^2}]^{\theta}\right\}^{\beta-1} .
	\end{equation}
	More information can be found on the reference mentioned above.

	\section{Beta-Lindley Distribution}
	Lindley distribution is often used to describe the lifetime of a system or a component. It is widely used in biology, engineering, and medicine.
	The PDF of Lindley distribution can be expressed as
	\begin{equation}
	\label{3.74}
	f(x)=\frac{\theta^2}{\theta+1}(1+x)e^{-\theta x}, \quad x>0,
	\end{equation}
	where  $\theta >0$ is a scale parameter. The corresponding CDF is of the form
	\begin{equation}
	\label{3.75}
	F(x)=1-\frac{\theta+1+\theta x}{\theta+1}e^{-\theta x} , \quad  x>0.
	\end{equation}
	The beta-Lindley distribution was studied in Merovci and Sharma (2011) [see also Mostafaee et al., 2015]. 
	According to (\ref{3.3})-(\ref{3.4}), the CDF of beta Lindley distribution is given by
	\begin{equation}
	\label{3.76}
	G(x)=I_{1-\frac{\theta+1+\theta x}{\theta+1}e^{-\theta x}} (\alpha,\beta), \quad x>0,
	\end{equation}
	while the PDF of this distribution can be written as
	\begin{equation}
	\label{3.77}
	g(x)=\frac{f(x)}{B(\alpha,\beta)}\left[1-\frac{\theta+1+\theta x}{\theta+1}e^{-\theta x}\right]^{\alpha-1} \left\{\frac{\theta+1+\theta x}{\theta+1}e^{-\theta x}\right\}^{\beta-1}, \quad x>0,
	\end{equation}
	with $f(x)$ as in (\ref{3.74}). The hazard rate function of the beta-Lindley distribution, is defined for $x>0$, is of the form
	\begin{equation}
	\label{3.78}
	h(x)=\frac{\theta^2(\theta+1)^{-1} (1+x)e^{-\beta \theta x}}{B(\alpha,\beta)-I_{1-\frac{\theta+1+\theta x}{\theta+1}e^{-\theta x}} (\alpha,\beta)}  \left[1-(1+\frac{\theta x}{\theta+1})e^{-\theta x}\right]^{\alpha-1} \left(1+\frac{\theta x}{\theta+1}\right)^{\beta-1}\;.
	\end{equation}
	More information about this model can be found in Merovci and Sharma (2011) and Mostafaee et al. (2015). 
	
	%				\section{Beta-Student F Distributions}
	%				
	%				Consider the $F(2a,2b)$ distribution with degree of freedom 2a and 2b and for $x>0,a>0,b>0$ the PDF of Student F distribution is given by
	%				\begin{equation}
	%				\label{ky1}
	%				f(x)=\frac{a^a x^{a-1}}{b^aB(a,b)(1+ax)^{a+b}}
	%				\end{equation}
	%				
	%				and The CDF can be expressed as 
	%				\begin{equation}
	%				\label{ky1}
	%				F(x)=I_{\frac{ax}{ax+b}}(\alpha,\beta) 
	%				\end{equation}
	%				
	%				
	%				According to Jones et all.(2004), \cite{c45},the Proposed PDF of the Beta-F distribution $BF(\alpha,\beta,2a,2b)$ with parameters $\alpha,\beta,2a,2b$ can be written as
	%				\begin{equation}
	%				\label{ky1}
	%				g(x)=\frac{a^a x^{a-1}}{b^a B(a,b)(1+ax)^{a+b} B(\alpha,\beta)} [I_{\frac{ax}{ax+b}}(\alpha,\beta)]^{\alpha-1}[1-I_{\frac{ax}{ax+b}}(\alpha,\beta)]^{\beta-1}
	%				\end{equation}
	%				
	%				
	%				\section*{Applications}
	%				
	%				F-distribution arises frequently as the null distribution of a test statistic (hypothesis testing) and used to develop confidence interval and in the Analysis of Variance (ANOVA) for comparison of several populations means. The F-distribution arises frequently as the null distribution of a test statistics most notably in ANOVA where for example the null hypothesis that two independent normal variances are equal and the observed sums of some appropriately selected squares are then examined to see whether their ratio is significantly incompatible with this null hypothesis.
	%				
	
	\section {The Beta Power distribution}
	The PDF of Power distribution is defined as follow :
	\begin{equation}
	\label{3.79}
	f(x)=a b^ax^{a-1},  \quad  0 < x < \frac{1}{b} ,
	\end{equation}
	This is a special case of beta distribution. The CDF of power distribution can be expressed as 
	\begin{equation}
	\label{3.80}
	F(x)=(bx)^a ,  \quad  0 \le x \le \frac{1}{b} ,
	\end{equation}
	where $a>0$ is a shape parameter and $b>0$ is a scale parameter.\\
	When we use equation (\ref{3.79})-(\ref{3.80}), in (\ref{3.3})-(\ref{3.4}) we obtain the beta power distribution studied by Cordeiro and Brito (2012) [see also McDonald and Richards, 1987].
	Thus, the PDF of beta-power distribution becomes
	\begin{equation}
	\label{3.81}
	g(x)=\frac{ab(bx)^{\alpha a-1}[1-(bx)^a]^{\beta-1}}{B(\alpha,\beta)},  \quad  0 < x < \frac{1}{b} ,
	\end{equation}
	while the CDF can be expressed as
	\begin{equation}
	\label{3.82}
	\begin{split}
	G(x) & =I_{(bx)^a}(\alpha,\beta) \\
	& =\frac{1}{B(\alpha,\beta)}\int_{0}^{(bx)^a} x^{\alpha-1}(1-x)^{\beta-1} dx \\ 
	& =\frac{(bx)^{\alpha a}}{\alpha B(\alpha,\beta)} {}_2 F_1(\alpha,1-\beta,\alpha+1;(bx)^a).
	\end{split}
	\end{equation}
	The corresponding hazard rate function, described in Cordeiro and Brito (2012), takes on the form 
	\begin{equation}
	\label{3.83}
	h(x)=\frac{ab(bx)^{\alpha a-1}[1-(bx)^a]^{\beta-1}}{B(\alpha,\beta)[1-I_{(bx)^a}(\alpha,\beta)]} .
	\end{equation}
	As shown by Cordeiro and Brito (2012), the $r^{th}$ moments (about zero) admits the representation
	\begin{equation}
	\label{3.84}
	E X^r=\frac{a\Gamma(\alpha+\beta)}{b^r \Gamma(\alpha)} \sum_{j=0}^{\infty} (-1)^j [\Gamma(\beta-j)[r+a(\alpha+j)]j!]^{-1} ,
	\end{equation}
	and simplify to 
	\begin{equation}
	\label{3.85}
	EX^r=\frac{B(\alpha+\frac{r}{a},\beta)}{b^r\; B(\alpha,\beta)}\;.
	\end{equation}
	We refer to Cordeiro and Brito (2012) for further information regarding this distribution.

	% \section*{Kummer Beta Distribution }

	%     \section*{Applications}
	
	\section{Beta-Dagum Distribution }
	The PDF of Dagum distribution is defined as follows :
	\begin{equation}
	\label{3.86}
	f(x)=\gamma \lambda \delta x^{-(\delta+1)}(1+\lambda x^{-\delta})^{-(\gamma+1)}, \quad x>0,
	\end{equation}
	where $\lambda$ is a scale parameter and $\gamma, \delta\; (\gamma,\delta >0)$ are shape parameters.
	The corresponding CDF of this distribution is given by
	\begin{equation}
	\label{3.87}
	F(x)=(1+\lambda x^{-\delta})^{-\gamma} , \quad x>0.
	\end{equation}
	A model of the form (\ref{3.3})-(\ref{3.4}) based on the above Dagum distribution was proposed by Domma and Condino (2013). The resulting CDF of beta-Dagum distribution can be written as
	\begin{equation}
	\label{3.88}
	G(x)=\sum_{j=o}^{\infty}\frac{\Gamma(\alpha+\beta)(-1)^j}{\Gamma(\alpha)\Gamma(\beta-j)j!(\alpha+j)}F(x)^{\alpha+j} ,
	\end{equation}
	with $F(x)$ as in (\ref{3.87}). The PDF of beta-Dagum distribution can be written as
	\begin{equation}
	\label{3.89}
	g(x)=\frac{1}{B(\alpha,\beta)}[(1+\lambda x^{-\delta})]^{-\gamma\alpha-1}[1-(1+\lambda x^{-\delta})^{-\gamma}]^{\beta-1} \gamma \lambda \delta x^{-(\delta+1)}.
	\end{equation}
	More information on the distribution can be found in Domma and Condino (2013).
	
	\section{Beta-Cauchy Distribution }
	The PDF of Cauchy distribution is of the form
	\begin{equation}
	\label{3.90}
	f(x)=\frac{1}{\lambda \pi[1+\left(\frac{x-\theta}{\lambda}\right)^2]}, \quad -\infty <x < \infty ,
	\end{equation}
	where $\theta>0$ is a location parameter and $\lambda>0$ is a scale parameter. The corresponding CDF becomes
	\begin{equation}
	\label{3.91}
	F(x)=\frac{1}{2}+\frac{1}{\pi} \tan^{-1}\left(\frac{x-\theta}{\lambda}\right) , \quad -\infty <x < \infty.
	\end{equation}
	A distribution of the form (\ref{3.3})-(\ref{3.4}) based on the above Cauchy distribution was studied by Alshawarbeh et al. (2012). The PDF of beta-Cauchy distribution can be written as
	\begin{equation}
	\label{3.92}
	g(x)=\frac{\lambda}{\pi B(\alpha,\beta)}\left[\frac{1}{2}+\frac{1}{\pi}\tan^{-1}\left(\frac{x-\theta}{\lambda}\right)\right]^{\alpha-1}\left[\frac{1}{2}-\frac{1}{\pi}\tan^{-1}\left(\frac{x-\theta}{\lambda}\right)\right]^{\beta-1}\frac{1}{\lambda^2+(x-\theta)^2} ,
	\end{equation}
	where the parameters $\alpha$ and $\beta$ are shape parameters, $\theta$ is a location parameter, and $\lambda$ is a scale parameter, with $0<\alpha,\beta,\lambda<\infty$ and $-\infty<\theta<\infty$.
	More information regarding this distribution can be found in the above literature.

	\section{Beta Nadarajah-Haghighi distribution }
	The CDF of Nadaraja-Haghighi distribution can be written as
	\begin{equation}
	\label{3.93}
	F(x)=1-e^{(1-(1+\lambda x)^{\sigma})} , \quad x>0,
	\end{equation}
	where $\lambda > 0$ is a scale parameter and $\sigma>0$ is a shape parameter.
	The corresponding PDF can be expressed as
	\begin{equation}
	\label{3.94}
	f(x)=\sigma \lambda (1+\lambda x)^{\sigma-1} e^{[1-(1+\lambda x)^\sigma]} , \quad x>0.
	\end{equation}
	The beta-Nadarajah-Haghighi distribution based on this distribution was proposed by Dias et al. (2016). Its CDF is given by
	\begin{equation}
	\label{3.95}
	G(x)=I_{1-e^{(1-(1+\lambda x)^{\sigma})}}(\alpha,\beta) , \quad x>0,
	\end{equation}
	while the corresponding PDF becomes
	\begin{equation}
	\label{3.96}
	g(x)=\frac{\sigma \lambda}{B(\alpha,\beta)}(1+\lambda x)^{\sigma-1}\left[1-e^{(1-(1+\lambda x)^{\sigma})}\right]^{\alpha-1}\left[e^{(1-(1+\lambda x)^{\sigma})}\right]^{\beta}, \quad x>0.
	\end{equation}
	The hazard rate function can be written as
	\begin{equation}
	\label{3.97}
	h(x)=\frac{\sigma \lambda}{B(\alpha,\beta)-I_{1-e^{(1-(1+\lambda x)^{\sigma})}}(\alpha,\beta)}(1+\lambda x)^{\sigma-1}\left[1-e^{(1-(1+\lambda x)^{\sigma})}\right]^{\alpha-1}\left[e^{(1-(1+\lambda x)^{\sigma})}\right]^{\beta}.
	\end{equation}
	See Dies et al. (2016) for more details.

	\section{Beta-Half Cauchy Distribution}
	The half Cauchy distribution is derived from the Cauchy distribution by taking only positive values. Its CDF is given by
	\begin{equation}
	\label{3.98}
	F(x)=\frac{2}{\pi}\tan^{-1}\left(\frac{x}{\phi}\right) , \quad  x>0 ,
	\end{equation}
	where $\Phi>0$ is a scale parameter. The corresponding PDF can be written as
	\begin{equation}
	\label{3.99}
	f(x)=\frac{2}{\pi \phi}\left[1+\left(\frac{x}{\phi}\right)^2\right]^{-1}, \quad  x>0.
	\end{equation}
	The beta-half Cauchy distribution was studied in Cordeiro and Lemonte (2011). According to (\ref{3.3})-(\ref{3.4}), the PDF of this distribution can be expressed as
	\begin{equation}
	\label{3.100}
	g(x)=\frac{2^{\alpha}}{\phi \pi^{\alpha} B(\alpha,\beta)}\left[1+\left(\frac{x}{\phi}\right)^2\right]^{-1}\left[\tan^{-1}\left(\frac{x}{\phi}\right)\right]^{\alpha-1} \left\{1-\frac{2}{\pi}\tan^{-1}\left(\frac{x}{\phi}\right) \right\}^{\beta-1} .
	\end{equation}
	The CDF and the hazard rate function of beta-half Cauchy distribution are given by,
	\begin{equation}
	\label{3.101}
	G(x)=I_{\frac{2}{\pi}\tan^{-1}(\frac{x}{\phi})}(\alpha,\beta),\quad x>0 ,
	\end{equation}
	and 
	\begin{equation}
	\label{3.102}
	h(x)=\frac{2^{\alpha}}{\phi \pi^{\alpha} B(\alpha,\beta)} \frac{[\tan^{-1}(\frac{x}{\phi})]^{\alpha-1}[1-\frac{2}{\pi}\tan^{-1}(\frac{x}{\phi})]^{\beta-1}}{[1+(\frac{x}{\phi})^2][1-I_{\frac{2}{\pi}\tan^{-1}(\frac{x}{\phi})}(\alpha,\beta)]}, \quad x>0, 
	\end{equation}
	respectively. See the above reference for more detailed information.\\
	This concludes our account of beta-generated probability distributions. The literature related to this consideration is summarized in Table (\ref{table2})-(\ref{table3}).

	\chapter{\textbf{Generalized Kumaraswamy Family of Distributions}}
	Kumaraswamy (1980) introduced a two-parameter family of distributions on $(0,1),$ that new bears his name. The CDF of Kumaraswamy distribution is defined as
	\begin{equation}
	\label{4.1}
	H(x)=1-(1-x^a)^b  ,\; x \in (0,1) ,
	\end{equation}
	where $a>0$ and $b>0$ are shape parameters. As discueed in Jones (2009), this distribution compares rather favorably in terms of simplicity with the beta CDF, which is given by the incomplete beta function ratio. 
	The PDF of this distribution is given by
	\begin{equation}
	\label{4.2}
	h(x)=ab\; x^{a-1}\;(1-x^a)^{b-1},  \;   x \in(0,1).
	\end{equation}
	The Kumaraswamy PDF (\ref{4.2}) has very similar basic shapes to that of beta distribution: $a>1$ and $b>1$ (unimodal), $a<1$ and $b<1 $ (uni-antimodal), $a> 1$ and $b\le 1$ (increasing), $a \le 1$ and $b>1$ (decreasing), $a=b=1$ (constant).\\
	According to Jones (2009), Kumaraswamy distribution has several advantages over the beta distribution. Normalizing constant of the Kumaraswamy distribution is very simple. Also, explicit formula for the distribution function is simple. This distribution has simple formula for random number generation, explicit formula for moments of order statistics and L-moments.\\
	For an arbitrary baseline CDF $F(x)$, the CDF $G(x)$ of the Kumaraswamy-generalized  distribution is defined by 
	\begin{equation}
	\label{4.3}
	G(x)=1-[1-F^a(x)]^b.
	\end{equation}
	If the distribution $F$ is continuous with the PDF $f$, the density of the generalized distribution takes the form
	\begin{equation}
	\label{4.4}
	g(x)=abf(x)F^{a-1}(x)[1-F^a(x)]^{b-1},
	\end{equation}
	where $a>0$ and $b>0$ are two shape parameters. If X is a random variable with the above PDF, we shall write $X \sim Kum-G(a,b)$. Many such generalized distributions have been introduced in the literature in recent years. We review few of them below.
	\begin{table}[ht]
		\caption{Summary of the literatures related to Kumaraswamy generated family of distributions}
		% title of Table
		\centering 
		% used for centering table
		\begin{tabular}{|p{1.5cm}|p{8.5cm}| p{6cm}|}
			% centered columns (4 columns)
			\hline                      %inserts double horizontal lines
			Number & Name of the distribution & Author(s) name \\[0.5ex]
			% inserts table 
			%heading
			\hline 
			% inserts single horizontal line
			1 & Kumaraswamy Weibull  &  Cordeiro et al. (2010) \\
			
			2 & Kumaraswamy generalized gamma & de Pascoa et al. (2011) \\ 
			3 & Kumaraswamy skew-normal  &  Kazemi et al. (2011) \\
			& & Mameli (2012) \\
			& & Mameli and Musio (2013) \\
			4 & Kumaraswamy Gumbel minimum & El-Sherpieny and Ahmed (2011) \\ 
			5 & Kumaraswamy log-logistic  & de Santana et al. (2012) \\ 
			&  & Muthulakshmi and Selvi (2013) \\
			6 & Kumaraswamy Gumbel  & Cordeiro et al. (2012) \\
			7 & Kumaraswamy Birnbaum-Sanders  & Saulo et al. (2012) \\
			8 & Kumaraswamy generalized half-normal  & Cordeiro et al. (2012) \\ 
			9 & Kumaraswamy inverse Weibull  &  Shahbaz et al. (2012) \\
			10 & Kumaraswamy normal  & Correa et al. (2012) \\
			11 & Kumaraswamy generalized inverse Weibull  & Yang (2012) \\
			12 & Kumaraswamy Pareto  & Bourguignion et al. (2013) \\
			13 & Kumaraswamy generalized Pareto  &  Nadarajah and Eljabri (2013) \\
			14 & Kumaraswamy Burr XII  & Parana\'iba et al. (2013) \\
			15 & Kumaraswamy generalized extreme value  & Eljabri (2013) \\
			16 & Kumaraswamy linear exponential  & Elbatal (2013) \\
			17 & Kumaraswamy generalized linear failure rate  &  Elbatal (2013) \\
			18 & Kumaraswamy exponentiated Pareto  & Elbatal (2013) \\
			19 & Kumaraswamy Lomax & Shams (2013) \\
			20 & Kumaraswamy modifid Weibull  &  Cordeiro et al. (2014) \\
			21 & Kumaraswamy generalized Rayleigh  & Gomes et al. (2014) \\
			22 &   Kumaraswamy-geometric  & Akinsete et al. (2014) \\
			23 & Kumaraswamy-Burr Type III  &  Behairy et al. (2016) \\
			24 &  Kumaraswamy-Lindley  &  Merovci and Sharma (2014) \\
			25 & Kumaraswamy GP  & Nadarajah and Eljabri (2013) \\
			26 & Kumaraswamy Gompertz  &  da Silva et al. (2015) \\
			27 & Kumaraswamy flexible Weibull   &  El-Damcese et al. (2016) \\
			28 & Kumaraswamy quasi Lindley &  Elbatal and Elgarhy (2013) \\
			29 &  Kumaraswamy-generalized
			exponentiated exponential  &  Mohammed (2014) \\
			30 & Kumaraswamy exponentiated
			gamma  &  Diab and Muhammed (2015) \\
			31 & Kumaraswamy Gompertz Makeham  & Chukwu and Ogunde (2016) \\
			32 & Kumaraswamy-half-Cauchy  &  Hamedani and Ghosh (2015)\\
			33 & Kumaraswamy Lindley Poisson  &  Pararai et al. (2015) \\
			34 & Kumaraswamy Kumaraswamy  &  El-Sayed et al. (2014) \\
			35 & Kumaraswamy generalized power Weibull  &  Selim  and Badr (2016) \\
			36 & Kumaraswamy exponentiated Rayleigh  &   Rashwan (2016) \\
			37 & Kumaraswamy Laplace  & Nassar (2016) \\
			& & Aryal and Zhang (2016)
			\\[1ex]
			% [1ex] adds vertical space
			\hline
			%inserts single line
		\end{tabular}
		\label{table4}
		% is usedto refer this table in the text
	\end{table}

	%	\begin{table}[ht]
	%		\caption{Table (\ref{table4} continue)}
	%		% title of Table
	%		\centering 
	%		% used for centering table
	%		\begin{tabular}{|p{1.5cm}|p{8.5cm}| p{6cm}|}
	%			% centered columns (4 columns)
	%			\hline                      %inserts double horizontal lines
	%			Number & Name of distribution & Author(s) name \\[0.5ex]
	%			% inserts table 
	%			%heading
	%			\hline 
	%		
	%			\\[1ex]
	%			% [1ex] adds vertical space
	%			\hline
	%			%inserts single line
	%		\end{tabular}
	%		\label{table5}
	%		% is usedto refer this table in the text
	%	\end{table}
	
	\section{Kumaraswamy Generalized Half Normal Distribution}
	The PDF of half normal distribution with shape parameter $\alpha >0$ and scale parameters $\theta>0$ is of the form
	\begin{equation}
	\label{4.5}
	f(x)=\sqrt{\frac{2}{\pi}}\left(\frac{\alpha}{x}\right)\left(\frac{x}{\theta}\right)^\alpha e^{-\frac{1}{2}(\frac{x}{\theta})^{2\alpha}} , \quad x>0.
	\end{equation}
	The corresponding CDF can be written in terms of error function
	\begin{equation}
	\label{4.6}
	F(x)=2\Phi \left(\frac{x}{\theta}\right)^{\alpha}-1  
	=erf\left(\frac{(\frac{x}{\theta})^{\alpha}}{\sqrt{2}}\right) ,\quad x>0,
	\end{equation}
	where
	\begin{equation}
	\label{4.7}
	\Phi(x)=\frac{1}{2}\left[1+erf\left(\frac{x}{\sqrt{2}}\right)\right],
	\end{equation}
	is the standard normal CDF and 
	\begin{equation}
	\label{4.8}
	erf(x)=\frac{2}{\sqrt{\pi}}\int_{0}^{x}e^{-t^2}dt,
	\end{equation}
	is the error special function.
	The Kumaraswamy generalized half normal distribution was studied by Cordeiro et al. (2012). This is four parameter family of distributions, given by the PDF
	\begin{equation}
	\label{4.9}
	g(x)=\sqrt{\frac{2}{\pi}}\left(\frac{\alpha}{x}\right)\left(\frac{x}{\theta}\right)^\alpha e^{-\frac{1}{2}(\frac{x}{\theta})^{2\alpha}} \left[2\Phi \left[\left(\frac{x}{\theta}\right)^{\alpha}\right]-1\right]^{a-1}[1-\left(2\Phi \left[\left(\frac{x}{\theta}\right)^{\alpha}\right]-1\right)^a]^{b-1},
	\end{equation}
	where $x>0,\; \alpha>0,\; \theta>0,\; a>0,\; b>0 $. This distribution is denoted by $ Kw-GHN(\alpha,\theta,a,b)$.
	The CDF of Kumaraswamy generalized half normal distribution is given by
	\begin{equation}
	\label{4.10}
	G(x)=1-\left[1-erf\left(\frac{(\frac{x}{\theta})^{\alpha}}{\sqrt{2}}\right)^a\right]^b,
	\end{equation}
	where $erf(\cdot)$ is the error function (\ref{4.6}). The hazard function is of the form
	\begin{equation}
	\label{4.11}
	h(x)=\frac{\sqrt{\frac{2}{\pi}}(\frac{\alpha}{x})(\frac{x}{\theta})^\alpha e^{-\frac{1}{2}(\frac{x}{\theta})^{2\alpha}} \left[2\phi \left[(\frac{x}{\theta})^{\alpha}\right]-1\right]^{a-1}[1-\left(2\phi \left[(\frac{x}{\theta})^{\alpha}\right]-1\right)^a]^{b-1}}{1-[1-\left(2\phi \left[(\frac{x}{\theta})^{\alpha}\right]-1\right)^a]^{b}} , \quad  x>0.
	\end{equation}
	Further, by setting $u=\left(\frac{x}{\theta}\right)^{\alpha}$, the $n^{th}$ moment of $X$ can be written as
	\begin{equation}
	\label{4.12}
	E(X^n)=\theta^n\sqrt{\frac{2}{\pi}}\sum_{k=0}^{\infty}t_k\;I\left(\frac{n}{\alpha},k\right),
	\end{equation}
	where
	$$I\left(\frac{n}{\alpha},k\right)=\int_{0}^{\infty} u^{\frac{n}{\alpha}} e^{-\frac{u^2}{2}}\left[erf\left(\frac{u}{\sqrt{2}}\right)\right] du. $$
	More information regarding this distribution can be found on the reference mentioned above.
	
	\section{The Kumaraswamy-Gumbel Distribution}
	The CDF of Gumbel distribution can be expressed as 
	\begin{equation}
	\label{4.13}
	F(x)=e^{-u(x)}  , \quad -\infty < x < \infty,
	\end{equation} 
	where
	\begin{equation}
	\label{4.14}
	u(x)=e^{-\left(\frac{x-\mu}{\sigma}\right)}  ,\quad -\infty < x < \infty,
	\end{equation}
	and $\mu \in R $, $\sigma>0 $ are location and scale parameters, respectively. The PDF of Gumbel distribution is
	\begin{equation}
	\label{4.15}
	f(x)=\sigma^{-1} u e^{-u(x)},\quad -\infty < x < \infty.
	\end{equation}
	According to Cordeiro et al. (2012), the PDF of the Kumaraswamy Gumbel distribution is given by
	\begin{equation}
	\label{4.16}
	g(x)=ab\sigma^{-1} u e^{-au}[1-e^{-au}]^{b-1}, \quad  -\infty < x < \infty,
	\end{equation}
	with $u(x) $ as in (\ref{4.14}). The CDF of the Kumaraswamy Gumbel distribution is given by
	\begin{equation}
	\label{4.17}
	G(x)=1-[1-e^{-au}]^b, \quad -\infty < x < \infty.
	\end{equation}
	The hazard rate function of Kumaraswamy Gumbel distribution is defined by
	\begin{equation}
	\label{4.18}
	h(x)=\frac{abue^{-au}}{\sigma[1-e^{-au}]}  ,\quad -\infty < x < \infty,
	\end{equation}
	where $u(x)$ is given by (\ref{4.14}). The $n^{th}$ moment of $X$ can be expressed as
	\begin{equation}
	\label{4.19}
	E(X^n)=a\Gamma(b+1)\sum_{k=0}^{\infty}\frac{(-1)^k}{\Gamma(b-k) k!}\int_{0}^{\infty}[\mu-\sigma \log(u)]^n e^{-(k+1)au} du.
	\end{equation}
	More information on this model can be found in Cordeiro et al. (2012).
	% \begin{equation}
	% \label{gum7}

	\section{The Kumaraswamy Weibull Distribution}
	The PDF of Weibull distribution is given by
	\begin{equation}
	\label{4.20}
	f(x)=c \; \lambda^c\; x^{c-1}e^{{-\left(\lambda x\right)}^c}, \quad x>0,
	\end{equation}
	where $\lambda>0$ is a scale parameter and $c>0$ is a shape parameter.
	The CDF is given by
	\begin{equation}
	\label{4.21}
	F(x)=1-e^{-{\left(\lambda x\right)}^c}, \quad x>0.
	\end{equation}
	The proposed PDF of the Kumaraswamy-Weibull distribution, given by Cordeiro et al. (2010) can be written as
	\begin{equation}
	\label{4.22}
	g(x)=ab\; c \lambda^c x^{c-1} e^{-(\lambda x)^c}[1-e^{-(\lambda x)^c}]^{a-1} (1-[1-e^{-(\lambda x)^c}]^{a-1} )^{b-1}, \quad x>0.
	\end{equation}
	The CDF of this distribution is
	\begin{equation}
	\label{4.23}
	G(x)=1-(1-[1-e^{{\left(-\lambda x\right)}^c}]^a)^b.
	\end{equation}
	The associated hazard rate function is
	\begin{equation}
	\label{4.24}
	h(x)=\frac{abc \lambda^c x^{c-1} e^{(-\lambda x)^c}[1-e^{(-\lambda x)^c}]^{a-1} }{1-[1-e^{(-\lambda x)^c}]^{a}}, \; x>0.
	\end{equation}
	More information on this model, denoted by $X \sim KumW(\lambda,c,a,b)$, can be found in Cordeiro et al. (2010).

	\section{The Kumaraswamy Laplace Distribution}
	A random variable $X$ has Laplace distribution with location parameter $\mu $ and scale parameter $\sigma >0 $ when its PDF is given by
	
	\begin{equation}
	\label{4.25}
	f(x)=\frac{1}{2\sigma}e^{-|\frac{x-\mu}{\sigma}|} , \quad \ -\infty<x<\infty.
	\end{equation}
	The CDF of Laplace distribution is  given by
	\begin{equation}
	\label{4.26}
	F(x)=\begin{cases}
	\frac{1}{2}e^{\frac{x-\mu}{\sigma}}, & x<\mu\\
	1-\frac{1}{2}e^{-\frac{x-\mu}{\sigma}}, & x\ge \mu.\\
	\end{cases}
	\end{equation}
	Thus, the CDF of Kumaraswamy-Laplace distribution, studied by Aryal and Zhang (2016) and Nassar (2016), is of the form
	\begin{equation}
	\label{4.27}
	G(x)=\begin{cases}
	1-(1-\left[\frac{1}{2}e^{\frac{x-\mu}{\sigma}}\right]^a)^b, & x<\mu  \\
	1-(1-\left[1-\frac{1}{2}e^{-\frac{x-\mu}{\sigma}}\right]^a)^b, & x\ge \mu.  \\
	\end{cases}
	\end{equation}
	The corresponding PDF can be expressed as
	\begin{equation}
	\label{4.28}
	g(x)=\begin{cases}
	\frac{2^{-a}ab[e^{\frac{x-\mu}{\sigma}}]^a[1-2^{-a}(e^{\frac{x-\mu}{\sigma}})^a]^{b-1}}{\sigma}, & x<\mu  \\
	
	\frac{abe^{\frac{x-u}{\sigma}}(1-\frac{1}{2}e^{\frac{x-\mu}{\sigma}})^{a-1}[ 1-(1-\left[\frac{1}{2}e^{-\frac{x-\mu}{\sigma}}\right]^a)^{b-1}]}{2\sigma}, & x\ge \mu. \\
	\end{cases} 
	\end{equation}
	As shown in Nassar (2016), moments of the Kumaraswamy Laplace distribution are given by 
	\begin{equation}
	\label{4.29}
	E(X^r)=m\int_{-\infty}^{0}x^re^{a(l+1)x}dx+k\int_{0}^{\infty}x^re^{-a(l+1)x}dx ,
	\end{equation}
	where
	$$m=\frac{ab}{2^a}\sum_{l=0}^{\infty}2^{-al}(-1)^l\binom{b-1}{l},$$
	and
	$$k=\frac{ab}{2}\sum_{j=0}^{\infty}\sum_{i=0}^{\infty}(-1)^{i+j}\binom{b-1}{j}\binom{a(j+1)-1}{i}2^{-i}.$$
	More information on this model can be found in Aryal and Zhang (2016) and Nassar (2016).
	
	% \section*{The Kumaraswamy-Geometric Distribution}
	
	% If X is a random variable having the geometric distribution with parameter P, then the probability    %density function may be written as
	
	% $$P(X=x)=pq^x , x=0,1,2,....., p+q=1 $$
	
	% Where p is the probability of success in a single Bernoulli trial.
	
	%The cdf of the geometric distribution is given by
	
	%$$p(X \le x)=1-q^{x+1}  , x=0,1,2....$$ 

	%The Proposed Probability density function for the Kumaraswamy-geometric distribution is
	
	%$$g(x)=[1-(1-q^{x})^a]^b-[1-(1-q^{x+1})^a]^b   ,x=0,1,2...a>0,b>0 $$
	
	%The cumulative density function of the Kumaraswamy-Geometric distribution is
	
	% $$G(x)=1-[1-(1-q^{x+1})^a]^b  ,x=0,1,2.....$$ 
	
	% The rth moment of KuG iswritten as
	
	% $$E(X^r)=\sum_{x=1}^{\infty}x^r[1-(1-q^x)^a]^b-\sum_{x=1}^{\infty}x^r[1-(1-q^{x+1})^a]^b $$ 
	
	% $$=\sum_{x=1}^{\infty}x^r \left[\left(\sum_{i=1}^{\infty} (-1)^{i-1} \binom{a}{i}q^{xi}\right)^b-\left(\sum_{i=1}^{\infty}(-1)^{i-1} \binom{a}{i}q^{(x+1)i}\right)^b\right]$$
	
	% and the corresponding Hazard Function is defined as
	
	%$$h(x)=\left(\frac{1-(1-q^x)^a}{1-(1-q^{x+1})^a}\right)^b-1 $$
	
	\section{The Kumaraswamy Lindley Distribution}
	A random variable $X$ has Lindley distribution if it has the PDF as follows :
	\begin{equation}
	\label{4.30}
	f(x)=\frac{\theta^2}{\theta+1}(1+x)e^{-\theta x},\quad x>0,
	\end{equation}
	where $\theta >0$ is a single parameter. The CDF of Lindley distribution can be written as
	
	\begin{equation}
	\label{4.31}
	F(x)=1-\frac{\theta+1+\theta x}{\theta+1}e^{-\theta x}, \quad x>0.
	\end{equation}
	The Kumaraswamy Lindley distribution was studied by Oluyede et al. (2015). The CDF of Kumaraswamy Lindley distribution becomes
	\begin{equation}
	\label{4.32}
	G(x)=1-[1-(1-\frac{\theta+1+\theta x}{\theta+1}e^{-\theta x})^a]^b, \quad x>0.
	\end{equation}
	The corresponding PDF, defined for $x>0$, can be written as
	\begin{equation}
	\label{4.33}
	g(x)=ab \frac{\theta^2}{\theta+1}(1+x)e^{-\theta x}\left[1-\frac{\theta+1+\theta x}{\theta+1}e^{-\theta x}\right]^{a-1}\left[1-(1-\frac{\theta+1+\theta x}{\theta+1}e^{-\theta x})^a\right]^{b-1}.
	\end{equation}
	The hazard rate function of the Kumaraswamy-Lindley distribution can be described as
	\begin{equation}
	\label{4.34}
	h(x)=\frac{ab \frac{\theta^2}{\theta+1}(1+x)e^{-\theta x}\left[1-\frac{\theta+1+\theta x}{\theta+1}e^{-\theta x}\right]^{a-1}}{1-(1-\frac{\theta+1+\theta x}{\theta+1}e^{-\theta x})^a}, \quad x>0.
	\end{equation}
	More information can be found in the above literature. 
	
	\section{The Kumaraswamy Burr Type-III Distribution}
	If X is a random variable with Burr-III distribution, then its CDF is given by
	\begin{equation}
	\label{4.35}
	F(x)=(1+x^{-c})^{-k} , \quad x>0, 
	\end{equation}
	where $c>0$ and $k>0$ are two shape parameters. The PDF of Burr-Type III distribution can be written as
	\begin{equation}
	\label{4.36}
	f(x)=ckx^{-(c+1)}(1+x^{-c})^{-(k+1)}, \quad x>0.
	\end{equation}
	The Kumaraswamy-Burr-Type III distribution was studied by Behairy et al. (2016). The PDF of this model is
	\begin{equation}
	\label{4.37}
	g(x)=ab\;c\; kx^{-(c+1)}(1+x^{-c}){-(ak+1)}(1-(1+x^{-c})^{-ak})^{b-1}, \quad x>0,
	\end{equation}
	while the CDF can be written as
	\begin{equation}
	\label{4.38}
	G(x)=1-(1-(1+x^{-c})^{-ak})^b, \quad x>0.
	\end{equation}
	The hazard rate function of the Kumaraswamy Burr Type-III distribution is given by
	\begin{equation}
	\label{4.39}
	h(x)=\frac{ab\;c \; kx^{-(c+1)}(1+x^{-c}){-(ak+1)}}{1-(1+x^{-c})^{-ak}}, \quad x>0.
	\end{equation}
	More information can be found in the above literature.
	
	% The rth moment of Kumaraswamy Burr Type-III Distribution becomes

	% where $s=ak(j+1)+1 $ and $w_j=(-1)^j \binom{b-1}{j}$

	\section{The Kumaraswamy-Log Logistic Distribution}
	The CDF of log-logistic distribution is
	\begin{equation}
	\label{4.40}
	F(x)=1-\left[1+\left(\frac{x}{\alpha}\right)^{\gamma}\right]^{-1} , \quad x>0,
	\end{equation}
	where $\alpha>0$ is a scale parameter and $\gamma>0$ is a shape parameter. The PDF of this distribution can be written as
	\begin{equation}
	\label{4.41}
	f(x)=\frac{\gamma}{\alpha^{\gamma}}x^{\gamma-1}\left[1+\left(\frac{x}{\alpha}\right)^{\gamma}\right]^{-2},\quad x>0.
	\end{equation}
	The four parameter Kumaraswamy-log-logistic distribution was studied by de Santana et al. (2012). The PDF of this distribution is given by
	\begin{equation}
	\label{4.42}
	g(x)=\frac{ab \gamma}{\alpha^{a \gamma}}x^{a \gamma-1} \left[1+\left(\frac{x}{\alpha}\right)^{\gamma}\right]^{-(a+1)} \left(1-\left[1-\frac{1}{1+(\frac{x}{\alpha})^{\gamma}}\right]^a\right)^{b-1}, \quad x>0,
	\end{equation}
	with the hazard rate of the form
	\begin{equation}
	\label{4.43}
	h(x)=\frac{\frac{ab \gamma}{\alpha^{a \gamma}}x^{a \gamma-1} \left[1+\left(\frac{x}{\alpha}\right)^{\gamma}\right]^{-(a+1)}}{\left(1-\left[1-\frac{1}{1+(\frac{x}{\alpha})^{\gamma}}\right]^a\right)}.
	\end{equation}
	For more information on this model, see de Santana et al. (2012).

	\section{The Kumaraswamy Gompertz Distribution}
	If $X$ has Gompertz distribution with parameters $\theta>0$ and $\gamma>0$, denoted by $X \sim KG(\theta,\gamma)$, than its the CDF is given by
	\begin{equation}
	\label{4.44}
	F(x)=1-e^{-\frac{\theta}{\gamma}(e^{\gamma x}-1)} , \quad x>0 ,
	\end{equation}
	while its PDF can be written as
	\begin{equation}
	\label{4.45}
	f(x)=\theta e^{\gamma x-\frac{\theta}{\gamma}(e^{\gamma x}-1)} ,\quad x>0.
	\end{equation}
	The four-parameter Kumaraswamy Gompertz distribution was first studied by da Silva et al. (2015). The CDF of this distribution is of the form
	\begin{equation}
	\label{4.46}
	G(x)=1-\left[1-(1-e^{-\frac{\theta}{\gamma}(e^{\gamma x}-1)})^a\right]^b , \quad x>0. 
	\end{equation}
	Here we have shape parameters $\theta,a,b$ and a positive scale parameter $\gamma$. The PDF of the Kumaraswamy Gompertz distribution is
	\begin{equation}
	\label{4.47}
	g(x)=ab\theta e^{\gamma x--\frac{\theta}{\gamma}(e^{\gamma x}-1)} \left[1-e^{-\frac{\theta}{\gamma}(e^{\gamma x}-1)}\right]^{a-1} \left[1-\left(1-e^{-\frac{\theta}{\gamma}(e^{\gamma x}-1)}\right)^a\right]^{b-1}, \quad x>0,
	\end{equation}
	while the corresponding hazard rate function becomes
	\begin{equation}
	\label{4.48}
	h(x)=\frac{ab\theta e^{\gamma x--\frac{\theta}{\gamma}(e^{\gamma x}-1)} \left[1-e^{-\frac{\theta}{\gamma}(e^{\gamma x}-1)}\right]^{a-1}}{1-\left(1-e^{-\frac{\theta}{\gamma}(e^{\gamma x}-1)}\right)^a}, \quad x>0.
	\end{equation}
	More information on this model can be found in da Silva et al. (2015).

	\section{The Kumaraswamy Birnbaum Saunders Distribution}
	A random variable $X$ follows a Birnbaum Saunders distribution with parameters $\alpha,\beta >0 $, denoted $BS(\alpha,\beta)$, if its CDF is of the form
	\begin{equation}
	\label{4.49}
	F(x;\alpha,\beta)=\Phi \left(\frac{1}{\alpha} \left[\left(\frac{x}{\beta}\right)^{\frac{1}{2}}-\left(\frac{\beta}{x}\right)^{\frac{1}{2}}\right]    \right), \quad x>0,
	\end{equation}
	where $\Phi$ denotes the standard normal distribution function. The Kumaraswamy Birnbaum-Saunders distribution was studied by Saulo et al. (2012). The corresponding PDF is given by
	\begin{equation}
	\label{4.50}
	f(x;\alpha,\beta)=k(\alpha,\beta)x^{-3/2}(x+\beta) e^{-\tau(x/\beta) 2\alpha^2} , \quad x>0,
	\end{equation}
	where $$k(\alpha,\beta)=\frac{e^{\alpha^{-2}}}{2\alpha\sqrt{2\pi \beta}},$$ and $$\tau(z)=z+z^{-1}.$$
	The CDF of this distribution is given by
	\begin{equation}
	\label{4.51}
	G(x;\alpha,\beta,a,b)=1-\left\{1-\Phi \left(\frac{1}{\alpha} \left[\left(\frac{x}{\beta}\right)^{\frac{1}{2}}-\left(\frac{\beta}{x}\right)^{\frac{1}{2}}\right]    \right)^a \right\}^b, \quad x>0,
	\end{equation}
	where $\beta$ is a scale parameter and the other positive parameters $\alpha,a,b $ are shape parameters.
	The corresponding PDF of the Kumaraswamy Birnbaum Saunders distribution becomes
	\begin{equation}
	\label{4.52}
	g(x;\alpha,\beta,a,b)=abk(\alpha,\beta)x^{-3/2}(x+\beta) e^{-\tau(x/\beta) 2\alpha^2}\phi(v)^{a-1}[1-\phi(v)^a]^{b-1}, \quad x>0.
	\end{equation}
	The hazard rate function can be written as
	\begin{equation}
	\label{4.53}
	h(x;\alpha,\beta,a,b)=\frac{abk(\alpha,\beta)x^{-3/2}(x+\beta) e^{-\tau(x/\beta) 2\alpha^2}\phi(v)^{a-1}}{1-\phi(v)^a}.
	\end{equation}
	More information on this distribution can be found in Saulo et al. (2012).
	
	\section{The Kumaraswamy-Kumaraswamy Distribution}
	The CDF of the Kumaraswamy distribution with two shape parameters $\alpha>0$ and $\beta>0$, respectively, is
	\begin{equation}
	\label{4.54}
	F(x)=1-(1-x^{\alpha})^{\beta} , \quad  0<x<1 , 
	\end{equation}
	and the corresponding PDF of this distribution is
	\begin{equation}
	\label{4.55}
	f(x)=\alpha \beta x^{\alpha-1}(1-x^{\alpha})^{\beta-1} , \quad 0<x<1 .
	\end{equation}
	Kumaraswamy-Kumaraswamy distribution was first studied by Sayed et al. (2014). 
	The proposed CDF of the Kumaraswamy-Kumaraswamy distribution is
	\begin{equation}
	\label{4.56}
	G(x)=1-\left(1-\left[1-(1-x^{\alpha})^{\beta}\right]^{a}\right)^{b} , \quad 0<x<1 , 
	\end{equation}
	while the PDF can be written as
	\begin{equation}
	\label{4.57}
	g(x)=ab\alpha \beta x^{\alpha-1}(1-x^{\alpha})^{\beta-1}  \left[1-(1-x^{\alpha})^{\beta}\right]^{a-1} \left(1-\left[1-(1-x^{\alpha})^{\beta}\right]^{a}\right)^{b-1}, \quad 0<x<1.
	\end{equation}
	The $r^{th}$ moment can be written as
	\begin{equation}
	\label{4.58}
	EX^r=\sum_{i,j=0}^{\infty}w_{i,j}\int_{0}^{1}x^{r+\alpha-1}(1-x^{\alpha})^{\beta(1+j)-1} dx ,
	\end{equation}
	where
	$$w_{i,j}=ab\alpha\beta\frac{(-1)^{i+j}}{j!}\frac{\Gamma(a(1+j))}{\Gamma(a(1+i)-j)}.$$
	More information on this model can be found in Sayed et al. (2014).

	\section{The Kumaraswamy Burr XII Distribution}
	The three-parameter Burr XII distribution is given by CDF
	\begin{equation}
	\label{4.59}
	F(x) =1-\left[1+\left(\frac{x}{s}\right)^c\right]^{-k}, \quad x>0,
	\end{equation}
	and the corresponding PDF is
	\begin{equation}
	\label{4.60}
	f(x)=cks^{-c}x^{c-1}\left[1+\left(\frac{x}{s}\right)^c\right]^{-k-1}, \quad x>0,
	\end{equation}
	where $k>0$ and $c>0$ are shape parameters and $s>0$ is a scale parameter. The Kumaraswamy Burr XII distribution was studied by Paranaiba et al. (2013). Its CDF is
	\begin{equation}
	\label{4.61}
	G(x)=1-\left[1-\left(1-\left[1+\left(\frac{x}{s}\right)^c\right]^{-k}\right)^a\right]^b, \quad x>0,
	\end{equation}
	where its corresponding PDF can be expressed as
	\begin{equation}
	\label{4.62}
	g(x)=\frac{abckx^{c-1}}{s^{c}\left[1+\left(\frac{x}{s}\right)^c\right]^{k+1}}\left[1-\left[1+\left(\frac{x}{s}\right)^c\right]^{-k}\right]^{a-1} \left[1-\left(1-\left[1+\left(\frac{x}{s}\right)^c\right]^{-k}\right)^a\right]^{b-1}, \; x>0.
	\end{equation}
	The hazard rate function for the Kumaraswamy Burr XII distribution is given by
	\begin{equation}
	\label{4.63}
	h(x)=\frac{abcks^{-c}x^{c-1}\left[1+\left(\frac{x}{s}\right)^c\right]^{-k-1} \left[1-\left[1+\left(\frac{x}{s}\right)^c\right]^{-k}\right]^{a-1}}{1-\left(1-\left[1+\left(\frac{x}{s}\right)^c\right]^{-k}\right)^a}, \quad x>0.
	\end{equation}
	More information on this model can be found in Paranaiba et al. (2013).
	
	\section{The Kumaraswamy Half Cauchy Distribution}
	
	If $X$ follows a half-Cauchy distribution with parameter $\delta$, then its PDF of the form
	\begin{equation}
	\label{4.64}
	f(x)=\frac{2}{\pi \delta}\left(1+[\frac{x}{\delta}]^2\right)^{-1}, \quad x>0, 
	\end{equation}
	and the corresponding CDF can be written as
	\begin{equation}
	\label{4.65}
	F(x)=\frac{2}{\pi}\tan^{-1}\left(\frac{x}{\delta}\right) , \quad x>0. 
	\end{equation}
	The Kumaraswamy half-Cauchy distribution was studied by Ghosh (2014). The PDF of Kumaraswamy-half Cauchy distribution can be written as
	\begin{equation}
	\label{4.66}
	g(x)=\frac{ab 2^a}{\delta \pi^a}(\tan^{-1}\left(\frac{x}{\delta})\right)^{a-1}(1-\left(\frac{x}{\delta}\right)^a)^{b-1}(1+\left(\frac{x}{\delta}\right)^2)^{-1}, \quad x>0.
	\end{equation}
	The CDF of this distribution can be written as
	\begin{equation}
	\label{4.67}
	G(x)=1-\left(1-\left[\frac{2}{\pi}\tan^{-1}\left(\frac{x}{\delta}\right)\right]^a\right)^b , \quad x>0.
	\end{equation}
	The hazard rate function associated with Kumaraswamy half-Cauchy distribution becomes
	\begin{equation}
	\label{ch5}
	h(x)=\frac{\frac{ab 2^a}{\delta \pi^a}(\tan^{-1}(\frac{x}{\delta}))^{a-1}(1+(\frac{x}{\delta})^2)^{-1}}{\left(1-\left[\frac{2}{\pi}\tan^{-1}\right]^a\right)^b } , \quad x>0.
	\end{equation}
	More information on this model can be found in Ghosh (2014). \\
	This concludes our account of Kumaraswamy-generated family of distributions. These, and other related papers on this topic, are summarized in Table (\ref{table4}).

	\chapter{\textbf{A New Generalized Asymmetric Laplace Distribution}}
	We would now introduce a new class of generalized distributions obtained via (\ref{eq1.6}) with $T$ having a mixture of two beta distributions, with the CDF of $T$ of the form 
	\begin{equation}
	\label{5.1}
	H(x)=p\; H_1(x)+(1-p)\; H_2(x)\; ,
	\end{equation}
	where $H_1$ and $H_2$ correspond to two special cases of beta distribution, where one of the parameters is equal to $1$. We start with reviewing special cases of beta distribution.\\
	First, recall that the PDF of Beta distribution is of the form 
	\begin{equation}
	\label{5.2}
	h(t)=\frac{\Gamma(\alpha+\beta)}{\Gamma(\alpha)\Gamma(\beta)} t^{\alpha-1}(1-t)^{\beta-1}  \; , t\in (0,1) , \;\alpha,\beta>0 .
	\end{equation}
	Particular special cases of this distribution are as follows :\\
	\textbf{Case I : $\alpha=\beta=1.$} Here, the PDF simplifies to
	\begin{equation} 
	\label{5.3}
	h(t) =\frac{\Gamma(2)}{\Gamma(1)} t^0 (1-t)^0 =1 , \quad t\in(0,1),
	\end{equation}
	and	we obtain standard uniform distribution.\\
	\textbf{Case II : $\beta=1.$} In this case, the beta PDF if of the form 
	\begin{equation} 
	\label{5.4}
	h(t) =\frac{\Gamma(\alpha+1)}{\Gamma(\alpha)} t^{\alpha-1} =\alpha t^{\alpha-1} , \quad t\in(0,1),
	\end{equation}
	with the corresponding CDF being
	\begin{equation} 
	\label{5.5}
	H(t) =\int_{0}^{t} h(x) dx =\int_{0}^{t} \alpha x^{\alpha-1} dx = t^{\alpha}, \quad t\in(0,1).
	\end{equation} 
	This particular beta distributions has an explicit form of the CDF, in contrast with general beta distribution. Because of the special form of the CDF, it is known as \textit{Power Function} distribution.\\
	\textbf{Remark:} If we generalize $F$ via (\ref{eq1.6}) with this particular variable $T$, the CDF of the generalized distribution will be of the form 
	\begin{equation} 
	\label{5.6}
	G(y) =\int_{0}^{F(y)} h(t) dt= [F(y)]^{\alpha}.
	\end{equation}
	\textbf{Case III : $\alpha=1.$} Similarly to the above case, here we have 
	\begin{equation} 
	\label{5.7}
	h(t) =\frac{\Gamma(\beta+1)}{\Gamma(\beta)} (1-t)^{\beta-1} =\beta (1-t)^{\beta-1} , \quad t\in(0,1).
	\end{equation}
	In turn, the CDF admits an explicit form as well,
	\begin{equation} 
	\label{5.8}
	H(t) =\int_{0}^{t} h(x) dx =\int_{0}^{t} \beta (1-x)^{\beta-1} dx = 1-(1-t)^{\beta}, \quad t\in(0,1).
	\end{equation}  
	\textbf{Remark:} If we generalize $F$ via (\ref{eq1.6}) with this particular variable $T$, the CDF of the generalization will be of the form :
	\begin{equation} 
	\label{5.9}
	G(y) =\int_{0}^{F(y)} h(t) dt =H(F(t))= 1-[1-F(y)]^{\beta}.
	\end{equation} 
	We now revisit the important property of this construction related to mixing, presented in Proposition (\ref{2.5}). Let $T_{1},T_{2} $ be two random variables with support on $(0,1)$, with the PDFs $h_{1}(t),h_{2}(t) $  respectively. The corresponding CDFs are $H_{1}(t), H_{2}(t)$, respectively. If these distributions are mixed with weights $p,1-p$, where $p\in [0,1]$, the PDF of the mixture will be of the form : 
	\begin{equation} 
	\label{5.10}
	h(t)  =p h_{1}(t)+(1-p) h_{2}(t)  , \; t\in(0,1).
	\end{equation}
	In turn, the corresponding CDF will be 
	\begin{equation} 
	\label{5.11}
	\begin{split}
	H(t) & =\int_{0}^{t} [p h_{1}(x)+(1-p) h_{2}(x)]dx \\
	& = p \int_{0}^{t} h_{1}(x) dx +(1-p) \int_{0}^{t} h_2(x) dx \\
	& = p H_{1}(t) +(1-p) H_{2}(t)  ,\; t\in(0,1).
	\end{split}
	\end{equation}	
	If we now generalized $F$ via (\ref{eq1.6}) with the above $H$, then, according to Proposition (\ref{2.5}), the generalized distribution will have the CDF of the form
	\begin{equation} 
	\label{5.12}
	G(y)=p\; G_1(y)+(1-p)\; G_2(y),
	\end{equation}
	where
	$$G_1(y)=H_1(F(y)),$$
	and
	$$G_2(y)=H_2(F(y)).$$
	We shall now utilize this constructions using $T_1$ and $T_2$ having special beta distributions discussed in the above special cases. Namely, we associate $T_1$ with the $T$ in special case II, so that
	\begin{equation}
	\label{5.13}
	H_1(t)=t^{\alpha}, \qquad h_1(t)=\alpha t^{\alpha-1},
	\end{equation}
	while $T_2$ is associated with the $T$ discussed in special case III, leading to 
	\begin{equation}
	\label{5.14}
	H_2(t)=1-(1-t)^{\beta}, \qquad h_2(t)=\beta (1-t)^{\beta-1}.
	\end{equation}
	Consequently, the PDF of $T$ in (\ref{5.10}) 
	\begin{equation}
	\label{5.15}
	h(t)=p\;\alpha t^{\alpha-1}+(1-p)\;\beta (1-t)^{\beta-1},\; t\in(0,1),
	\end{equation}
	with the corresponding CDF in (\ref{5.11}) being of the form
	\begin{equation}
	\label{5.16}
	H(t)=p\;t^{\alpha}+(1-p)\;[1-(1-t)^{\beta}],\; t\in(0,1).
	\end{equation}
	Consequently, the generalized CDF $G$, obtained from $F$ and the above $H$ via (\ref{eq1.6}), will be of the form
	\begin{equation}
	\label{5.17}
	G(x)=p[F(x)]^{\alpha}+(1-p)[1-(1-F(x))^{\beta}]\; .
	\end{equation}
	Note that in contrast with generalized beta distribution, here the CDF will have an explicit form. The PDF corresponding to (\ref{5.17}) will be
	\begin{equation}
	\label{5.18}
	g(x)=\left[p\alpha(F(x))^{\alpha-1}+(1-p)\beta(1-F(x))^{\beta-1}\right] f(x) \; ,
	\end{equation}
	where $f(\cdot)$ is the PDF corresponding to the base CDF $F$. In the sequel, we shall use the notation $BM(\alpha,\beta)$ to denote the distribution with the PDF and CDF as in (\ref{5.15}) and (\ref{5.16}), respectively. The term $BM$ connects with the fact that this is a mixture of beta distributions (so $BM$ stands for beta mixture). Similarly, a generalized distribution with the PDF (\ref{5.18}) based on $F$ shall be denoted as $BM-F$.
	
	\section{A New Asymmetric Laplace Distribution}
\begin{figure}[H]
		\centering
		\includegraphics[width=.8\textwidth]{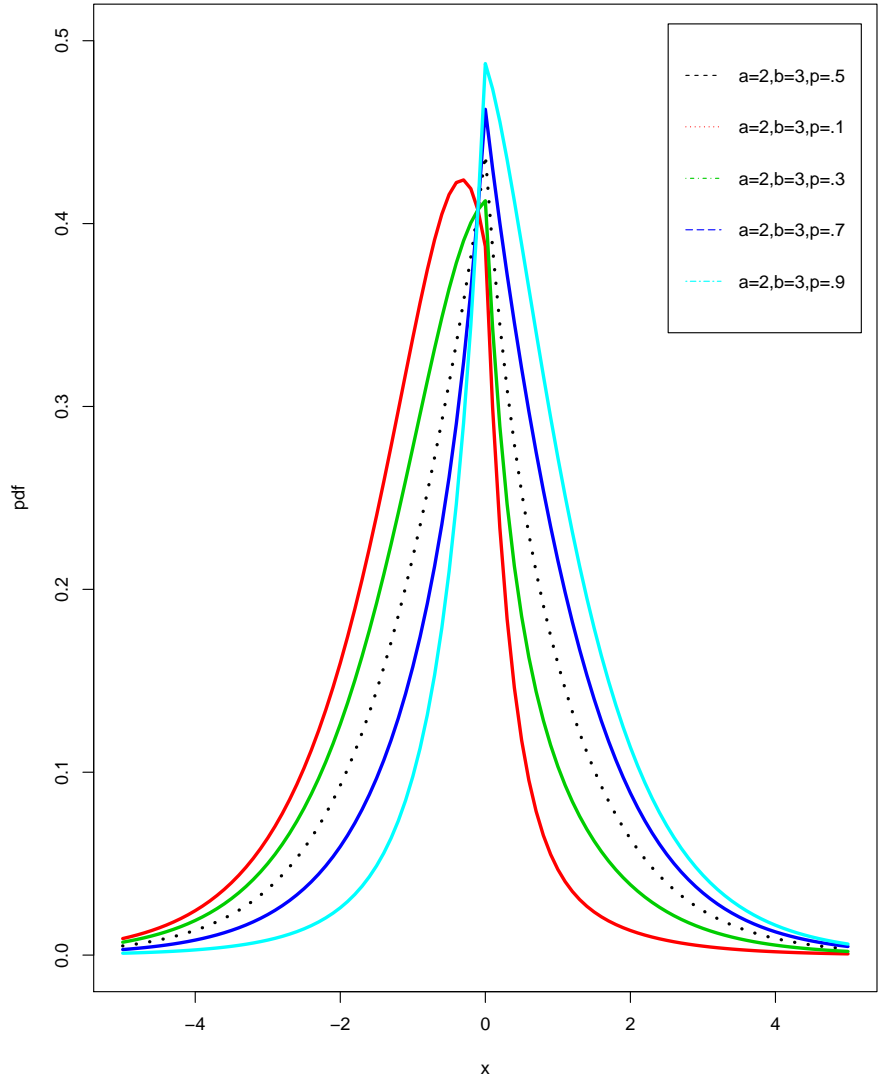}
		\caption{Selected PDFs of $BML$ distributions with $\alpha=2,\beta=3,$ and selected values of $p$.}
		\label{6.1}
	\end{figure}

\begin{figure}[H]
		\centering
		\includegraphics[width=.8\textwidth]{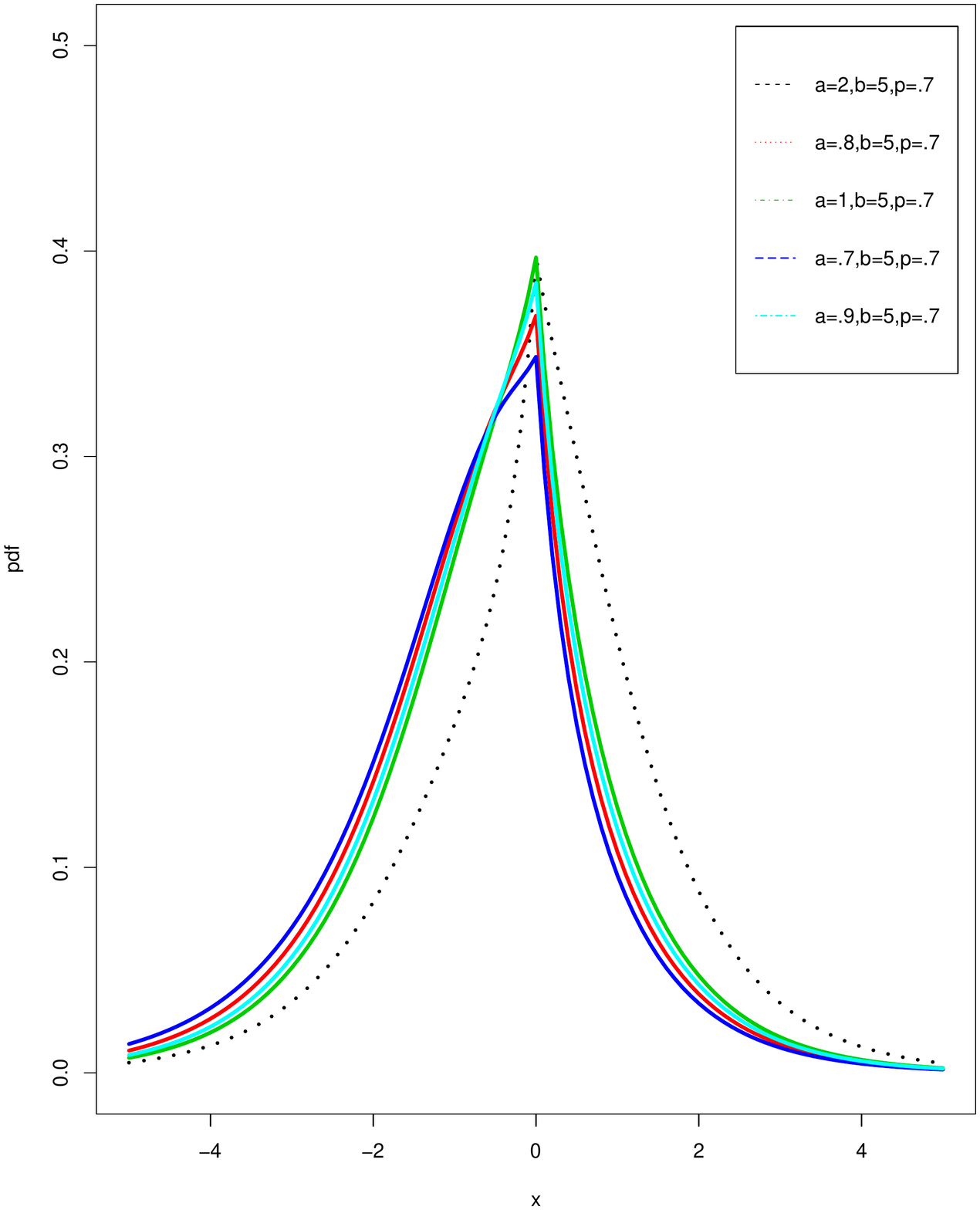}
		\caption{Selected PDFs of $BML$ distributions with $\beta=5, p=.7$, and selected values of $\alpha$.}
		\label{6.2}
	\end{figure}

\begin{figure}[H]
		\centering
		\includegraphics[width=.8\textwidth]{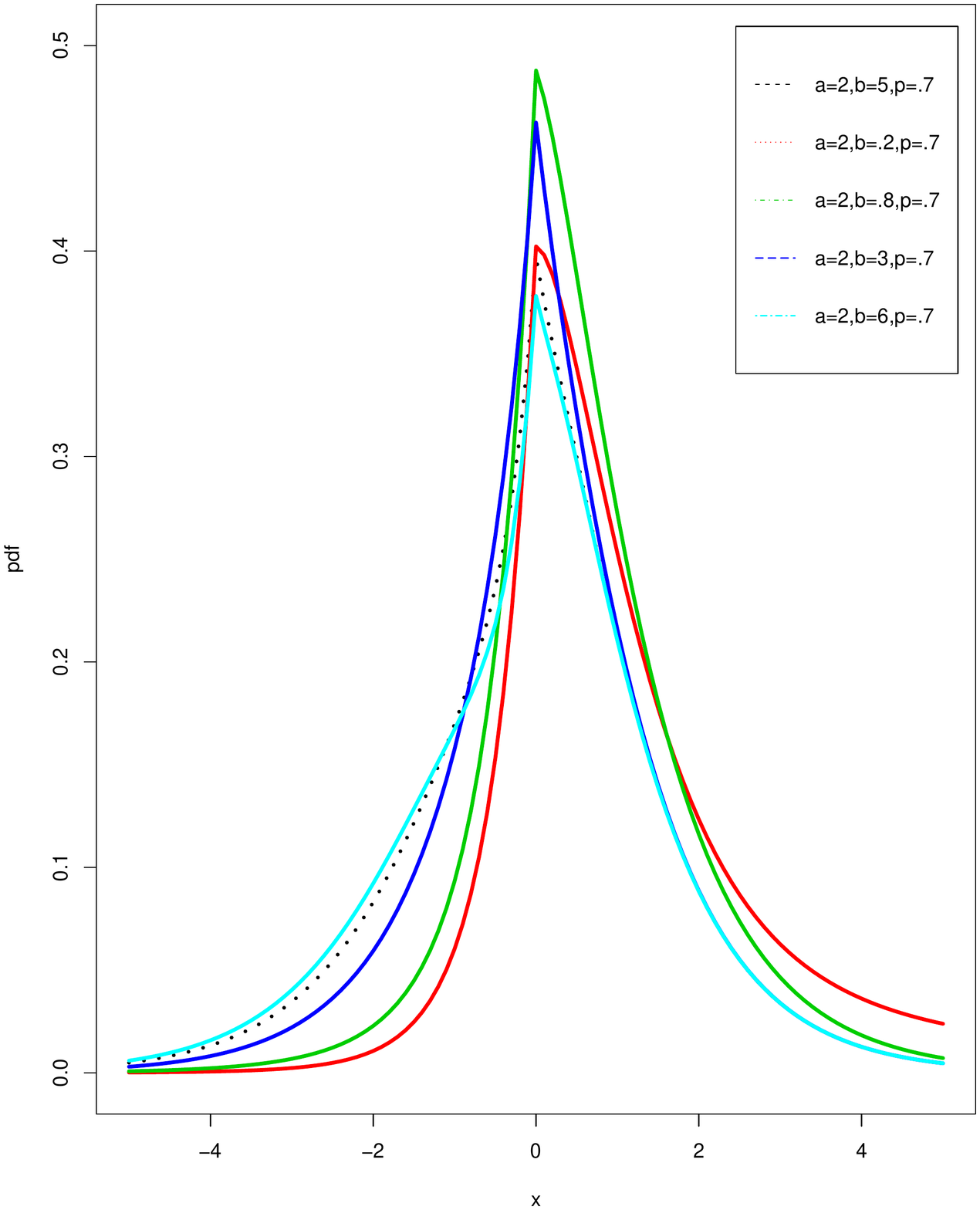}
		\caption{Selected PDFs of $BML$ distributions with $\alpha=2, p=.7$, and selected values of $\beta$.}
		\label{6.3}
	\end{figure}

	We now follow up on the ideas set up above to obtain a new generalization of the Laplace distribution. Recall that, Laplace distribution has the PDF and CDF specified as
	\begin{equation}
	\label{5.19}
	f(x)=\frac{1}{2\sigma}e^{-|\frac{x-\mu}{\sigma}|} , \quad \ -\infty<x<\infty,
	\end{equation}
	and
	\begin{equation}
	\label{5.20}
	F(x)=\begin{cases}
	\frac{1}{2}e^{\frac{x-\mu}{\sigma}}, & x<\mu\\
	1-\frac{1}{2}e^{-\frac{x-\mu}{\sigma}}, & x\ge \mu,\\
	\end{cases}
	\end{equation}
	respectively. Upon standardization, $$Z=\frac{x-\mu}{\sigma},$$ the PDF and CDF of $Z$ reduce to
	\begin{equation}
	\label{5.21}
	f(z)=\frac{1}{2}e^{-|z|} ,\quad \ -\infty<z<\infty,
	\end{equation}
	and
	\begin{equation}
	\label{5.22}
	F(z)=\begin{cases}
	\frac{1}{2}e^{z}, & z<0\\
	1-\frac{1}{2}e^{-z}, & z\ge 0,\\
	\end{cases}
	\end{equation}
	respectively. We now apply (\ref{eq1.6}) with $H$ as in (\ref{5.16}) and $h$ as in (\ref{5.15}) to obtain a generalized distribution, which has PDF of the following form :
	\begin{equation}
	\label{5.23}
	g(x)=\begin{cases}
	p \alpha (\frac{1}{2}e^{x})^{\alpha}+(1-p) \beta [1-\frac{1}{2}e^{x}]^{\beta-1} \frac{1}{2}e^{x},  & x<0\\
	p \alpha (1-\frac{1}{2}e^{-x})^{\alpha-1} \frac{1}{2}e^{-x}+(1-p) \beta (\frac{1}{2}e^{-x})^{\beta}, & x\ge 0.\\
	\end{cases}
	\end{equation}
	We shall refer to this distribution as the $BML$ model, which stands for {\bf b}eta {\bf m}ixture {\bf L}aplace, and denote this distribution by $BML(\alpha,\beta,p)$. Figure (\ref{6.1})-(\ref{6.3}) presented selected PDFs from this new stochastic model.\\  
	\textbf{Remark.} A more generalized scale-location model can be defined through 
	\begin{equation*}
		g_{\mu,\sigma}(x)=\frac{1}{\sigma}g(\frac{x-\mu}{\sigma}), \quad x \in R,
	\end{equation*}
	with $\mu \in R,\; \sigma>0$ and $g$ is given by (\ref{5.23}). Although we focus on the standard model with $\mu=0$ and $\sigma=1$, the properties of the general case can be easily obtained from those for the standard case.
	% Plot for different P values
	
\begin{figure}[H]
		\centering
		\includegraphics[width=.8\textwidth]{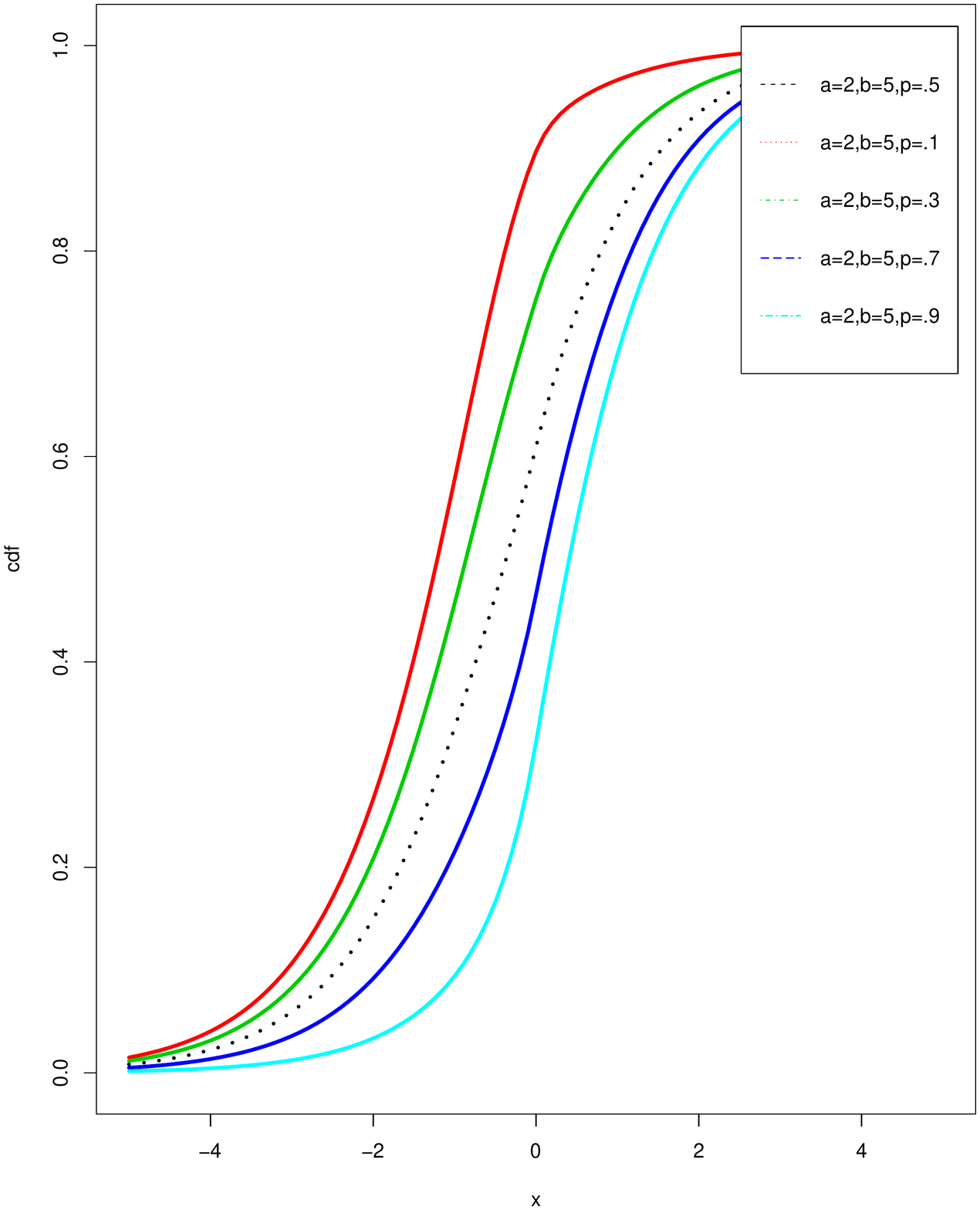}
		\caption{Selected CDFs of $BML$ distributions with $\alpha=2,\beta=5,$ and selected values of $p$.}
		\label{6.4}
	\end{figure}

\begin{figure}[H]
		\centering
		\includegraphics[width=.8\textwidth]{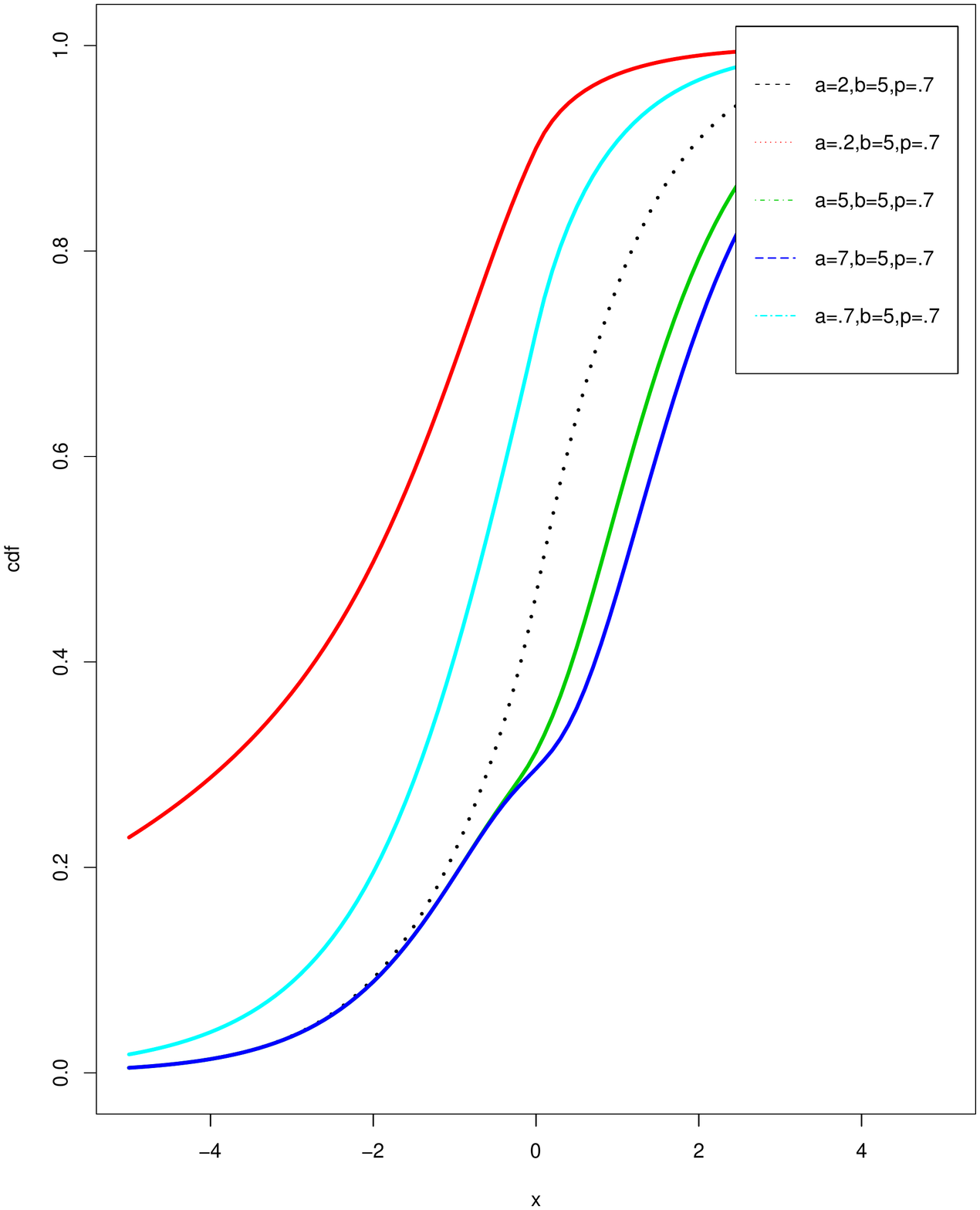}
		\caption{Selected CDFs of $BML$ distributions with $\beta=5, p=.7$, and selected values of $\alpha$.}
		\label{6.5}
	\end{figure}

\begin{figure}[H]
		\centering
		\includegraphics[width=.8\textwidth]{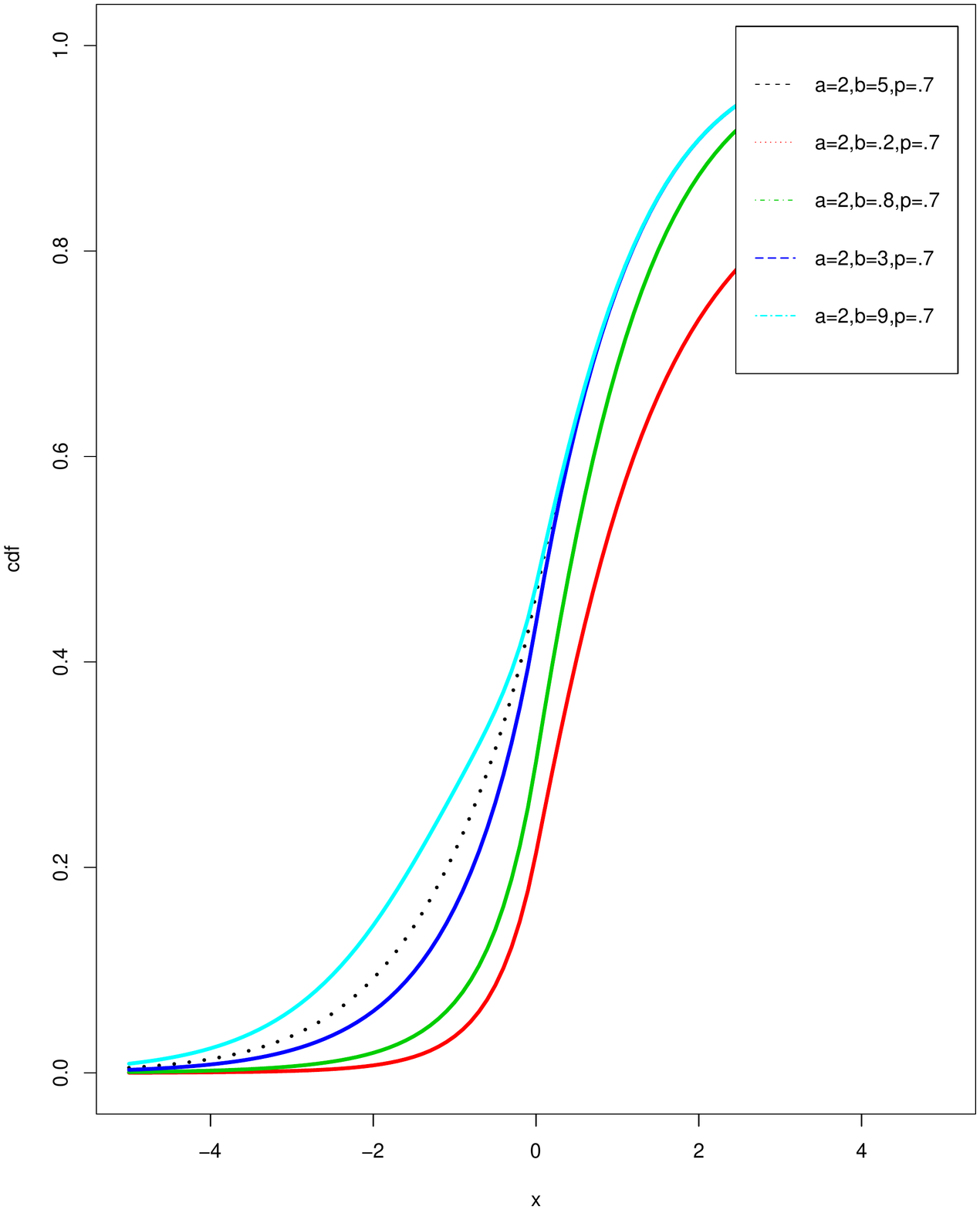}
		\caption{Selected CDFs of $BML$ distributions with $\alpha=2, p=.7$, and selected values of $\beta$.}
		\label{6.6}
	\end{figure}

	The CDF of the $BML(\alpha,\beta,p)$ distribution can be written as
	\begin{equation}
	\label{5.24}
	G(x)=\begin{cases}
	p(\frac{1}{2}e^{x})^{\alpha}+(1-p)[1-(1-\frac{1}{2}e^{x})^{\beta}], & x<0\\
	p(1-\frac{1}{2}e^{-x})^{\alpha}+(1-p)[1-(\frac{1}{2}e^{-x})^{\beta}], & x\ge 0.\\
	\end{cases}
	\end{equation}
	Figure (\ref{6.4})-(\ref{6.6}) shows the CDFs with selected parameters.
	
	%	\section{Limiting Properties of Generalized Asymmetric Laplace distribution}
	%	For both $x <0$ and $x \ge 0$ we can say,
	%	\begin{equation}
	%	\label{lim}
	%	\lim_{x\to 0} g(x)= p \alpha (1/2)^{\alpha}+(1-p) \beta (1/2)^{\beta} 
	%	\end{equation}
	%	\justify
	%	and for other cases, lets say x goes to $\pm \infty$,
	%	
	%	\begin{equation}
	%	\label{lim1}
	%	\lim_{x\to \infty} g(x)=\begin{cases}
	%	\infty & x<0 \\
	%	0 & x\ge 0
	%	\end{cases}
	%	\end{equation}
	%	\justify
	%	Similarly,
	%	
	%	\begin{equation}
	%	\label{lim2}
	%	\lim_{x\to -\infty} g(x)=\begin{cases}
	%	0 & x<0 \\
	%	\infty & x\ge 0
	%	\end{cases}
	%	\end{equation}
	%	\justify
	%	
	\section{Special Cases of the $BML$ Model}
	Let us note several special cases of the $BML$ model. If $p=1$, the the CDF (\ref{5.24}) reduces to 
	\begin{equation}
	\label{5.25}
	G_1(x)=	\begin{cases}
	\left(\frac{1}{2}e^{x}\right)^{\alpha},  & x<0\\
	\left(1-\frac{1}{2}e^{-x}\right)^{\alpha},  & x \ge 0,\\
	\end{cases}
	\end{equation}
	and the distribution coincides with beta-Laplace distribution, studied in section (3.5), built upon beta skewing mechanism with $\beta=1$. It also coincide with Kumaraswamy-Laplace distribution, studied in section (4.4), built upon Kumaraswamy skewing mechanism with $a=\alpha$ and $b=1$.
	
	Similarly, if $p=0$, the CDF (\ref{5.24}) turns into
	\begin{equation}
	\label{5.26}
	G_2(x)=	\begin{cases}
	1-\left(1-\frac{1}{2}e^{x}\right)^{\beta},  & x<0\\
	1-\left(\frac{1}{2}e^{-x}\right)^{\beta},  & x \ge 0.\\
	\end{cases}
	\end{equation}
	This turns out to be a special case of beta-Laplace and Kumaraswamy-Laplace distributions as well, with $\alpha=1$ (in the beta-Laplace case) and $a=1$ (in the Kumaraswamy-Laplace case).\\
	\textbf{Remark}: In the general case $p\in [0,1]$, the $BML$ distribution is a mixture of these two distributions, that is the $BML$ CDF (\ref{5.24}) can be written as
	\begin{equation}
	\label{5.27}
	G(x)=p\; G_1(x)+(1-p)\;G_2(x),
	\end{equation}
	with the above $G_1$ and $G_2$.\\
	We now turn to the parameters $\alpha$ and $\beta$. First observe that when $\alpha=\beta=1$, then the resulting distribution is the Laplace distribution itself.
	
	Next, if $\alpha=1$, then the resulting PDF
	\begin{equation}
	\label{5.28}
	g(x)=\begin{cases}
	p (\frac{1}{2}e^{x})+(1-p) \beta [1-\frac{1}{2}e^{x}]^{\beta-1} \frac{1}{2}e^{x},  & x<0\\
	p(\frac{1}{2}e^{-x})+ (1-p) \beta (\frac{1}{2}e^{-x})^{\beta}, & x\ge 0.\\
	\end{cases}
	\end{equation}
	In addition to $\alpha=1$ we also have $p=1$, we recover the Laplace distribution itself, with the PDF (\ref{5.19}).
	
	Furthre, if $\beta=1$, then the resulting PDF becomes
	\begin{equation}
	\label{5.29}
	g(x)=\begin{cases}
	p \alpha (\frac{1}{2}e^{x})^{\alpha}+(1-p) \frac{1}{2}e^{x},  & x<0\\
	p\alpha (1-\frac{1}{2}e^{-x})^{\alpha-1} \frac{1}{2}e^{-x}+(1-p) \frac{1}{2}e^{-x}, & x\ge 0.\\
	\end{cases}
	\end{equation}
	In addition we also have $p=1$, then (\ref{5.29}) turns into
	\begin{equation}
	\label{5.30}
	g(x)=\begin{cases}
	\alpha (\frac{1}{2}e^{x})^{\alpha},  & x<0\\
	\alpha (1-\frac{1}{2}e^{-x})^{\alpha-1} \frac{1}{2}e^{-x}, & x\ge 0.\\
	\end{cases}
	\end{equation}
	If instead $p=0$, we obtain the standard Laplace distribution.
	
		\section{Series Representations}
		From the generalized binomial theorem, we know that, if $\alpha$ is a positive real number and $|v| < 1$, then 
		$$ (1-v)^{\alpha}=\sum_{i=0}^{\infty}(-1)^{i} \binom{\alpha}{i} v^{i} .$$
	    Applying this theorem to a CDF $F$ or $L-F$, where $F$ is strictly between 0 and 1, we can get the following series expression of the PDF $g(x)$ in (\ref{5.18}), assuming that $\alpha,\beta \ge 1$ :
		\begin{equation*} 
		\begin{split}
		g(x) & =[p \alpha F(x)^{\alpha-1}+(1-p) \beta (1-F(x))^{\beta-1}] f(x) \\
		& = p \alpha f(x) F(x)^{\alpha-1}+(1-p) \beta f(x) (1-F(x))^{\beta-1} \\
		& = p \alpha f(x) F(x)^{\alpha-1}+(1-p) \beta f(x) \sum_{i=0}^{\infty}(-1)^i \binom{\beta-1}{i} F(x)^i \\
		& = p \alpha f(x) [1-{1-F(x)}]^{\alpha-1}+(1-p) \beta f(x) \sum_{i=0}^{\infty}(-1)^i \binom{\beta-1}{i} F(x)^i \\
		& = p \alpha f(x) \sum_{j=0}^{\infty} (-1)^j \binom{\alpha-1}{j}[1-F(x)]^j+(1-p) \beta f(x) \sum_{i=0}^{\infty}(-1)^i \binom{\beta-1}{i} F(x)^i \\
		& = p \alpha f(x) \sum_{j=0}^{\infty} (-1)^j \binom{\alpha-1}{j} \sum_{k=0}^{j} (-1)^k \binom{j}{k} F(x)^k +(1-p) \beta f(x) \sum_{i=0}^{\infty}(-1)^i \binom{\beta-1}{i} F(x)^i \\
		&= f(x) [\sum_{j=0}^{\infty} \sum_{k=0}^{j} w_{j,k} F(x)^k + \sum_{i=0}^{\infty} w_i F(x)^i],
		\end{split}
		\end{equation*}
		where $$ w_{j,k}=(-1)^{j+k} p \alpha \binom{\alpha-1}{k} \binom{j}{k} ,$$
		and
		$$w_i=(-1)^i (1-p) \beta \binom{\beta-1}{i}.$$
	\section{Moment Generating Function and Moments }
	We now derive the moment generating function of $BML$ distribution, as well as moments. We begin with the case of the special case $p=1$, in which case the $BML(\alpha,\beta,p)$ distribution is given by the CDF (\ref{5.25}), with the corresponding PDF given by
	\begin{equation}
	\label{5.31}
	g(x)=\begin{cases}
	\alpha (\frac{1}{2}e^{x})^{\alpha},  & x<0\\
	\alpha (1-\frac{1}{2}e^{-x})^{\alpha-1} \frac{1}{2}e^{-x}, & x\ge 0.\\
	\end{cases}
	\end{equation}
	Thus, the moment generating function (MGF) of $X \sim BML(\alpha,\beta,1)$ becomes
	\begin{equation}
	\label{5.32}
	M_1(x)=E\;e^{tx}=\frac{\alpha}{2^{\alpha}}\int_{-\infty}^{0} e^{x(t+\alpha)}dx + \int_{0}^{\infty} \frac{\alpha}{2} e^{-x(1-t)}(1-\frac{1}{2}e^{-x})^{\alpha-1}dx.
	\end{equation}
	The first integral above, which converges whenever $t>-\infty $, becomes
	\begin{equation}
	\label{5.33}
	\int_{-\infty}^{0} e^{x(t+\alpha)}dx=\frac{1}{t+\alpha}, \quad t>-\infty .
	\end{equation}
	The second integral can be handled through a substitution 
	$$u=1-\frac{1}{2}e^{-x},$$
	where after some algebra, it reduces to
	\begin{equation}
	\label{5.34}
	\frac{\alpha}{2^{t}} \int_{1/2}^{1} u^{\alpha-1}(1-u)^{-t} du,
	\end{equation}
	which converges only for $t<1$. Put this all together, we obtain the following expression for the MGF of $X$ 
	\begin{equation}
	\label{5.35}
	M_1(x)=\frac{\alpha}{2^{\alpha}}\frac{1}{(t+\alpha)}+ \frac{\alpha}{2^{t}}\int_{1/2}^{1} u^{\alpha-1}(1-u)^{-t} du,
	\end{equation}
	valid for $-\infty<t<1$. The integral in the above expression generally does not admit an explicit form. However, it can be related to the incomplete beta function ratio 
	\begin{equation}
	\label{5.36}
	I_x(\alpha,\beta)=\frac{1}{B(\alpha,\beta)}\int_{0}^{x} u^{\alpha-1}(1-u)^{\beta-1} du, \quad x\in (0,1),
	\end{equation}
	where
	$$B(\alpha,\beta)=\frac{\Gamma(\alpha) \Gamma(\beta)}{\Gamma(\alpha+\beta)},\; \alpha,\beta>0, $$
	is the beta function. Indeed, we have
	\begin{equation}
	\label{5.37}
	\int_{1/2}^{1} u^{\alpha-1}(1-u)^{-t} du=\int_{0}^{1} u^{\alpha-1}(1-u)^{-t} du-\int_{0}^{1/2} u^{\alpha-1}(1-u)^{-t} du,
	\end{equation}
	and since
	\begin{equation}
	\label{5.38}
	\int_{0}^{1} u^{\alpha-1}(1-u)^{1-t-1} du=B(\alpha,1-t),
	\end{equation}
	we obtain
	\begin{equation}
	\label{5.39}
	\int_{1/2}^{1} u^{\alpha-1}(1-u)^{-t} du=B(\alpha,1-t)\{1-I_{1/2}(\alpha,1-t)\}.
	\end{equation}
	We now insert (\ref{5.39}) into (\ref{5.35}), we get our final expression for the MGF, which is presented in the following result.
	\begin{proposition}
		\label{prop 5.1}
		If $X\sim BML(\alpha,\beta,1)$, then the MGF of $X$ is
		\begin{equation}
		\label{5.40}
		M_1(x)=\frac{\alpha}{2^{\alpha}}\frac{1}{(t+\alpha)}+ \frac{\alpha \; B(\alpha,1-t)}{2^{t}}\{1-I_{1/2}(\alpha,1-t)\}, \quad -\alpha<t<1.
		\end{equation} 
	\end{proposition}
	\textbf{Remark.} In the special case when $\alpha=m \ge 1$ is an integer, the integral in (\ref{5.35}) will admit an explicit form under a substitution 
	$$w=1-u,$$
	followed by the well-known binomial expansion
	\begin{equation}
	\label{5.41}
	(1-w)^{m-1}=\sum_{j=0}^{m-1} \binom{m-1}{j} (-1)^j w^j.
	\end{equation}
	Following straight forward integration of an exponential function, we obtain
	\begin{equation}
	\label{5.42}
	M_1(t)=\frac{1}{2^m}\frac{m}{m+t}+\frac{m}{2}\sum_{j=0}^{m-1}\binom{m-1}{j} (-1)^j \left(\frac{1}{2}\right)^{j}\frac{1}{1+j-t}\; .
	\end{equation}
	By taking the derivatives of the above expression and inserting $t=0$, we can then recover the moments of $X$ of an integer of order $K$, leading to
	\begin{equation}
	\label{5.43}
	E X^k=k! \left\{\frac{(-1)^k}{2^m m^k}+\frac{m}{2} \sum_{j=0}^{m-1} \binom{m-1}{j}\frac{(-1)^j}{2^j (j+1)^{k+1}}\right\}.
	\end{equation}
	We can follow similar calculations in case of $X\sim BML(\alpha,\beta,0)$, where the CDF of $X$ is given by $G_2$ in (\ref{5.26}), with the corresponding PDF
	\begin{equation}
	\label{5.44}
	g_2(x)=\begin{cases}
	\beta \frac{1}{2}e^{x} [1-\frac{1}{2}e^{x}]^{\beta-1} ,  & x<0\\
	\beta (\frac{1}{2}e^{-x})^{\beta}, & x\ge 0.\\
	\end{cases}
	\end{equation}
	We shall omit stnadard derivations leading to the following result,
	\begin{proposition}
		\label{prop5.2}
		If $X\sim BML(\alpha,\beta,0)$, then the MGF of $X$ is
		\begin{equation}
		\label{5.45}
		M_2(x)=\frac{\beta}{2^{\beta}}\frac{1}{(\beta-t)}+ \frac{\beta \; B(\beta,1+t)}{2^{-t}}\{1-I_{1/2}(\beta,1+t)\}, \quad -1<t<\beta.
		\end{equation} 
	\end{proposition}
	\textbf{Remark.} Again, in the special case when $\beta=n \ge 1$ is an integer, the MGF simplifies as follows :
	\begin{equation}
	\label{5.46}
	M_2(t)=\frac{1}{2^n}\frac{n}{n-t}+\frac{n}{2}\sum_{j=0}^{n}\binom{n-1}{j} (-1)^j \left(\frac{1}{2}\right)^{j}\frac{1}{1+j+t},\; -1<t<n .
	\end{equation}
	In turn, by taking derivatives of the above expression (\ref{5.46}) and setting them to zero, we obtain
	\begin{equation}
	\label{5.47}
	E X^k=k! \left\{\frac{1}{2^n n^k}+\frac{n(-1)^k}{2} \sum_{j=0}^{n-1} \binom{n-1}{j}\frac{(-1)^j}{2^j (j+1)^{k+1}}\right\}.
	\end{equation}
	To get the MGF for the general case of on $BML$ distribution with $p\in [0,1]$, we use the mixture representation (\ref{5.27}), which leads to the following result,
	\begin{proposition}
		\label{prop5.3}
		If $X\sim BML(\alpha,\beta,p)$ then the MGF of $X$ is given by
		\begin{equation}
		\label{5.48}
		M(x)=e^{tx}=p M_1(x)+(1-p)M_2(x),
		\end{equation}
		where
		$$-\min(1,\alpha) < x < \min(1,\beta),$$
		and $M_1(x)$ and $M_2(x)$ are given by (\ref{5.40}) and (\ref{5.45}) respectively.
	\end{proposition}
	Similarly, we can get formulas for the moments of a general $X\sim BML(\alpha,\beta,p)$ with integer valued $\alpha$ and $\beta$, by combining expression (\ref{5.43}) and (\ref{5.47}), leading to the following result,
	\begin{proposition}
		\label{prop5.4} 
		If $X\sim BML(\alpha,\beta,p)$, where $m,n \in N$, then
		\begin{equation}
		\begin{aligned}
		\label{5.49}
		E X^k= {} & k! \left\{\frac{p (-1)^k}{2^m m^k}+\frac{1-p}{2^n n^k} \right\} \\
		& + k! \left\{\frac{pm}{2} \sum_{j=0}^{m-1} \binom{m-1}{j}\frac{(-1)^j}{2^j (j+1)^{k+1}} + \frac{(1-p) n(-1)^k}{2} \sum_{j=0}^{n-1} \binom{n-1}{j}\frac{(-1)^j}{2^j (j+1)^{k+1}}\right\}.
		\end{aligned}
		\end{equation}
	\end{proposition}

	\section{Reliability Analysis}
\begin{figure}[H] 
		\centering
		\includegraphics[width=.8 \textwidth]{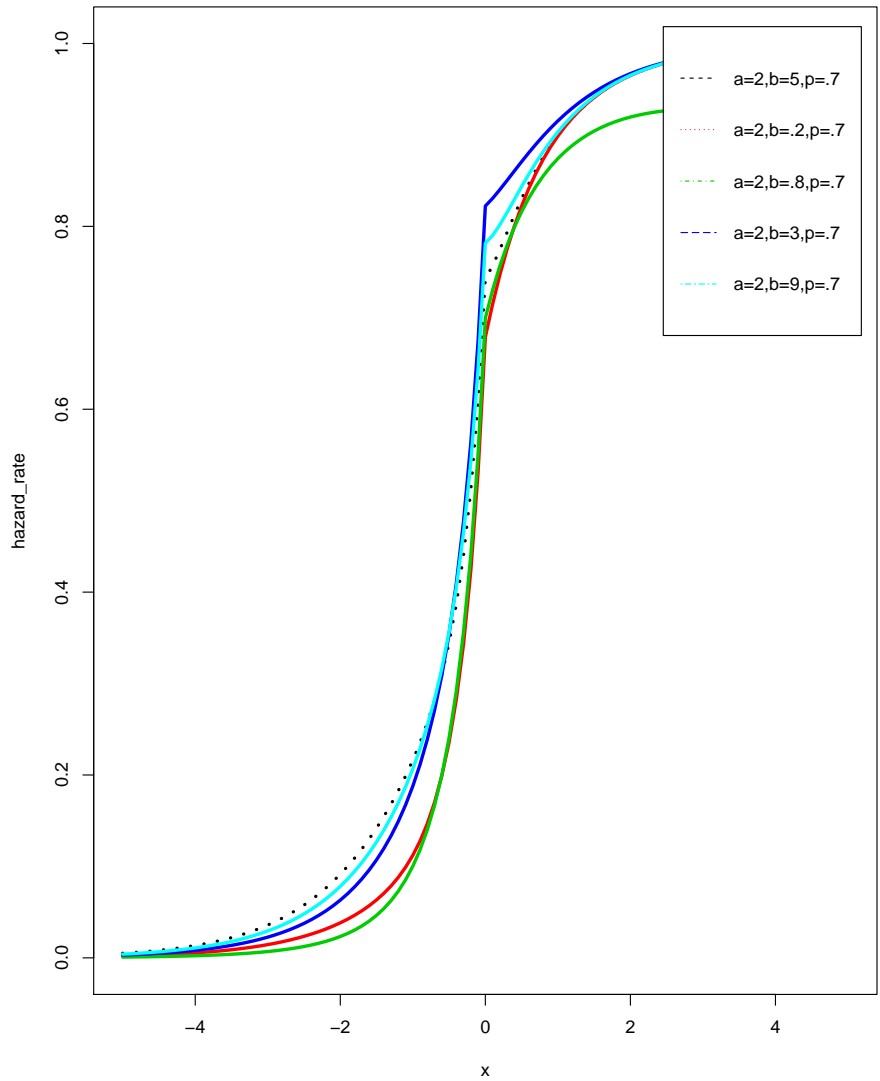}
		\caption{$h(x) $ of $BML$ distribution with $\alpha=2,p=0.7$, and selected values of $\beta$.}
		\label{hazard1}
	\end{figure}

\begin{figure}[H] 
	\centering
	\includegraphics[width=.8 \textwidth]{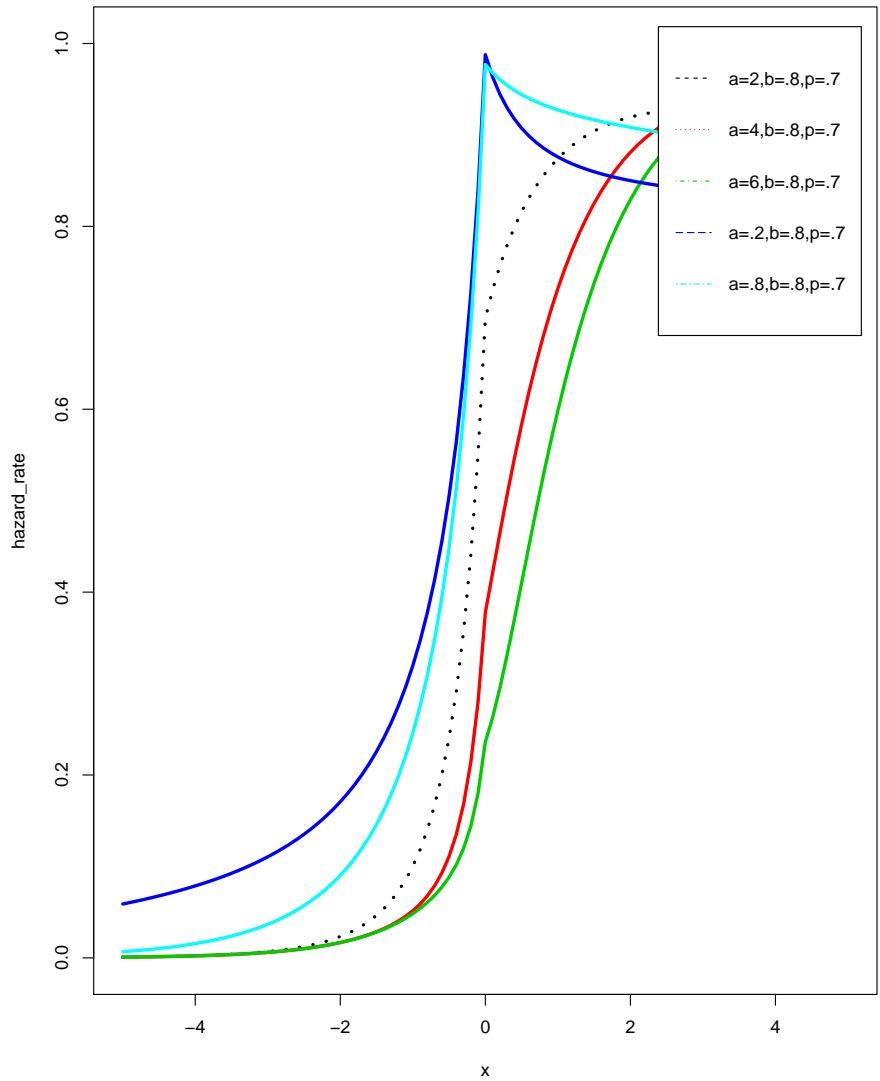}
	\caption{$h(x) $ of $BML$ distribution with $\beta=0.8,p=0.7$, and selected values of $\alpha$.}
	\label{hazard2}
\end{figure}

\begin{figure}[H] 
	\centering
	\includegraphics[width=.8 \textwidth]{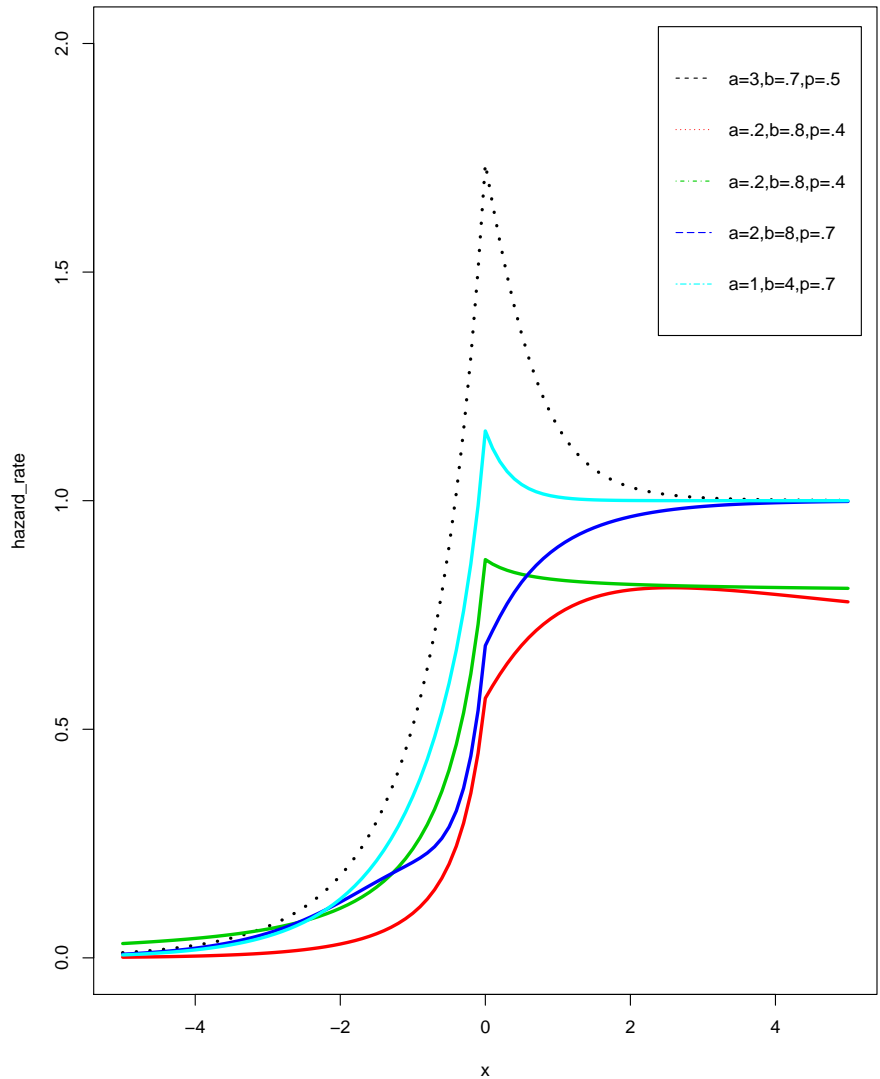}
	\caption{Selected graphs of $h(x)$ for $BML$ distribution with selected parameters.}
	\label{hazard3}
\end{figure}

	The reliability function of the $BML$ distribution, also known as the survival function and denoted by $R(t)=1-G(t)$ and given by
	\begin{equation}
	\label{5.50}
	R(x)=\begin{cases}
	1-\left[p(\frac{1}{2}e^{x})^{\alpha}+(1-p)[1-(1-\frac{1}{2}e^{x})^{\beta}]\right], & x<0\\
	1-\left[p(1-\frac{1}{2}e^{-x})^{\alpha}+(1-p)[1-(\frac{1}{2}e^{-x})^{\beta}]\right], & x\ge 0.\\
	\end{cases}
	\end{equation}
	The hazard rate function, which plays an important role in reliability and other areas, takes on the form
	\begin{equation}
	\label{5.51}
	\begin{split}
	h(x)=\frac{g(x)}{1-G(x)}
	=\begin{cases}
	\frac{p \alpha (\frac{1}{2}e^{x})^{\alpha}+(1-p) \beta [1-\frac{1}{2}e^{x}]^{\beta-1} \frac{1}{2}e^{x}}{1-\left[p(\frac{1}{2}e^{x})^{\alpha}+(1-p)[1-(1-\frac{1}{2}e^{x})^{\beta}]\right]}, & x<0\\
	\frac{p \alpha (1-\frac{1}{2}e^{-x})^{\alpha-1} \frac{1}{2}e^{-x}+(1-p) \beta (\frac{1}{2}e^{-x})^{\beta}}{1-\left[p(1-\frac{1}{2}e^{-x})^{\alpha}+(1-p)[1-(\frac{1}{2}e^{-x})^{\beta}]\right]}, & x\ge 0.\\
	\end{cases}
	\end{split}
	\end{equation}
	The hazard rate function can take a variety of shapes as illustrates in figures (\ref{hazard1})-(\ref{hazard3}).
	
	The limit of the hazard rate function when $x \to 0^{+}$ is as follows :
	\begin{equation}
	\label{5.52}
	\lim_{x\to 0^{+}} h(x)= \frac{p \alpha (\frac{1}{2})^{\alpha}+(1-p) \beta (\frac{1}{2})^{\beta}}{1-\frac{p}{2^\alpha}-(1-p)[1-\frac{1}{2} \beta]}. 
	\end{equation}
	We also see that
	\begin{equation}
	\label{lim1}
	\lim_{x\to \pm \infty} h(x)=0.
	\end{equation}

	\section{Random Variate Generation}
	In this section, we discuss techniques of random variate generation from $BML$ distribution. Since the CDF of this distribution is
	$$G(y)=H(F(y))=p [F(y)]^\alpha+(1-p)[1-(1-F(y)^\beta)],$$
	so, $Y \sim BML(\alpha,\beta,p)$ is a mixture of $BML(\alpha,\beta,1)$ distribution, with the CDF $[F(y)]^\alpha $, and $BML(\alpha,\beta,0)$ distribution, with the CDF $[1-(1-F(y)^\beta)]$.\\
	To generate a random variate $Y_\alpha$, which follows the first distribution, we write $Y_\alpha= F^{-1}(T_1),$ where $T_1\sim Beta(\alpha,1)$ and $F$ is the CDF of Laplace distribution,
	\begin{equation}
	\label{5.54}
	F(x)=\begin{cases}
	\frac{1}{2}e^x, & x<0 \\
	1-\frac{1}{2}e^{-x}, & x\ge 0.
	\end{cases}
	\end{equation}
	The inverse function, $F^{-1}(u)$, can be obtained as follows : \\
	For $u< \frac{1}{2}$, we have
	\begin{equation*}
	\begin{aligned}
	u=F(x)= {} &\frac{1}{2}e^x \\
	e^x & =2u \\
	x & =\log(2u).
	\end{aligned}
	\end{equation*}
	Similarly, for $u> \frac{1}{2}$,
	\begin{equation*}
	\begin{aligned}
	u=F(x)= {} & 1-\frac{1}{2}e^{-x} \\
	\frac{1}{2}e^{-x} & =1-u \\
	e^{-x} & =2(1-u) \\
	x & =-\log(2(1-u)).
	\end{aligned}
	\end{equation*}
	Combining these facts are obtained the quantile function of the Laplace distribution, $Q(u)=F^{-1}(u)$, as follows :
	\begin{equation}
	\label{5.55}
	Q(u)=\begin{cases}
	\log(2u), & 0<u<\frac{1}{2} \\
	-\log(2(1-u)), & \frac{1}{2}\le u<1.
	\end{cases}
	\end{equation}
	To generate a random variate $Y_{\beta}$, which follows the second distribution, we write $Y_{\beta}=F^{-1}(T_2)$, where $T_2\sim Beta(1,\beta)$ and $F(x)$ is as before.\\
	Now, to generate a variate that follows $BML$ distribution, one can generate $Y_{\alpha}$ and $Y_{\beta}$, as discussed above, and the first with probability $p$ and the second one with probability $1-p$. The following algorithm provides the necessary steps. 
	
	\textbf{Random variate generation from $BML(\alpha,\beta)$ distribution}.
	
	Step 1: Generate a standard uniform random variable $U$.\\
	Step 2: If $U<p$, then generate a random variate $B$ from $Beta(\alpha,1)$ distribution. Else, generate random variate $B$ from $Beta(1,\beta)$ distribution.\\
	Step 3: Calculate
	\begin{equation}
	\label{5.56}
	Y=\begin{cases}
	\log(2B), & 0<B<\frac{1}{2} \\
	-\log(2(1-B)), & \frac{1}{2}\le B<1.
	\end{cases}
	\end{equation}
	Step 4: Return $Y$.

	\section{Statistical Inference}
	
	Here we consider the issues of estimating the parameters $\alpha$ and $\beta$ from $BML(\alpha,\beta,p)$ distribution. Let $X_1,X_2,\cdots X_n$ be a random sample from this distribution, so that the PDF is given by
	\begin{equation}
	\label{5.57}
	g(x|\alpha,\beta)=\begin{cases}
	p \alpha (\frac{1}{2}e^{x})^{\alpha}+(1-p) \beta [1-\frac{1}{2}e^{x}]^{\beta-1} \frac{1}{2}e^{x},  & x<0\\
	p \alpha (1-\frac{1}{2}e^{-x})^{\alpha-1} \frac{1}{2}e^{-x}+(1-p) \beta (\frac{1}{2}e^{-x})^{\beta}, & x\ge 0.\\
	\end{cases}
	\end{equation}
	The likelihood function is then of the form
	\begin{equation}
	\label{5.58}
	L(\alpha,\beta)=\prod_{i=1}^{n} g(x_i|\alpha,\beta).
	\end{equation}
	We shall initially assume that the sample size is $n=1$, so that we have just one observation $X$, with the likelihood function of the form :
	\begin{equation}
	\label{5.59}
	L(\alpha,\beta)=\begin{cases}
	p \alpha (\frac{1}{2}e^{x})^{\alpha}+(1-p) \beta [1-\frac{1}{2}e^{x}]^{\beta-1} \frac{1}{2}e^{x},  & x<0\\
	p \alpha (1-\frac{1}{2}e^{-x})^{\alpha-1} \frac{1}{2}e^{-x}+(1-p) \beta (\frac{1}{2}e^{-x})^{\beta}, & x\ge 0.\\
	\end{cases}
	\end{equation}
	Our objective is to find the values of $\alpha$ and $\beta$ that maximize the function $L(\alpha,\beta)$ in (\ref{5.59}).

	\section{A special case $n=1$ }
	Our estimation procedure shall depend on the sign of the single observation $X \sim BML(\alpha,\beta,p)$. Thus, we have two distinct cases, discussed below
	
	\textbf{ Case I: $x < 0$.} In this case, the likelihood function $L(\alpha,\beta)$ in (\ref{5.59}) becomes
	\begin{equation}
	\label{5.60}
	L(\alpha,\beta)=L_1(\alpha)+L_2(\beta),
	\end{equation}
	where
	\begin{equation}
	\label{5.61}
	L_1(\alpha)=p \alpha (\frac{1}{2}e^{x})^{\alpha},
	\end{equation}
	and
	\begin{equation}
	\label{5.62}
	L_2(\beta)= (1-p) \beta (1-\frac{1}{2}e^{x})^{\beta-1}\frac{1}{2}e^{x}.  
	\end{equation}
	Thus, $L(\alpha,\beta)$ can be maximized separately for $\alpha$ and $\beta$, by maximizing $L_1$ with respect to $\alpha$ and maximizing $L_2$ with respect to $\beta$. We start with the problem of maximizing $L_1$ with respect to $\alpha$. By taking the logarithm of both sides of equation (\ref{5.61}) we get,
	\begin{equation}
	\label{5.63}
	\log L_1(\alpha)=\log(p)+\log(\alpha)+ \alpha \log(\frac{1}{2}e^{x}).
	\end{equation}
	Upon differentiating the functions in (\ref{5.63}) with respect to $\alpha$, we obtain
	\begin{equation}
	\label{5.64}
	\frac{d \log (L_1(\alpha))}{d \alpha} =\frac{1}{\alpha}+\log(\frac{1}{2} e^x).
	\end{equation}
	By setting the right hand side of (\ref{5.64}) to zero, we easily find the critical number to be   
	\begin{equation}
	\label{5.65}
	\hat{\alpha}= -\frac{1}{\log(\frac{1}{2} e^x)}.
	\end{equation}
	To determine whether the function has a maximum or minimum value at the critical point, we use the first derivative test. It follows that
	\begin{equation}
	\label{5.66}
	\frac{d \; \log L_1(\alpha)}{d \alpha}  =\frac{1}{\alpha}+\log\left(\frac{1}{2} e^x\right)>0 \quad if \quad
	\alpha < \hat{\alpha},
	\end{equation}
	and 
	\begin{equation}
	\label{5.67}
	\frac{d\; \log L_1(\alpha)}{d \alpha}  =\frac{1}{\alpha}+\log\left(\frac{1}{2} e^x\right)< 0 \quad if \quad \alpha > \hat{\alpha}.
	\end{equation}
	Therefore, the log likelihood function  (\ref{5.63}) is increasing when $\alpha < \hat{\alpha}$ and it is decreasing when $\alpha > \hat{\alpha}$. With the largest value occurring at $\hat{\alpha}$ given by (\ref{5.65}).
	%	There, for each $x_i$, i=1,2,..n, the MLE of $\alpha$ is:
	%	\begin{equation}
	%	\label{zz3}
	%	\hat{\alpha_i}=-\frac{1}{\ln(\frac{1}{2} e^{x_{i}})} ,i=1,2,...n
	%	\end{equation}
	%	\justify
	%	Thus the estimated value of $\alpha$ will be given by
	%	
	%	\begin{equation}
	%	\label{alpha 2}
	%	\hat{\alpha}=\sum_{i=1}^{n} w_i\hat{\alpha_i} ,i=1,2,...n
	%	\end{equation}
	%	\justify
	%%%%%%%%%%%%%%%%%%%%%%%%%%%
	We now turn to estimate $\beta$ based on a single observation of $X=x$. Here, we need to maximize the function $L_2(\beta)$ given by (\ref{5.62}). Upon taking the logarithm on both sides, we get
	\begin{equation}
	\label{5.68}
	\log\; L_2(\beta)=\log(1-p)+\log(\beta)+ (\beta-1) \log(1-\frac{1}{2}e^{x}) +\log(\frac{1}{2}e^{x}).
	\end{equation}
	Now, when we take the derivative of the function in (\ref{5.68}) with respect to $\beta$, we obtain
	\begin{equation}
	\label{5.69}
	\frac{d \; \log (L_2(\beta))}{d \beta} = \frac{1}{\beta}+\log(1-\frac{1}{2} e^{x}).
	\end{equation}
	To find the critical value of $\beta$ for which the likelihood function has max/min we set the function in (\ref{5.69}) equal to zero and solve for $\beta$, which results in a unique value
	\begin{equation}
	\label{5.70}
	\hat{\beta} = -\frac{1}{\ln(1-\frac{1}{2} e^{x})}.
	\end{equation}
	To determine whether the likelihood function has a maximum or minimum at the critical point, we will use the first derivative test and conclude that
	\begin{equation}
	\label{5.71}
	\frac{d\; \log L_2(\beta)}{d \beta}  =\frac{1}{\beta}+\log\left(1-\frac{1}{2} e^{x}\right)>0 \quad if \quad \beta < \hat{\beta},
	\end{equation}
	and 
	\begin{equation}
	\label{5.72}
	\frac{d\; \log L_2(\beta)}{d \beta}  =\frac{1}{\beta}+\log\left(1-\frac{1}{2} e^{x}\right)< 0 \quad if \quad \beta > \hat{\beta}.
	\end{equation}
	Therefore, the function in (\ref{5.68}) is increasing when $\beta < \hat{\beta}$ and it is decreasing when $\beta > \hat{\beta}$, so that the maximum occurs at $\hat{\beta}$.

	\textbf{Case 2: $x > 0$}.
	The procedure is quite similar when the sample value is positive. Here, the likelihood function in (\ref{5.59}) becomes 
	\begin{equation}
	\label{5.73}
	L(\alpha,\beta)=W_1(\alpha)+W_2(\beta),
	\end{equation}
	where
	\begin{equation}
	\label{5.74}
	W_1(\alpha)=p \alpha (1-\frac{1}{2}e^{-x})^{\alpha-1}\frac{1}{2}e^{-x},
	\end{equation}
	and
	\begin{equation}
	\label{5.75}
	W_2(\beta)=(1-p) \beta (\frac{1}{2}e^{-x})^{\beta}. 
	\end{equation}
	Once again, the function $L(\alpha,\beta)$ can be maximized separately for $\alpha>0$ and $\beta>0$. we shall start with estimation of $\alpha$. By taking the logarithm function on both sides of formula (\ref{5.74}) we get,
	\begin{equation}
	\label{5.76}
	\log W_1(\alpha)=\log(p)+\log(\alpha)+ (\alpha-1) \log(1-\frac{1}{2}e^{-x}) +\log(\frac{1}{2}e^{-x}).
	\end{equation}
	The derivative of the above function with respect to $\alpha$ becomes
	\begin{equation}
	\label{5.77}
	\frac{d \log (W_1(\alpha))}{d \alpha} =\frac{1}{\alpha}+\log \left(1-\frac{1}{2} e^{-x}
	\right).
	\end{equation}
	
	To find the critical value of $\alpha$, we set the derivative in (\ref{5.77}) equal to zero and solve for $\alpha$, which results in
	\begin{equation}
	\label{5.78}
	\hat{\alpha} = -\frac{1}{\log(1-\frac{1}{2} e^{-x})}.
	\end{equation}
	%	\justify
	%	leading to
	%	
	%	$$\frac{1}{\alpha}+\ln(1-\frac{1}{2} e^{-x})=0$$
	%	
	%	$$\alpha= -\frac{1}{\ln(1-\frac{1}{2} e^{-x})}$$
	%	
	%	\justify
	%	Therefore, the critical value of the likelihood function is  
	%	
	To determine whether this corresponds to a maximum or a minimum value, we will use the first derivative test. Here,
	\begin{equation}
	\label{5.79}
	\frac{d \log W_1(\alpha)}{d \alpha}  =\frac{1}{\alpha}+\log\left(1-\frac{1}{2} e^{-x}\right)>0 \quad  if \quad
	\alpha < \hat{\alpha}.
	\end{equation}
	and 
	\begin{equation}
	\label{5.80}
	\frac{d \log W_1(\alpha)}{d \alpha}  =\frac{1}{\alpha}+\log\left(1-\frac{1}{2} e^{-x}\right)< 0 \quad if \quad \alpha > \hat{\alpha}.
	\end{equation}
	Therefore, the maximum occurs at $\hat{\alpha}$. \\
	We now consider estimation of $\beta$, where we need to maximize the function $W_2(\beta)$ in (\ref{5.75}).
	%	Therefore,  The log likelihood function  (\ref{alp1}) is increasing when $\alpha < -\frac{1}{\ln(1-\frac{1}{2} e^{-x})}$ and it is decreasing when $\alpha > -\frac{1}{\ln(1-\frac{1}{2} e^{-x})}$ . So the Log-likelihood function (\ref{alp1}) has an extrema  at (\ref{alphahat1}) so does the likelihood function (\ref{alp}).
	%	\justify
	%	There, for each $x_i$, i=1,2,..n, the MLE of $\alpha$ is:
	%	\begin{equation}
	%	\label{alp3}
	%	\hat{\alpha_i}=-\frac{1}{\ln(1-\frac{1}{2} e^{-x_{i}})} ,i=1,2,...n
	%	\end{equation}
	%	\justify
	%	Thus the estimated value of $\alpha$ will be given by
	%	
	%	\begin{equation}
	%	\label{alpha 2}
	%	\hat{\alpha}=\sum_{i=1}^{n} w_i\hat{\alpha_i} ,i=1,2,...n
	%	\end{equation}
	%	\justify
	%	%%%%%%%%%%%%%%%%%%%%%%%%%%%%%%%%%%%%%%%%%%%%%%%%%%%
	%	When $\alpha $ is fixed and $x> 0$, the likelihood function based on a single observation of $X=x$ is 
	%	
	%	\begin{equation}
	%	\label{beta}
	%	L(\beta)=g(x|\beta)=(1-p) \beta (\frac{1}{2}e^{-x})^{\beta}  
	%	\end{equation}
	%	\justify
	Taking the logarithm on both sides of  (\ref{5.75}) we get,
	\begin{equation}
	\label{5.81}
	\log W_2(\beta)=\log(1-p)+\log(\beta)+ \beta \log\left(\frac{1}{2}e^{-x}\right).
	\end{equation}
	Now, we take the derivative of the function (\ref{5.81}) with respect to $\beta$, resulting in
	\begin{equation}
	\label{5.82}
	\frac{d\log (W_2(\beta))}{d \beta} =\frac{1}{\beta}+\log\left(\frac{1}{2} e^{-x}\right).
	\end{equation}
	To find the critical value of $\beta$ for which this function attains max/min, we set the derivative equal to zero and solve for $\beta$, resulting in
	\begin{equation}
	\label{5.83}
	\hat{\beta}= -\frac{1}{\ln(\frac{1}{2} e^{-x})}.
	\end{equation}
	To determine whether this is a maximum or minimum, we use the first derivative test,
	\begin{equation}
	\label{5.84}
	\frac{d\log W_2(\beta)}{d \beta}  =\frac{1}{\beta}+\log\left(\frac{1}{2} e^{-x}\right)>0 \quad if \quad \beta < \hat{\beta}.
	\end{equation}
	and 
	\begin{equation}
	\label{5.85}
	\frac{d\log W_2(\beta)}{d \beta}  =\frac{1}{\beta}+\log\left(\frac{1}{2} e^{-x}\right)< 0 \quad  if \quad \beta > \hat{\beta}.
	\end{equation}
	Therefore, the function is increasing when $\beta < \hat{\beta}$ and it is decreasing when $\beta > \hat{\beta}$. So the function (\ref{5.75}) has a maximum value at $\hat{\beta}$ given by (\ref{5.83}).

	We summarize the above calculations through the following result.
	\begin{proposition}
		\label{prop}
		Let $X$ be a single observation from $BML(\alpha,\beta,p)$ distribution. Then there exist an unique MLEs of $\alpha$ and $\beta$, and they are given by
		\begin{equation}
		\label{5.86}
		\hat{\alpha}=\begin{cases}
		-\frac{1}{\log\left(\frac{e^x}{2}\right)}, & x<0  \\
		-\frac{1}{\log\left(1-\frac{1}{2} e^{x}\right)}, & x\ge 0. 
		\end{cases}
		\end{equation}
		and 
		\begin{equation}
		\label{5.87}
		\hat{\beta}=\begin{cases}
		-\frac{1}{\log\left(1-\frac{e^x}{2}\right)}, & x<0  \\
		-\frac{1}{\log\left(\frac{1}{2} e^{x}\right)}, & x\ge 0. 
		\end{cases}
		\end{equation}
	\end{proposition}
	
	\section{General Case}
	We now go back to the general case, where we have a random sample $X_1,X_2, \cdots, X_n$ from $BML(\alpha,\beta,p)$ distribution. The likelihood function is of the form (\ref{5.58}) with $g(\cdot)$ given by (\ref{5.57}). Since maximum likelihood estimation involving maximization of (\ref{5.58}) is neither complex, we suggest a recommendation of Hossain et al. (2016) and combine the estimates obtained from individual $\{X_i\}$ into a weighted average to obtain the final estimate. Specifically our procedure is as follows :
	
	Step 1: Estimate $\alpha$ and $\beta$ according to (\ref{5.86})-(\ref{5.87}) in Proposition (\ref{prop}) from each sample point $X_i$, leading to $n$ pairs
	\begin{equation}
	\label{5.88}
(\hat{\alpha}_1,\hat{\beta}_1),\; (\hat{\alpha}_2,\hat{\beta}_2),\cdots,(\hat{\alpha}_n,\hat{\beta}_n).
	\end{equation}
	Step 2: Combine the estimates (\ref{5.88}) via the weighted averages,
	\begin{equation}
	\label{5.89}
\hat{\alpha}=\sum_{i=1}^{n} \hat{\alpha}_i w_i,
	\end{equation}
	\begin{equation}
	\label{5.90}
\hat{\beta}=\sum_{i=1}^{n} \hat{\beta}_i w_i,
	\end{equation}
	where the weights, proportional to the likelihood evaluated at the $i^{th}$ estimate, are given by
	\begin{equation}
	\label{5.91}
w_i=\frac{L(\hat{\alpha}_i,\hat{\beta}_i)}{\sum\limits_{j=1}^{n} L(\hat{\alpha}_j,\hat{\beta}_j)} ,
	\end{equation}
	with $L$ as in (\ref{5.58}). 
	
	\section{Simulation Result}
	The following tables from (\ref{tab1})-(\ref{tab4}) is presented the estimated values of $BML$ distribution using weighted likelihood estimation method.
	
	\justify
	%TABLE STARTED HERE %%%%%%%%%%%%%%%%%%%%%%%%%%%%%%%%%%%%%%%%%%%%%%%%%
	%%% TABLE FOR INTEGER PARAMETER

		\begin{table}[H] 
		\caption{Estimated parameters of $BML$ distribution for $\alpha,\beta \ge 1$.} % title of Table 
		\centering % used for centering table 
		\begin{tabular}{|p{1cm}|p{1cm} | p{1cm}|p{2cm}|p{2cm}|p{1cm}|p{2cm}|p{2cm}|}% centered columns (4 columns) 
			\hline \ %inserts double horizontal lines 
			n & k &  $\alpha$  &  $\hat{\alpha}$ & MSE & $\beta$ & $\hat{\beta}$ & MSE($\hat{\beta}$) \\ [0.6ex] % inserts table 
			%heading 
			\hline  % inserts single horizontal line 
			1 & 10000  & 1  & 1.722293 & 13.21667 & 2 & 8.1652 & 48.8299 \\ 
			[0.6ex]
			\hline  % inserts single horizontal line 
			10 & 10000  & 1 & 1.241041 & 0.1723607 & 2 &  2.510154 & 11.50961 \\ 
			[0.6ex]
			\hline
			50    & 10000  & 1   & 1.114591 & 0.03346608 & 2   & 2.03774 & 0.1127001 \\
			[0.6ex]
			\hline
			80    & 10000  & 1   & 1.095235 & 0.0214523 & 2   & 2.029866 & 0.06523504 \\
			[0.6ex]
			\hline
			100   & 10000   & 1   & 1.087628 & 0.01728452 & 2   & 2.02997 & 0.0508898 \\ 
			[0.6ex]
			\hline
			
			%[1ex] adds vertical space 
			\hline %inserts single line 
		\end{tabular} 
		\label{tab1} % is used to refer this table in the text 
	\end{table}

		\begin{table}[H] 
		\caption{Estimated parameters of $BML$ distribution for $\alpha\ge 1$ and $ 0<\beta<1$.} % title of Table 
		\centering % used for centering table 
		\begin{tabular}{|p{1cm}|p{1cm} | p{1cm}|p{2cm}|p{2cm}|p{1cm}|p{2cm}|p{2cm}|}% centered columns (4 columns) 
			\hline \ %inserts double horizontal lines 
			n & k &  $\alpha$  &  $\hat{\alpha}$ & MSE & $\beta$ & $\hat{\beta}$ & MSE($\hat{\beta}$) \\ [0.6ex] % inserts table 
			%heading 
			\hline  % inserts single horizontal line 
			1 & 10000  & 2  & 41.66873 & 296.9668 & 0.9 & 1.370306 & 5.019549 \\ [0.6ex]
			\hline  % inserts single horizontal line 
			10 & 10000  & 2 & 2.521222 & 1.530615 & 0.9 &  1.149045 & 0.2211706 \\ [0.6ex]
			\hline
			50    & 10000  & 2   & 2.225727 & 0.1730746 & 0.9   & 1.028014 & 0.03258948 \\[0.6ex]
			\hline
			80   & 10000   & 2   & 2.204947 & 0.1131908 & 0.9   & 1.017985 & 0.02332787 \\[0.6ex]
			\hline
			100   & 10000   & 2   & 2.197811 & 0.09450559 & 0.9   & 1.015069 & 0.02066828 \\ [0.6ex]
			\hline
			
			%[1ex] adds vertical space 
			\hline %inserts single line 
		\end{tabular} 
		\label{tab2} % is used to refer this table in the text 
	\end{table}

		\begin{table}[H] 
		\caption{Estimated parameters of $BML$ distribution for $ 0<\alpha<1$ and $\beta\ge 1$.} % title of Table 
		\centering % used for centering table 
		\begin{tabular}{|p{1cm}|p{1cm} | p{1cm}|p{2cm}|p{2cm}|p{1cm}|p{2cm}|p{2cm}|}% centered columns (4 columns) 
			\hline \ %inserts double horizontal lines 
			n & k &  $\alpha$  &  $\hat{\alpha}$ & MSE & $\beta$ & $\hat{\beta}$ & MSE($\hat{\beta}$) \\ [0.6ex] % inserts table 
			%heading 
			\hline  % inserts single horizontal line 
			1 & 10000  & 0.9  & 1.398884 & 13.12406 & 2 & 15.15379 & 259.9138 \\ [0.6ex]
			\hline  % inserts single horizontal line 
			10 & 10000  & 0.9 & 1.142916 & 0.2069947 & 2 & 2.530062 & 1.510668 \\ [0.6ex]
			\hline
			50    & 10000  & 0.9   & 1.028462 & 0.03280372 & 2   & 2.224754 & 0.1729768 \\[0.6ex]
			\hline
			80    & 10000  & 0.9   & 1.019391 & 0.02359852 & 2   & 2.200844 & 0.1093862 \\[0.6ex]
			\hline
			100   & 10000   & 0.9   & 1.016459 & 0.02087915 & 2   & 2.193817 & 0.09244837 \\ [0.6ex]
			\hline
			
			%[1ex] adds vertical space 
			\hline %inserts single line 
		\end{tabular} 
		\label{tab3} % is used to refer this table in the text 
	\end{table}

	\begin{table}[H] 
		\caption{Estimated parameters of $BML$ distribution for $\alpha,\beta < 1$.} % title of Table 
		\centering % used for centering table 
		\begin{tabular}{|p{1cm}|p{1cm} | p{1cm}|p{2cm}|p{2cm}|p{1cm}|p{2cm}|p{2cm}|}% centered columns (4 columns) 
			\hline \ %inserts double horizontal lines 
			n & k &  $\alpha$  &  $\hat{\alpha}$ & MSE & $\beta$ & $\hat{\beta}$ & MSE($\hat{\beta}$) \\ [0.6ex] % inserts table 
			%heading 
			\hline  % inserts single horizontal line 
			1 & 10000  & 0.8  & 39.97228 & 2969750 & 0.9 & 29.29763 & 216377.2\\ [0.6ex]
			\hline  % inserts single horizontal line 
			10 & 10000  & 0.8 & 1.337916 & 1.788733 & 0.9 &  4.045057 & 341.8288 \\ [0.6ex]
			\hline
			50    & 10000  & 0.8   & 1.014848 & 0.0890033 & 0.9   & 2.420546 & 2.714263\\[0.6ex]
			\hline
			80    & 10000  & 0.8   & 0.9973678 & 0.06406065 & 0.9   & 2.36158 & 2.364395 \\[0.6ex]
			\hline
			100   & 10000   & 0.8   & 0.9911185 & 0.05594836 & 0.9   & 2.34409 & 2.264043 \\ [0.6ex]
			\hline
			
			%[1ex] adds vertical space 
			\hline %inserts single line 
		\end{tabular} 
		\label{tab4} % is used to refer this table in the text 
	\end{table}

The estimated values $\hat{\alpha}$ and $\hat{\beta}$ from the tables (\ref{tab1})-(\ref{tab4}) are the average values based on $k=10000$ estimated values. The mean square errors are also calculated on the basis of this k iterations as well.\\
From the above tables, it is clear that the estimated values $\hat{\alpha}$ and $\hat{\beta}$ are converging to the true values of $\alpha$ and $\beta$ as the sample size increases.
	
%\section{Summary And Future Research}
%In this thesis, we considered a family of generalized distribution	

	%%%%%%%%%%%%%%%%%%%%%%%%%%%%%%%%%%%%%%%%%%%
	% Appendices
	% Appendices are just chapters that are numbered A, B, C,...
	% Delete or modify as needed.
	%%%%%%%%%%%%%%%%%%%%%%%%%%%%%%%%%%%%%%%%%%%
	
	\appendix% Starts chapter numbering with A, B, C, etc.
	% Puts  "Appendices" in Table of Contents
	\addtocontents{toc}{\noindent Appendices\par}
	\chapter{R code used in this thesis paper}\label{appA}
	
	\section{Plot the probability density of beta- mixture Laplace distribution using different P values. see figure (\ref{6.1})}
	\begin{verbatim}
	###########################################################################
	## Mixture laplace distribution
	## Probability density plot of beta mixture laplace distribution
	## Input : x, alpha, beta, p
	## Output: density Plot for different P values when alpha and beta are fixed 
	###########################################################################
	
	density <- function(x, alpha,beta,p) {
	if(x < 0) {
	p*alpha*(0.5*exp(x))^(alpha)+(1-p)*beta*(1-0.5*exp(x))^(beta-1)*(0.5*exp(x))
	} else if(x >= 0) {
	p*alpha*(1-0.5*exp(-x))^(alpha-1)*(0.5*exp(-x))+(1-p)*beta*(0.5*exp(-x))^(beta)
	}
	}
	
	#x <- seq(-1,1,0.1)
	x <-seq(-5,5,0.1)
	plot(x, sapply(x, density, alpha=2, beta=3,p=0.5),ylim=c(0,0.5),type = "l" ,
	lwd=3,lty=3)
	lines(x, sapply(x, density, alpha=2, beta=3,p=0.1),type = "l" ,lwd=3,col="2")
	lines(x, sapply(x, density, alpha=2, beta=3,p=0.3),type = "l" ,lwd=3,col="3")
	lines(x, sapply(x, density, alpha=2, beta=3,p=0.7),type = "l" ,lwd=3,col="4")
	lines(x, sapply(x, density, alpha=2, beta=3,p=0.9),type = "l" ,lwd=3,col="5")
	legend("topright",inset=0.02,legend=c("a=2,b=3,p=.5","a=2,b=3,p=.1","a=2,b=3,
	p=.3","a=2,b=3,p=.7","a=2,b=3,p=.9"),col=c("black","2","3","4","5"),lty=2:6)
	
	\end{verbatim}
	
	\section{Plot the probability density of beta- mixture Laplace distribution using different  $\alpha$ values. see figure (\ref{6.2})}
	
	\begin{verbatim}
	###########################################################################
	## Mixture laplace distribution
	## Probability density plot of beta mixture laplace distribution
	## Input : x, alpha, beta, p
	## Output: density Plot for different alpha values when p and beta are fixed 
	###########################################################################
	
	
	density <- function(x, alpha,beta,p) {
	if(x < 0) {
	p*alpha*(0.5*exp(x))^(alpha)+(1-p)*beta*(1-0.5*exp(x))^(beta-1)*(0.5*exp(x))
	} else if(x >= 0) {
	p*alpha*(1-0.5*exp(-x))^(alpha-1)*(0.5*exp(-x))+(1-p)*beta*(0.5*exp(-x))^(beta)
	}
	}
	
	#x <- seq(-1,1,0.1)
	x <-seq(-5,5,0.1)
	plot(x, sapply(x, density, alpha=2, beta=5,p=0.7),ylim=c(0,0.5),type = "l" ,
	lwd=3,lty=3)
	lines(x, sapply(x, density, alpha=.8, beta=5,p=0.7),type = "l" ,lwd=3,col="2")
	lines(x, sapply(x, density, alpha=1, beta=5,p=0.7),type = "l" ,lwd=3,col="3")
	lines(x, sapply(x, density, alpha=.7, beta=5,p=0.7),type = "l" ,lwd=3,col="4")
	lines(x, sapply(x, density, alpha=.9, beta=5,p=0.7),type = "l" ,lwd=3,col="5")
	legend("topright",inset=0.02,legend=c("a=2,b=5,p=.7","a=.8,b=5,p=.7","a=1,b=5,
	p=.7","a=.7,b=5,p=.7","a=.9,b=5,p=.7"),col=c("black","2","3","4","5"),lty=2:6)
	
	\end{verbatim}
	
	\section{Plot the probability density of beta- mixture Laplace distribution using different  $\beta$ values. see figure (\ref{6.3})}
	\begin{verbatim}
	
	###########################################################################
	## Mixture laplace distribution
	## Probability density plot of beta mixture laplace distribution
	## Input : x, alpha, beta, p
	## Output: density Plot for different beta values when alpha and P are fixed 
	###########################################################################
	
	density <- function(x, alpha,beta,p) {
	if(x < 0) {
	p*alpha*(0.5*exp(x))^(alpha)+(1-p)*beta*(1-0.5*exp(x))^(beta-1)*(0.5*exp(x))
	} else if(x >= 0) {
	p*alpha*(1-0.5*exp(-x))^(alpha-1)*(0.5*exp(-x))+(1-p)*beta*(0.5*exp(-x))^(beta)
	}
	}
	
	#x <- seq(-1,1,0.1)
	x <-seq(-5,5,0.1)
	plot(x, sapply(x, density, alpha=2, beta=5,p=0.7),ylim=c(0,.5),type = "l" ,
	lwd=3,lty=3)
	lines(x, sapply(x, density, alpha=2, beta=.2,p=0.7),type = "l" ,lwd=3,col="2")
	lines(x, sapply(x, density, alpha=2, beta=.8,p=0.7),type = "l" ,lwd=3,col="3")
	lines(x, sapply(x, density, alpha=2, beta=3,p=0.7),type = "l" ,lwd=3,col="4")
	lines(x, sapply(x, density, alpha=2, beta=6,p=0.7),type = "l" ,lwd=3,col="5")
	legend("topright",inset=0.02,legend=c("a=2,b=5,p=.7","a=2,b=.2,p=.7","a=2,b=.8,
	p=.7","a=2,b=3,p=.7","a=2,b=6,p=.7"),col=c("black","2","3","4","5"),lty=2:6)
	
	\end{verbatim}
	
	\section{Plot the cumulative distribution function of beta- mixture Laplace distribution using different P values. see figure (\ref{6.4})}
	
	\begin{verbatim}
	###########################################################################
	## Mixture laplace distribution
	## Cumulative distribution function plot of beta mixture laplace distribution
	## Input : x, alpha, beta, p
	## Output: Plot for different P values when alpha and beta are fixed 
	###########################################################################
	
	density <- function(x, alpha,beta,p) {
	if(x < 0) {
	p*(0.5*exp(x))^(alpha)+(1-p)*(1-(1-0.5*exp(x))^(beta))
	} else if(x >= 0) {
	p*(1-0.5*exp(-x))^(alpha)+(1-p)*(1-(0.5*exp(-x))^(beta))
	}
	}
	x=10
	density(x,10,10,0.5)
	#x <- seq(-1,1,0.1)
	x <-seq(-5,5,0.1)
	plot(x, sapply(x, density, alpha=2, beta=5,p=0.5),ylim=c(0,1),type = "l" ,
	lwd=3,lty=3)
	lines(x, sapply(x, density, alpha=2, beta=5,p=0.1),type = "l" ,lwd=3,col="2")
	lines(x, sapply(x, density, alpha=2, beta=5,p=0.3),type = "l" ,lwd=3,col="3")
	lines(x, sapply(x, density, alpha=2, beta=5,p=0.7),type = "l" ,lwd=3,col="4")
	lines(x, sapply(x, density, alpha=2, beta=5,p=0.9),type = "l" ,lwd=3,col="5")
	legend("topright",inset=0.02,legend=c("a=2,b=5,p=.5","a=2,b=5,p=.1","a=2,b=5,
	p=.3","a=2,b=5,p=.7","a=2,b=5,p=.9"),col=c("black","2","3","4","5"),lty=2:6)
	
	\end{verbatim}
	
	\section{Plot the cumulative distribution function of beta- mixture Laplace distribution using different $\alpha$ values. see figure (\ref{6.5})}
	
	\begin{verbatim}
	###########################################################################
	## Mixture laplace distribution
	## Cumulative distribution function plot of beta mixture laplace distribution
	## Input : x, alpha, beta, p
	## Output: Plot for different alpha values when P and beta are fixed 
	###########################################################################
	
	density <- function(x, alpha,beta,p) {
	if(x < 0) {
	p*(0.5*exp(x))^(alpha)+(1-p)*(1-(1-0.5*exp(x))^(beta))
	} else if(x >= 0) {
	p*(1-0.5*exp(-x))^(alpha)+(1-p)*(1-(0.5*exp(-x))^(beta))
	}
	}
	
	#x <- seq(-1,1,0.1)
	x <-seq(-5,5,0.1)
	plot(x, sapply(x, density, alpha=2, beta=5,p=0.7),ylim=c(0,1),type = "l" ,
	lwd=3,lty=3)
	lines(x, sapply(x, density, alpha=.2, beta=5,p=0.7),type = "l" ,lwd=3,col="2")
	lines(x, sapply(x, density, alpha=5, beta=5,p=0.7),type = "l" ,lwd=3,col="3")
	lines(x, sapply(x, density, alpha=7, beta=5,p=0.7),type = "l" ,lwd=3,col="4")
	lines(x, sapply(x, density, alpha=0.7, beta=5,p=0.7),type = "l" ,lwd=3,col="5")
	legend("topright",inset=0.02,legend=c("a=2,b=5,p=.7","a=.2,b=5,p=.7","a=5,b=5,
	p=.7","a=7,b=5,p=.7","a=0.7,b=5,p=0.7"),col=c("black","2","3","4","5"),lty=2:6)
	
	\end{verbatim}
	
	\section{Plot the cumulative distribution function of beta- mixture Laplace distribution using different $\beta$ values. see figure (\ref{6.6})}
	
	\begin{verbatim}
	###########################################################################
	## Mixture laplace distribution
	## Cumulative distribution function plot of beta mixture laplace distribution
	## Input : x, alpha, beta, p
	## Output: Plot for different beta values when alpha and beta are fixed 
	###########################################################################
	
	density <- function(x, alpha,beta,p) {
	if(x < 0) {
	p*(0.5*exp(x))^(alpha)+(1-p)*(1-(1-0.5*exp(x))^(beta))
	} else if(x >= 0) {
	p*(1-0.5*exp(-x))^(alpha)+(1-p)*(1-(0.5*exp(-x))^(beta))
	}
	}
	
	#x <- seq(-1,1,0.1)
	x <-seq(-5,5,0.1)
	plot(x, sapply(x, density, alpha=2, beta=5,p=0.7),ylim=c(0,1),type = "l" ,
	lwd=3,lty=3)
	lines(x, sapply(x, density, alpha=2, beta=.2,p=0.7),type = "l" ,lwd=3,col="2")
	lines(x, sapply(x, density, alpha=2, beta=.8,p=0.7),type = "l" ,lwd=3,col="3")
	lines(x, sapply(x, density, alpha=2, beta=3,p=0.7),type = "l" ,lwd=3,col="4")
	lines(x, sapply(x, density, alpha=2, beta=9,p=0.7),type = "l" ,lwd=3,col="5")
	legend("topright",inset=0.02,legend=c("a=2,b=5,p=.7","a=2,b=.2,p=.7","a=2,b=.8,
	p=.7","a=2,b=3,p=.7","a=2,b=9,p=.7"),col=c("black","2","3","4","5"),lty=2:6)
	
	
	\end{verbatim}
	
	\section{Plot the hazard rate function of beta- mixture Laplace distribution using different parameters see figure (\ref{hazard1})}
	
	\begin{verbatim}
	###########################################################################
	## Mixture laplace distribution
	## hazard rate function plot of beta mixture laplace distribution
	## Input : x, alpha, beta, p
	## Output: Plot for different parameter. 
	###########################################################################
	
	density <- function(x, alpha,beta,p) {
	if(x < 0) {
	(p*alpha*(0.5*exp(x))^(alpha)+(1-p)*beta*(1-0.5*exp(x))^(beta-1)*(0.5*exp(x)))
	/(1-(p*(0.5*exp(x))^(alpha)+(1-p)*(1-(1-0.5*exp(x))^(beta))))
	} else if(x >= 0) {
	(p*alpha*(1-0.5*exp(-x))^(alpha-1)*(0.5*exp(-x))+(1-p)*beta*(0.5*exp(-x))^(beta)) /(1-(p*(1-0.5*exp(-x))^(alpha)+(1-p)*(1-(0.5*exp(-x))^(beta))))
	}
	}
	
	#x <- seq(-1,1,0.1)
	x <-seq(-5,5,0.1)
	plot(x, sapply(x, density, alpha=2, beta=5,p=0.7),ylim=c(0,1),type = "l" ,
	lwd=3,lty=3)
	lines(x, sapply(x, density, alpha=2, beta=6,p=0.9),type = "l" ,lwd=3,col="2")
	lines(x, sapply(x, density, alpha=2, beta=.8,p=0.7),type = "l" ,lwd=3,col="3")
	lines(x, sapply(x, density, alpha=2, beta=3,p=0.7),type = "l" ,lwd=3,col="4")
	lines(x, sapply(x, density, alpha=2, beta=4,p=0.7),type = "l" ,lwd=3,col="5")
	legend("topright",inset=0.02,legend=c("a=2,b=5,p=.7","a=2,b=.2,p=.7","a=2,b=.8,
	p=.7","a=2,b=3,p=.7","a=2,b=9,p=.7"),col=c("black","2","3","4","5"),lty=2:6)
	
	###################################################################
	\end{verbatim}
	
	\section{Simulation study and random number generator. see table (\ref{tab1})-(\ref{tab4})}.
	
	\begin{verbatim}
	###########################################################################
	## Mixture laplace distribution
	## Simulation studt of beta mixture laplace distribution
	## Input : n, k, x, alpha, beta, p
	## Output: estimated alpha and beta and their MSE 
	###########################################################################
	
	## Random number Generation
	
	myfun=function(n,p,alpha,beta){
	u=runif(n,0,1)
	u1=qbeta(u,alpha,1)
	x1=ifelse(u1<0.5,log(2*u1),-log(2*(1-u1))) 
	u2=qbeta(u,1,beta)
	x2=ifelse(u2<0.5,log(2*u2),-log(2*(1-u2)))
	xvalues=c(x1,x2)
	x=ifelse(u<p,x1,x2)
	}
	
	
	## alpha and beta parameter estimation
	
	
	estimate <- function(n,p,alpha,beta,k){
	alphahat=1:k
	betahat=1:k
	for (i in 1:k) {
	x=myfun(n,p,alpha,beta)
	#x=c(-0.2,-0.6,.4,.9)
	a=ifelse(x<0, -1/(log(0.5*exp(x))),-(1/(log(1-(0.5*exp(-x))))))
	b=ifelse(x<0,-(1/(log(1-(0.5*exp(x))))),-(1/(log(0.5*exp(-x)))))
	df <- data.frame(a,b) #cbind(a,b)
	df=as.matrix(df)
	M=matrix(0,length(df[,1]),1)
	for(j in 1:length(df[,1]))
	{
	aa=df[j,1]
	bb=df[j,2]
	#M[j,1]=sum(log(ifelse(x<0,p*aa*(0.5*exp(x))^(aa)+(1-p)*bb*
	(1-(0.5*exp(x)))^(bb-1)*(0.5*exp(x)),p*aa*(1-(0.5*exp(-x)))^(aa-1)*(0.5*exp(-x))
	+(1-p)*bb*(0.5*exp(-x))^(bb))))
	
	M[j,1]=prod(ifelse(x<0,p*aa*(0.5*exp(x))^(aa)+(1-p)*bb*(1-(0.5*exp(x)))^(bb-1)
	*(0.5*exp(x)),p*aa*(1-(0.5*exp(-x)))^(aa-1)*(0.5*exp(-x))+(1-p)*bb*
	(0.5*exp(-x))^(bb)))
	}
	M
	#print(M)
	W=M/(sum(M))
	#print(W)
	alphahat[i]=sum(W*a)
	#print(alphahat)
	betahat[i]=sum(W*b)
	#print(betahat)
	}
	#hist(alphahat,probability=TRUE, main='Histogram of estimated alpha values')
	#hist(betahat,probability=TRUE, main='Histogram of estimated beta values')
	MSE1=(1/k)*sum((alphahat-alpha)^2)
	MSE2=(1/k)*sum((betahat-beta)^2)   
	print('True alpha')
	print(alpha)
	estimated.alpha=mean(alphahat)
	print('Estimated alpha')
	print(estimated.alpha)
	print('MSE of alpha')
	print(MSE1)
	print('True beta')
	print(beta)
	estimated.beta=mean(betahat)
	print('Estimated beta')
	print(estimated.beta)
	print('MSE of beta')
	print(MSE2)
	}
	
	
	\end{verbatim}

\end{document}